\documentclass[%
 reprint,
 amsmath,amssymb,
 aps,
]{revtex4-2}

\usepackage{booktabs}
\usepackage{xcolor}
\usepackage{comment}
\usepackage{graphicx}% Include figure files
\usepackage{dcolumn}% Align table columns on decimal point
\usepackage{bm}% bold math
\usepackage{multirow}

\usepackage{makecell}
\usepackage{subcaption}

\begin{document}

\preprint{APS/123-QED}

\title{The Interplay of Heterogeneity and Product Detachment in Templated Polymer Copying}% Force line breaks with \\

\author{Jeremy E. B. Guntoro}%
\email{jeg16@ic.ac.uk}%
\affiliation{Department of Bioengineering, Imperial College London, Exhibition Road, South Kensington, London}%

\author{Benjamin J. Qureshi}%
\email{bq16@ic.ac.uk}%
\affiliation{Department of Bioengineering, Imperial College London, Exhibition Road, South Kensington, London}%

\author{Thomas E. Ouldridge}%
\email{t.ouldridge@imperial.ac.uk}%
\affiliation{Department of Bioengineering and Centre for Synthetic Biology, Imperial College London, Exhibition Road, South Kensington, London}%

\date{\today}% It is always \today, today,
             %  but any date may be explicitly specified

\begin{abstract}

Templated copolymerization, in which information stored in the sequence of a heteropolymer template is copied into another polymer product, is the mechanism behind all known methods of genetic information transfer. A key aspect of templated copolymerization is the eventual detachment of the product from the template. A second key feature of natural biochemical systems is that the template-binding free energies of both correctly-matched and incorrect monomers are heterogeneous. Previous work has considered the thermodynamic consequences of detachment, and the consequences of heterogeneity for polymerisation speed and accuracy, but the interplay of both separation and heterogeneity remains unexplored. In this work, we investigate a minimal model of templated copying that simultaneously incorporates both detachment from behind the leading edge of the growing copy and heterogeneous interactions.
We first extend existing coarse-graining methods for models of polymerisation to allow for heterogeneous interactions. We then show that heterogeneous copying systems with explicit detachment do not exhibit the subdiffusive behaviour observed in the absence of detachment when near equilibrium. Next, we show that heterogeneity in correct monomer interactions tends to result in slower, less accurate copying, while heterogeneity in incorrect monomer interactions tends to result in faster, more accurate copying, due to an increased roughness in the free energy landscape of either correct or incorrect monomer pairs. Finally, we show that heterogeneity can improve on known thermodynamic efficiencies of homogeneous copying, but these increased thermodynamic efficiencies do not always translate to increased efficiencies of information transfer. 

%\begin{description}
%\item[Usage]
%Secondary publications and information retrieval purposes.
%\item[Structure]
%You may use the \texttt{description} environment to structure your abstract; use the optional argument of the \verb+\item+ command to give the category of each item. 
%\end{description}
\end{abstract}

%Use showkeys class option if keyword
                              %display desired
\maketitle

%\tableofcontents

%Word subscripts to ctrl-f check: pol; pol,eq; eff; Therm; Inf; abs
%regex forQ: Q_\{\\mathbf\{x\},l[+-]?1?\}\(m_
\section{\label{sec:Introduction}Introduction}

Genetic information transfer, through the processes of translation, transcription and replication, underpins all life \cite{Crick1970}. Underlying these processes is the same fundamental motif, wherein a long information-carrying heteropolymer, known as a template, has the information stored in its sequence of monomers copied into another polymer, known as the copy. The copy itself may or may not be composed of the same types of monomers. As demonstrated by the sheer breadth of biological form and function, this templated polymer copying motif is capable of specifically assembling a massive space of structures from a small handful of subunits. To give a brief idea of the scale of the challenge, consider the converse approach of self-assembly without an underlying template. In this counterfactual biology, a cell would need to obtain all its proteins (tens of thousands in humans), in their correct proportions, by mixing together the 20 natural amino acids. This task would be impossible as there is not enough information stored in the interactions between the 20 amino acids to create anything but a mess of partially formed or malformed assemblies \cite{Sartori2020}.

Given the central role of templated polymer copying in biology, understanding its fundamental principles will likely play a crucial role in attempts at building artificial life \cite{Guindani2022} or understanding life's origins \cite{Preiner2020}. A key test of our understanding of these principles is our ability to build synthetic copying systems, without relying on highly-evolved enzymes such as polymerases. Our progress here has been far slower than in template-free self-assembly \cite{Rothemund2006,Woods2019,Ke2012,Mohammed2013,Young2020,Videbaek2022}, highlighting the difficulties that come with engineering templated copying systems. The earliest attempts at building synthetic copiers resulted in copies that  remained stuck to the template, rendering the template unable to catalyze the production of more copies \cite{Tjivikua1990}. Later designs tended to utilize time-varying external conditions such as temperature and pressure to induce separation \cite{Mast2010,Schulman2012,Kreysing2015,NunezVillanueva2019}. 

The need to separate copies from templates is in fact the crux of the difficulty in engineering synthetic polymer copying systems. Accurately producing a state where the correct copy monomers are bound to their correct positions along the template is relatively straightforward, simply requiring a large free-energy advantage for correct monomer pairings \cite{Rothemund2006,Woods2019,Ke2012,Mohammed2013,Young2020,Videbaek2022}. However, the copy and template will then tend to remain bound due to cooperative interactions, preventing the production of further copies. Cells are able to supply the free energy needed to separate chemically, without requiring changes in conditions external to the cell, through the use of ATP-consuming enzymes. This separation occurs after the completion of copying in the case of replication \cite{Meselson1958} or during copying in the case of transcription \cite{Browning2004} and translation \cite{Dever2016}. Mimicking this autonomy without enzymes is challenging, but approaches based on DNA nanotechnology \cite{Garcia2021, Garcia2023, Mukherjee2024} and organic chemistry \cite{Osuna2019,Lewandowski2023} that use the free energy of backbone formation to drive the product off the template have shown some promise.

Limited progress in the development of artificial copying systems highlights a lack of understanding of the basic theory of polymer copying. To remedy this knowledge gap, high-level models have been devised that emphasise the thermodynamic consequences of separation from the template \cite{Ouldridge2017b,Poulton2021,Qureshi2024,Genthon2024}. The fundamental challenge in templated copying is the production of a polymer or ensemble of polymers with low sequence entropy. Although specific bonds form during the production of a copy, they are transient, and thus cannot supply the free energy to compensate for this drop in entropy relative to random sequences. Hence, extra chemical work is required to produce low entropy polymer ensembles and an efficiency can be defined by comparing this chemical work to the free energy stored in the low entropy ensemble 
 \cite{Ouldridge2017b,Poulton2021}. Furthermore, maintaining such an ensemble in a non-equilibrium, low entropy steady state imposes thermodynamic constraints on the network that maintains the steady state \cite{Qureshi2024,Genthon2024}.
 
 More detailed models have also been made that focus on the polymerisation mechanism, typically discrete state Markov models. The simplest ones involve one step per added monomer \cite{Bennett1979,Sartori2013,Gaspard2014,Gaspard2016,Gaspard2017,Poulton2019,Gaspard2021}; more complex ones have multiple steps \cite{Juritz2021} and sometimes added cycles \cite{Hopfield1974,Ninio1975,Rao2015,Murugan2012,Murugan2014,Banerjee2017,Chiuchiu2019,Qureshi2023}. Some models explicitly consider microscopic reversibility  \cite{Bennett1979,Andrieux2008,Sartori2013,Gaspard2014,Gaspard2016,Gaspard2017,Poulton2019,Gaspard2021,Qureshi2023}, while others incorporate irreversible steps \cite{Hopfield1974,Ninio1975,Rao2015,Murugan2014,Banerjee2017,Chiuchiu2019}. Another distinction is whether they assume a symmetry between all correct pairings and, separately, all incorrect pairings (homogeneous) \cite{Hopfield1974,Ninio1975,Bennett1979,Andrieux2008,Sartori2013,Murugan2012,Murugan2014,Rao2015,Gaspard2016,Banerjee2017,Chiuchiu2019,Poulton2019,Juritz2021,Qureshi2023} or consider explicit differences between them (heterogeneous) \cite{Gaspard2017,Gaspard2021}. Finally, some approaches are not preoccupied with copy-template separation \cite{Hopfield1974,Ninio1975,Bennett1979,Andrieux2008,Murugan2012,Murugan2014,Rao2015,Gaspard2014,Gaspard2016,Chiuchiu2019,Gaspard2021}. On the other hand, other works consider models in which the copy explicitly separates from the template \cite{Banerjee2017,Gaspard2016b}, including some that attempt to understand the thermodynamic consequences of separation \cite{Sartori2013, Poulton2019,Juritz2021,Qureshi2023}.

 Although there are multiple theoretical frameworks to treat heterogeneity in the context of separating copolymerization systems \cite{Gaspard2016b, Gaspard2017,Li2019,Li2021} (in particular, \cite{Gaspard2016b} explicitly applies their framework to RNA polymerase), existing work has not considered both separation and heterogeneity in models with thermodynamic self-consistency. Thermodynamically self-consistent models have been constructed for homogeneous systems \cite{Sartori2013, Poulton2019, Juritz2021, Qureshi2023}, but it remains unclear if and how the effects of heterogeneity interplay with the thermodynamic consequences of separation. We fill this gap by incorporating heterogeneity into a known thermodynamically consistent model of polymer copying that incorporates separation from behind the growing tip of the product, analogous to transcription and translation \cite{Poulton2019}. The addition of each monomer in our model consists of multiple steps, following \cite{Juritz2021}. We first introduce a method of coarse-graining the model to a single-step description, allowing existing methods (\cite{Gaspard2017} and \cite{Qureshi2023}) of analysis to be used. Using these methods, we  demonstrate qualitative differences between heterogeneous copying with and without separation. We then identify regimes where heterogeneity facilitates or hinders fast, accurate and efficient copying. 

We organize our work as follows. In Sections \ref{sec:CoarseGrainModel} and \ref{sec:FineGrainModel}, we introduce our coarse-grained and fine-grained Markov copying models with separation. We briefly describe Gaspard's \cite{Gaspard2017} and Qureshi et al's \cite{Qureshi2023} analysis methods in Section \ref{sec:GaspardMethod} and our extended iteration for state visits in Section \ref{sec:VisitIter}. In Section \ref{sec: Parameterization}, we describe the parameter sets we initially consider. We begin our results in Section \ref{sec:Velocity} by comparing velocities of heterogeneous copying with and without separation. In Section \ref{sec:ErrorTime}, we investigate the effect of heterogeneity on error and completion time for our limiting regimes. Finally, in Section \ref{sec:Entropy}, we show that in the low driving regime, the thermodynamic efficiency of copying can be improved by heterogeneity.

\section{Methods}

\subsection{General Coarse-Grained Model of Templated Polymer Copying with Heterogeneity and Product Separation \label{sec:CoarseGrainModel}}

\begin{figure*}
\begin{subfigure}[b]{\textwidth}
  \includegraphics[width=0.8\linewidth]{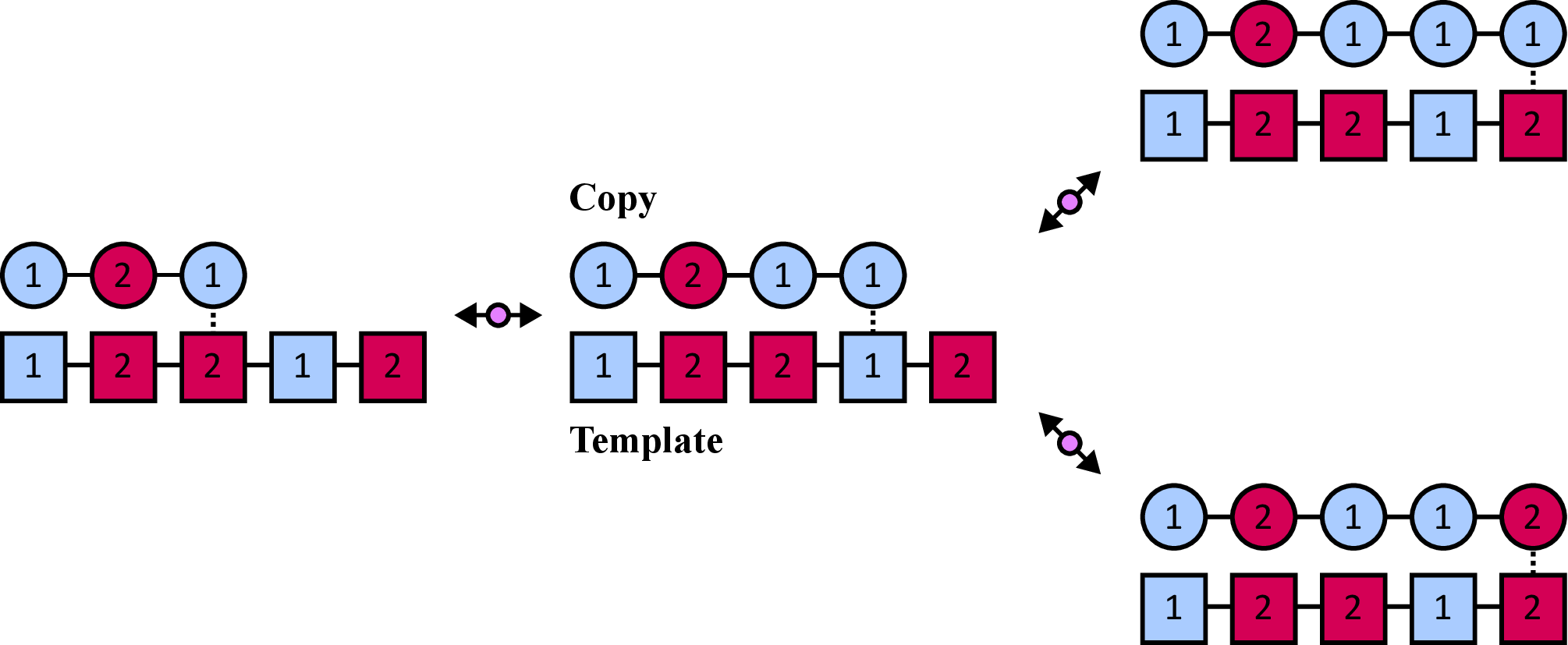}
  \caption{\label{fig:CoarseGrained}}
\end{subfigure}
\begin{subfigure}[b]{\textwidth}
  \includegraphics[width=0.8\linewidth]{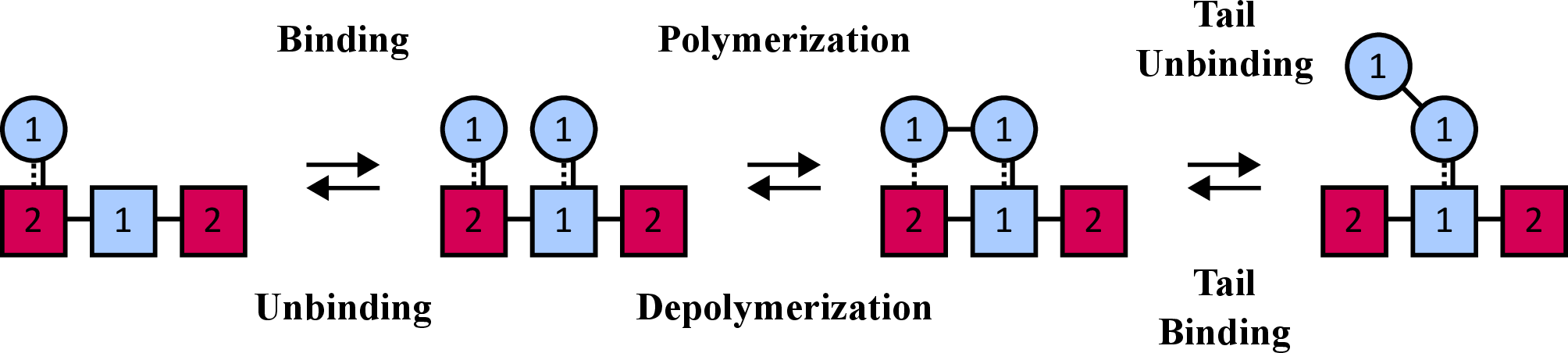}
  \caption{\label{fig:FineGrained}}
  \includegraphics[width=0.8\linewidth]{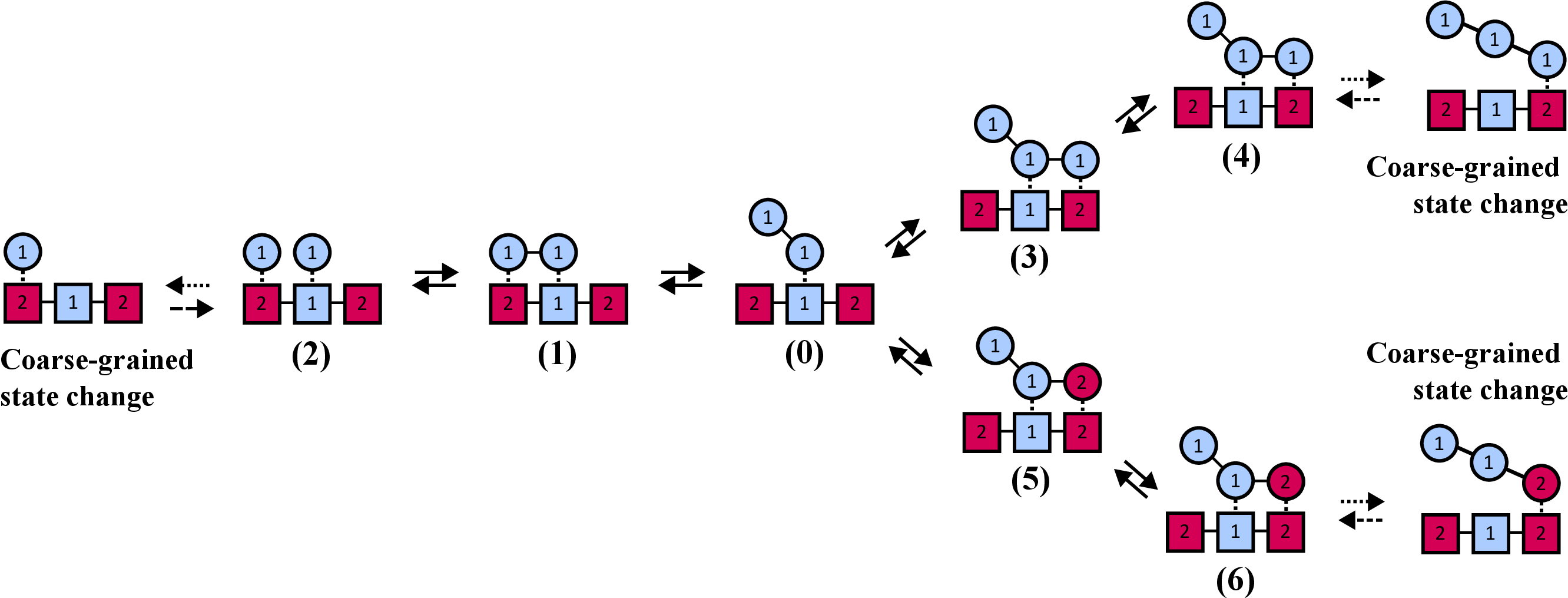}
  \caption{\label{fig:FineGrainedStates}}
\end{subfigure}
\caption{ Polymer copying with continuous separation of the product from the template. (a) Illustration of the coarse-grained model for a two-monomer system. Each monomer type is assigned a unique number (1 or 2 in this case) and color (light blue for 1 and red for 2). Square monomers correspond to the template polymer, while round monomers correspond to the copy polymer. A coarse-grained state is represented by the full sequence of the copy and the template polymers. Purple dotted arrows represent coarse-grained state transitions corresponding to either monomer addition (forward arrow) or monomer removal (backward arrow).  Each coarse-grained transition may be further subdivided into fine-grained steps. (b) Illustration of the fine-grained steps that occur between coarse-grained states. Specifically, the fine-grained steps corresponding to the leftmost purple dot in subfigure (a) are depicted (only monomers in the final 3 positions of the copy and template are depicted). First, a new monomer binds to the template with both a generic (dashed line) and sequence-specific (solid line) bond. Then, the newly bound monomer is covalently incorporated into the growing polymer, breaking the generic bond in the process. Finally, the previously added monomer unbinds from the template. (c) Labelling of fine-grained states for a given coarse-grained state, with indices $f$ denoted by bold bracketed numbers. Arrows here correspond to fine-grained state transitions. Transitions to other states of index $f = 0$ (denoted here by the dotted arrows) change the coarse-grained state $(\mathbf{x},\mathbf{y})$. Note that these fine-grained state indices $f$ are not valid when returning from a different coarse-grained state (via a dashed arrow) as $f$ is dependent on the last completed state visited. For simplicity, general and sequence-specific interactions are not depicted separately in subfigures (a) and (c) }
\end{figure*}

 We consider the coarse-grained model of polymer copying in Figure \ref{fig:CoarseGrained}, well-known in the literature \cite{Andrieux2008,Gaspard2014,Gaspard2017,Poulton2019}. In this section, we will define the model in a general sense, without enforcing separation or thermodynamic consistency (we shall define a parameterization of this model that fulfills these conditions in Section \ref{sec: Parameterization}). The state of the model $(\mathbf{x},\mathbf{y})$ is defined by the two sequences $\mathbf{x} = n_1n_2...n_L$ of the template polymer and $\mathbf{y} = m_1m_2...m_l$ of the copy polymer (Note that $L$ and $l$ are indices of the last monomer in $\mathbf{x}$ and $\mathbf{y}$, respectively). The sequences draw their elements from finite template and copy alphabets $\{1,...,N\}$ and $\{1,2,...,M\}$, respectively. We list a few notational conventions before we proceed. First, for both $\mathbf{x}$ and $\mathbf{y}$, we use $n_{i:i'}$ and $m_{i:i'}$ to refer to the subsequences of $\mathbf{x}$ and $\mathbf{y}$ between indices $i$ and $i'$, inclusive. Second, we will often use $\&$ as a shorthand for $m_{1:l-2}$ for convenience, writing $\mathbf{y} = \&m_{l-1}m_l$.

 From each state $(\mathbf{x},\&m_{l-1}m_l)$, two types of transitions are allowed. Monomer addition at a propensity $\Phi^+(m_{l+1},\mathbf{x},\&m_{l-1}m_l)$ results in the state $(\mathbf{x},\&m_{l-1}m_lm_{l+1})$, while monomer removal at a propensity $\Phi^-(\mathbf{x},\&m_{l-1}m_l)$ results in the state $(\mathbf{x},\&m_{l-1})$ (we use the term propensity at the coarse-grained level to distinguish  $\Phi^{\pm}$ from the rates of the fine-grained process introduced in Section~\ref{sec:FineGrainModel}).  Hence, changes to the template are forbidden in the model, as are monomer additions and removals away from the tip of the copy. From this generic framework, the only  additional constraint imposed is that propensities of monomer addition and removal are only dependent on copy monomers within a small locality of the copy tip, %\Ben{[the props will depend on the template monomers through the tip monomers - These are assuming dependence on only the two monomers at the tip, should specify here that we focus on that]} {\color{purple} [Assumption of template tip dependence is inserted later on in the parameterization; both Gaspard's method and our new extension don't require this so not included here]}
 such that $\Phi^+(m_{l+1},\mathbf{x},\&m_{l-1}m_l) = \Phi^+_l(\mathbf{x},m_lm_{l+1})$ and $\Phi^-(\mathbf{x},\&m_{l-1}m_l) = \Phi^-_l(\mathbf{x},m_{l-1}m_l)$ \cite{Andrieux2008,Gaspard2014,Gaspard2017,Poulton2019}. These forms are physically justified by our mechanism in Section \ref{sec: Parameterization}, but for now it is sufficient to note that these forms are an appropriate first-order approximation for processes like transcription and translation that are composed of reactions that occur in a small neighbourhood of the tip of the growing polymer \cite{Gaspard2014,Gaspard2017}. Hence, local transitions and propensities from $(\mathbf{x},\mathbf{y})$ are fully determined by $(\mathbf{x}, m_{l-1} m_l,l)$, known as the tip state. 

For finite $L$, once the copy reaches the  length of the template $l = L$, the copy has a certain propensity of detaching from the template. Detachment terminates copying, and hence a detached copy is an absorbing state of the model. Formally, we need to distinguish between a complete copy still attached to the template and one completely detached; purely as a trick of notation, we append the integer $0$ to the end of a detached copy sequence (resulting in $l = L+1$).

\subsection{Fine-Grained Reaction Steps \label{sec:FineGrainModel}}
The monomer addition and removal steps in the previous section are not elementary chemical reactions, and in a general model will be described through a series of smaller, fine-grained steps. Properties of the fine-grained system can be preserved in the coarse-grained system if an appropriate coarse-graining procedure is followed (we cover this procedure for our case in Sections \ref{sec:GaspardMethod} and \ref{sec:VisitIter}; refer to Qureshi et al. \cite{Qureshi2023} for more detail). Fine-grained steps for our system follow from earlier work \cite{Juritz2021} on a copying system that separates as it is being copied; in brief, monomer binding, polymerization and unbinding of the copy from the template are each treated as separate steps. Note that although we use the term fine-grained, these states should still be understood as biochemical macrostates \cite{Ouldridge2017c}.

To define the state space, we assume that no more than two monomers are bound to the template at any given time, and fully incorporating a new monomer onto the copy requires breaking the existing template-copy bond. The exact process of monomer addition, summarized in Figure \ref{fig:FineGrained} and Table \ref{table:summary_fine_grain}, consists of three steps. First, a monomer binds to the next empty site on the template. Then, a covalent bond is formed between the new monomer and the tip of the copy. Finally, the previous tip detaches from the template. The reverse processes occur for monomer removal. Copy-template bonds have a generic component  and a sequence-specific component. The generic bond is broken during polymerization; this mechanism was found to reduce product inhibition in \cite{Juritz2021}, and so we incorporate it here. 

 We use the notation $(\mathbf{x},\mathbf{y},f)$ to specify the fine-grained states, consisting of the coarse-grained state as well as an index $f \in \{0,..,6\}$ that specifies the fine-grained state given the coarse-grained state (Figure \ref{fig:FineGrainedStates}). We treat the fine-grained process as a series of sub-processes in which the system transitions between ``completed states'' with $f=0$, via ``transitory states'' with $f \neq 0$, as in \cite{Qureshi2023}. The coarse-grained label $(\mathbf{x},\mathbf{y})$ is given by the completed state that was visited most recently, with $f$ then defined relative to that state.   
 % Entry into fine-grained states corresponding to $f = 0$ (we call such a state a ``completed state" as in \cite{Qureshi2023}, and non-completed states are called ``transitory states") is associated with a change in coarse-grained state. 
Hence, biochemical macrostates with $f \neq 0$ have two different fine-grained state descriptors $\mathbf{y}$ and $f$ depending on whether they were approached ``from the back" or ``from the front". For example, starting from $(\mathbf{x},\&m_l,0)$, binding and then polymerization of a type $1$ monomer would yield $(\mathbf{x},\&m_l,3)$. On the other hand, the same state would be labelled $(\mathbf{x},\&m_l1,1)$ if $(\mathbf{x},\&m_l1,0)$ was the last completed state visited.

\begin{table*}[]
\begin{tabular}{@{}lll@{}}
\toprule
\textbf{Event}                                               & \textbf{Forward Rate} & \textbf{Backward Rate} \\ \midrule
Binding a monomer to the template                            & $K^+_{\textnormal{bind}}(n_{l}n_{l+1},m_{l}m_{l+1})$                     & $K^-_{\textnormal{bind}}(n_{l-1}n_l,m_{l-1}m_l)$ \\
Polymerization of the tip monomer to the newly bound monomer & $K^+_{\textnormal{pol}}(n_{l}n_{l+1},m_{l}m_{l+1})$                     & $K^-_{\textnormal{pol}}(n_{l-1}n_l,m_{l-1}m_l)$ \\
Detachment of the previous tip monomer (tail unbinding)    & $K^-_{\textnormal{tail}}(n_{l}n_{l+1},m_{l}m_{l+1})$                     & $K^+_{\textnormal{tail}}(n_{l-1}n_l,m_{l-1}m_l)$  \\ \bottomrule
\end{tabular}
\caption{A summary of the fine-grained events and their associated rates. \label{table:summary_fine_grain}}
\end{table*}

\subsection{Solving for Distributions of Complete Polymers \label{sec:GaspardMethod}}
 We will cover here and in Section \ref{sec:VisitIter} the analysis methods used in this work. Readers interested only in our physical results may skip to Section \ref{sec: Parameterization}.
 To be able to discuss our methods and results precisely, we must first distinguish between the various stochastic processes that arise from our models and clarify the relationships between them. Let $t$ refer to (continuous) time, and $k$ refer to the discrete number of steps taken (number of state transitions). For a fixed template $X = {\bf x} $, the stochastic processes $Y(t)$ and $F(t)$, giving the coarse-grained and fine-grained index of the copy, respectively, together define  a Markov process $S^\mathbf{x}_F(t) = (Y(t),F(t)|X = \mathbf{x})$. From $S^{\mathbf x}_F(t)$, we can extract an embedded discrete-time Markov chain 
 %stochastic process
 $S^{\mathbf{x}}_F[k]$ representing the sequence of states visited. If we remove any information about the fine-grained index $F(t)$, and only record $Y(t)$, we obtain the continuous, non-Markovian stochastic process $\tilde{S}^{\mathbf{x}}_F(t)$ and the embedded discrete sequence of states $\tilde{S}^{\mathbf{x}}_F[k]$. If we are only interested in the distribution of final copies, it is sufficient to consider $\tilde{S}^{\mathbf{x}}_F[k]$, since it records any change in the composition of the copy. 

There are many alternative ways that $S^{\mathbf x}_F(t)$ can be coarse-grained with a view to avoiding the need of handling non-Markovian processes. Qureshi et al.'s coarse-graining procedure \cite{Qureshi2023} can be applied if the full state space of the system can be partitioned into transitory and completed states such that every transitory state is only encountered during a transition between two unique completed states. In such systems, between any two completed states is a `petal' of transitory states. Each petal between completed states can then be analyzed independently. 

For our model, transitory states $(\mathbf{x},\&m_{l}m_{l+1},f)$ for $f \in \{1,2\}$  (Figure \ref{fig:FineGrainedStates}) are accessible only during transitions from completed state $(\mathbf{x},\&m_{l}m_{l+1},0)$ to $(\mathbf{x},\&m_{l},0)$. All other transitory states are accessible only during transitions from $(\mathbf{x},\&m_{l}m_{l+1},0)$ to $(\mathbf{x},\&m_{l}m_{l+1}m_{l+2},0)$ for some $m_{l+2}$. Hence, the entire set of states and transitions in Figure \ref{fig:FineGrained}, which connect two completed states together, is a `petal' that can be analyzed independently of others. Qureshi et al.'s procedure \cite{Qureshi2023} is based on a graphical interpretation of discrete-state markov processes, where states in a petal are nodes in a directed weighted graph and transitions between states are directed edges in the graph with weights equal to transition rates. Qureshi et al. then consider rooted spanning trees of this graph for a single petal - subgraphs without cycles where the out-degree of every node is equal to 1 except for the root node, which has out-degree zero (we say the spanning tree is ``rooted at" the root node). Given a starting state $(\mathbf{x},\mathbf{y,0})$ and target state $(\mathbf{x},\mathbf{y}^*,0)$, the key quantity $\Lambda^+_{\mathbf{x}}(\mathbf{y,y^*})$ can be calculated by multiplying the rates of all transitions in each spanning tree rooted at $(\mathbf{x},\mathbf{y}^*,0)$ and then summing over these spanning trees. $\Lambda^-_{\mathbf{x}}(\mathbf{y,y^*})$ can be obtained by doing the same over spanning trees rooted at $(\mathbf{x},\mathbf{y},0)$. Summing over spanning trees rooted at $(\mathbf{x},\mathbf{y},0)$ for a modified process where all transitions to $(\mathbf{x},\mathbf{y}^*,0)$ are redirected back to $(\mathbf{x},\mathbf{y},0)$ yields a second key quantity $A_{\mathbf{x}}(\mathbf{y,y^*})$. Qureshi et al.'s procedure \cite{Qureshi2023} assigns the coarse-grained backward and forward propensities from $\mathbf{y}$ to $\mathbf{y^*}$ as follows:
\begin{equation}
\Phi^{\pm}_{\mathbf{x}}(\mathbf{y},\mathbf{y^*}) = \frac{\Lambda^{\pm}_{\mathbf{x}}(\mathbf{y,y^*})}{A_{\mathbf{x}}(\mathbf{y,y^*})}
\label{eq:Ben}
.\end{equation}
This coarse-graining procedure, yielding both the Markov process $S^\mathbf{x}_C(t) = (Y_C(t)|X = \mathbf{x})$ and the embedded sequence of states $S^\mathbf{x}_C[k]$, has two advantages. First, the resultant addition and removal rates obey the constraint $\Phi^+(m_{l+1},\mathbf{x},\mathbf{y}) = \Phi^+_l(\mathbf{x},m_lm_{l+1})$ and $\Phi^-(\mathbf{x},\mathbf{y}) = \Phi^-_l(\mathbf{x},m_{l-1}m_{l})$, facilitating the use of Gaspard's iterated function system \cite{Gaspard2017}, which we will describe shortly. Second, it preserves the distribution of the sequence of coarse-grained  states visited: $S^\mathbf{x}_C[k] = \tilde{S}^\mathbf{x}_F[k]$. Hence, if we are interested in the distribution of complete polymers (defined as the distribution of polymers obtained after complete detachment from the template) and derivative quantities like error rate, then we can work with the Markovian coarse-grained processes  $S^\mathbf{x}_C(t)$ and $S^\mathbf{x}_C[k]$, while completely ignoring the fine-grained steps other than when calculating $\Lambda_{\mathbf{x}}^{\pm}(\mathbf{y,y^*})$ and $A_{\mathbf{x}}(\mathbf{y,y^*})$. However, in general $S^{\bf x}_C(t) \neq \tilde{S}^{\bf x}_F(t)$, which means that polymer completion times and velocities are not preserved by this mapping. In the presence of cycles in the fine-grained steps, free-energy consumption will also not be preserved (this fact is less important for us as our fine-grained system has no cycles). Qureshi et al. describes a method for recovering these properties in the homogeneous case \cite{Qureshi2023}, but we need to derive an additional result (Section \ref{sec:VisitIter}) in order to adapt his method to the heterogeneous case. 

\begin{figure*}
\includegraphics[width=0.8\textwidth]{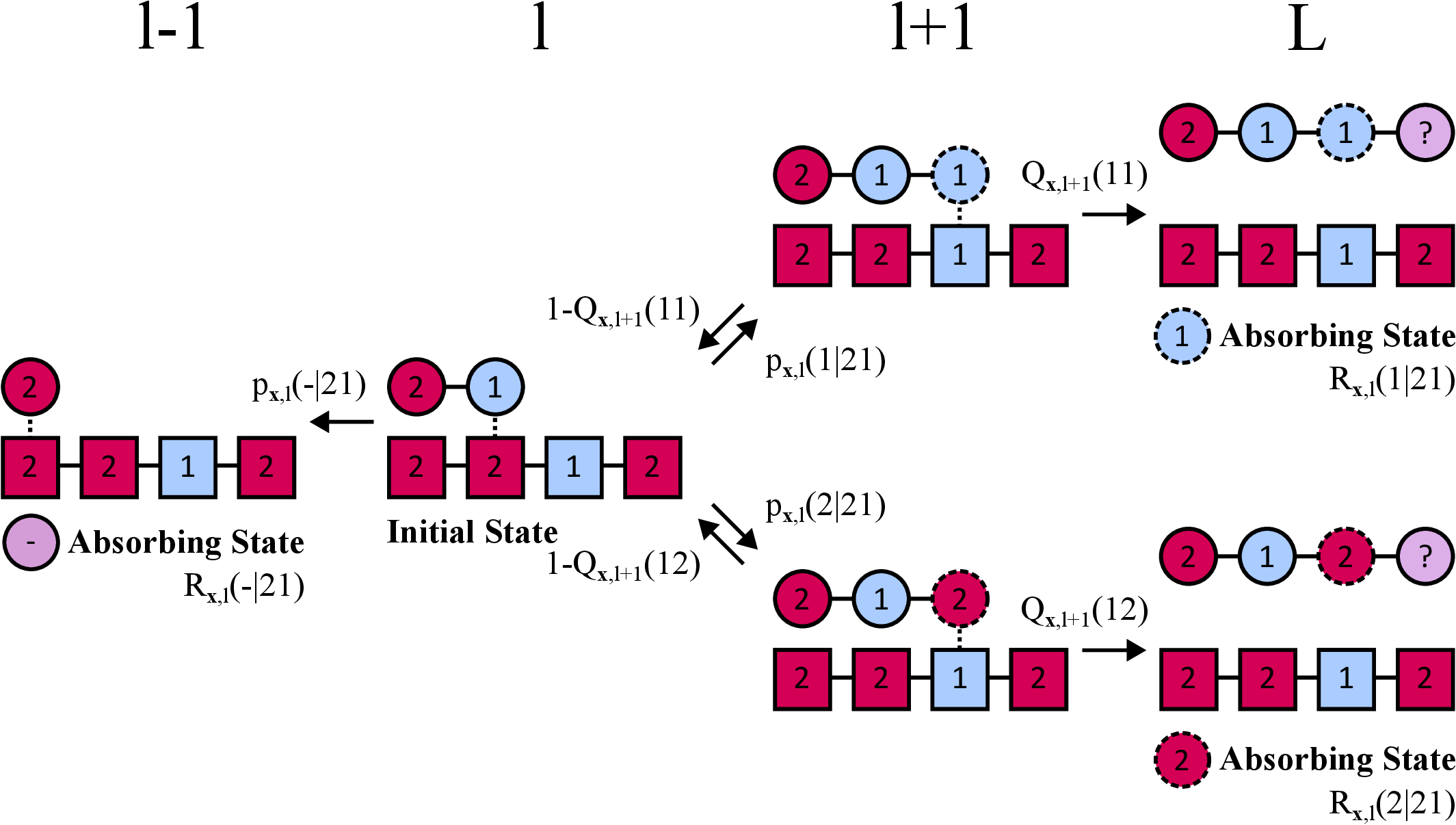}
\caption{Markov chain absorption probability problem for the calculation of complete polymer distributions with a length $l$ product polymer as an initial state. Single arrows (without corresponding reverse counterparts) represent transitions to absorbing states (monomer removal to a copy of length $l-1$ and copy completion to some complete polymer for some given moomer at position $l+1$), and the probability of absorption to each absorbing state is considered. The length of the copy for each state is given at the top of the figure (note the jump from l+1 to L, indicating absorption into a complete and fully detached polymer). The $p_{\mathbf{x},l}(m_{l+1}|m_{l-1}m_l)$ variables are local transition probabilities that can be obtained by considering coarse-grained propensities $\Phi^+_l$ or $\Phi^-_l$. The $Q$ variables are local probabilities of absorption (into some arbitrary complete polymer) occurring before monomer removal. The absorption probabilities $R_{\mathbf{x},l}(m_{l+1}|m_{l-1} m_l)$ follow from values of $Q$ for the next position index $Q_{\mathbf{x},l+1} (m_lm_{l+1})$. The $Q$ values (since these are the probabilities of absorption before removal) for the current iteration step can then be calculated by summing over all the non-backward absorption probabilities, in this case $Q_{\mathbf{x},l}(21) = R_{\mathbf{x},l}(1|21)+R_{\mathbf{x},l}(2|21)$. With this current $Q$ value, the next step of the backward iteration may proceed. \label{fig:Markov}}
\end{figure*}

Having argued that quantities of the fine-grained model can be obtained from solutions of a Markovian coarse-grained model, we now proceed to discuss the method we use to actually solve for quantities of the coarse-grained model. Gaspard and Andrieux showed that the distribution of consecutive copy monomers, conditioned on the entire template, is Markovian if transition rates are only dependent on tip states \cite{Gaspard2014,Gaspard2017}. Moreover, the conditional probability distributions $P_{\mathbf{x},l}(m_l|m_{l-1})$ for complete polymers given a fixed template $\mathbf{x}$ and position indices $l$ can be obtained through the solution of an iterated function system. We discuss here a brief intuition for Gaspard's iterated function system \cite{Gaspard2017} in terms of absorbing probabilities of Markov chains, and perform a full derivation for our system in Appendix \ref{app:Gaspard} (refer to \cite{Gaspard2017} for a full treatment).

Consider a system at a state $(\mathbf{x},\&m_{l-1}m_l)$. We can construct a Markov chain that includes all possible coarse-grained transitions from $(\mathbf{x},\&m_{l-1}m_l)$, including monomer addition to form $(\mathbf{x},\&m_{l-1}m_lm_{l+1})$ and monomer removal to form $(\mathbf{x},\&m_{l-1})$. $(\mathbf{x},\&m_{l-1})$ is treated as an absorbing state. Trajectories that reach each of the monomer-added states $(\mathbf{x},\&m_{l-1}m_lm_{l+1})$ must eventually either  get absorbed into a complete polymer $(\mathbf{x},\&m_{l-1}m_lm_{l+1}..m_L0)$ with a probability $Q$, or undergo monomer removal to $(\mathbf{x},\&m_{l-1}m_l)$ with a probability $1-Q$. Due to the local nature of propensities $\Phi^+$ and $\Phi^-$, $Q$ cannot depend on $\&m_{l-1}$ and can be written as a function $Q_{\mathbf{x},l+1}(m_lm_{l+1})$. Refer to Figure \ref{fig:Markov} for an example. Assuming that we know all the values of $Q_{\mathbf{x},l+1}(m_lm_{l+1})$ for this Markov chain centred on $(\mathbf{x},\&m_{l-1}m_l)$, we can calculate the absorption probability into the backward absorbing state $R_{\mathbf{x},l}(-|m_{l-1} m_l)$ and each of the forward absorbing states $R_{\mathbf{x},l}(m_{l+1}|m_{l-1} m_l)$ through standard methods (details for a generic chain of the form in Figure \ref{fig:Markov} are given in Appendix \ref{app:Gaspard}). Then, 
\begin{equation}
Q_{\mathbf{x},l}(m_{l-1}m_l) = \Sigma_{m_{l+1}} R_{\mathbf{x},l}(m_{l+1}|m_{l-1} m_l)
\label{eq:oneIter}
.\end{equation}

As each $R_{\mathbf{x},l}(m_{l+1}|m_{l-1} m_l)$ is wholly determined by local propensities $\Phi^+_l(\mathbf{x},m_l m_{l+1})$ and $\Phi^-_l(\mathbf{x},m_{l-1}m_l)$ as well as $Q_{\mathbf{x},l+1}(m_lm_{l+1})$, equation \ref{eq:oneIter} implies that $Q_{\mathbf{x},l}(m_{l-1}m_l)$ is obtainable from $Q_{\mathbf{x},l+1}(m_lm_{l+1})$ and the local propensities. From this construction, the entire ensemble of probabilities $Q_{\mathbf{x},l}(m_{l-1}m_l)$ can be obtained through a backward iteration. These variables can in turn be used to obtain conditional monomer probabilities (in the ensemble of complete polymers) $P_{\mathbf{x},l+1}(m_{l+1}|m_l)$ as follows.
\begin{align}
P_{\mathbf{x},l+1}&(m_{l+1}|m_l) \nonumber\\ &= \frac{\Phi^+_l(\mathbf{x},m_lm_{l+1}) Q_{\mathbf{x},l+1}(m_lm_{l+1})} {\Sigma_{m_{l+1}'} \Phi^+_l(\mathbf{x},m_l m_{l+1}') Q_{\mathbf{x},l+1}(m_lm_{l+1}')}  
.\end{align}
This form for $P_{\mathbf{x},l+1}(m_{l+1}|m_l)$ comes from analyzing a modified version of the Markov chain in Figure \ref{fig:Markov} that completely omits the probability of moving back to $(\mathbf{x},\&m_{l-1})$. This procedure works because the rate of production of  complete copies $..m_lm_{l+1}..0$ is the rate of addition of $m_{l+1}$ after $m_l$ such that $m_l$ is never removed again (Appendix \ref{app:Gaspard} and  \cite{Gaspard2017} have further details).

For the very first monomer, analysis of these Markov chains results in a row vector of probabilities $\mathbf{P}_{\mathbf{x},1}$ representing the distribution of monomers in the first index for complete polymers. Conditional distributions for other template positions are defined by matrices $\mathcal{P}_{\mathbf{x},l}$. The absolute distributions $\mathbf{P}_{\mathbf{x},l}$ can therefore be obtained by serial multiplication of the conditional probability matrices as follows.

\begin{equation}
\mathbf{P}_{\mathbf{x},l} = \mathbf{P}_{\mathbf{x},1} \prod_{j=2}^{l} \mathcal{P}_{\mathbf{x},j}
.\end{equation}

\noindent Once we define correct and incorrect monomer pairings, 
these absolute distributions can then be used to define an error probability. In this work, we will consider systems where the copy (excluding the end of copy character $0$) and template monomers draw from the same set $\{1,2\}$ and pairings are considered correct if the copy monomer at a given location equals the template monomer.

\subsection{Recovering Further Properties of the Fine-Grained System from the Coarse-Grained System \label{sec:VisitIter}}

While the probabilities $\mathbf{P}_{\mathbf{x},l}$ can be directly obtained from analysis of the coarse-grained system, other properties of interest such as the average free-energy consumption and time to completion cannot be directly obtained in this manner, as they are dependent on the details of the fine-grained system. However, if an observable $O(\mathbf{x})$ is extensive in the number of visits to coarse-grained tip states $(\mathbf{x},m_{l-1}m_l,l)$ (for instance, the time elapsed and the free energy consumed), then we can evaluate its expectation by first performing a count of the local tip states visited. Let $V_{\mathbf{x},l}(m_{l-1} m_l)$ be a random variable representing the number of visits to a tip state $(\mathbf{x},m_{l-1}m_l,l)$, and let $O_{\mathbf{x},l}(m_{l-1} m_l)$ be a random variable representing the value of the observable $O$ for a given visit of the tip state $(\mathbf{x}, m_{l-1} m_l,l)$. Throughout this paper, we use angled brackets $\langle \rangle$ to refer to per-site averages (i.e. averages over $l$) in the $L \rightarrow \infty$ limit for a stationary distribution of templates $X$ (that is, joint probability distributions of monomers in $X$ at some fixed distance from each other do not change with polymer length; in practice we use Bernoulli template distributions throughout, which are always stationary) and assuming a self-averaging variable. On the other hand, we use the expectation operator $\mathbb{E}[]$ (without any subscripts) acting on a variable with dependencies on $\mathbf{x}$ or $l$ of the form $O_{\mathbf{x},l}$ refers to expected values associated with a particular position $l$ for some fixed $\mathbf{x}$. $\mathbb{E}_V[]$ is an expectation for each tip state visit. Then \cite{Qureshi2023},

\begin{align}
\langle O \rangle = \lim_{L \rightarrow \infty} \frac{1}{L} \Sigma_{l=1}^L\Sigma_{m_{l-1}m_l} &\mathbb{E}[V_{\mathbf{x},l}(m_{l-1} m_l)] \nonumber\\
&\mathbb{E}_V[O_{\mathbf{x},l}(m_{l-1} m_l)]
.\end{align}

\noindent Since our fine-grained model lacks cycles, we do not need to perform this procedure to calculate the free energy consumed for each visit. In Appendix \ref{app:Ben}, we show how the expected waiting time for each visit of a given coarse-grained copy tip state $\mathbb{E}_V[\theta_{\mathbf{x},l}(m_{l-1} m_l)]$ can be calculated for our specific case. 

\begin{figure*}
\includegraphics[width=0.8\textwidth]{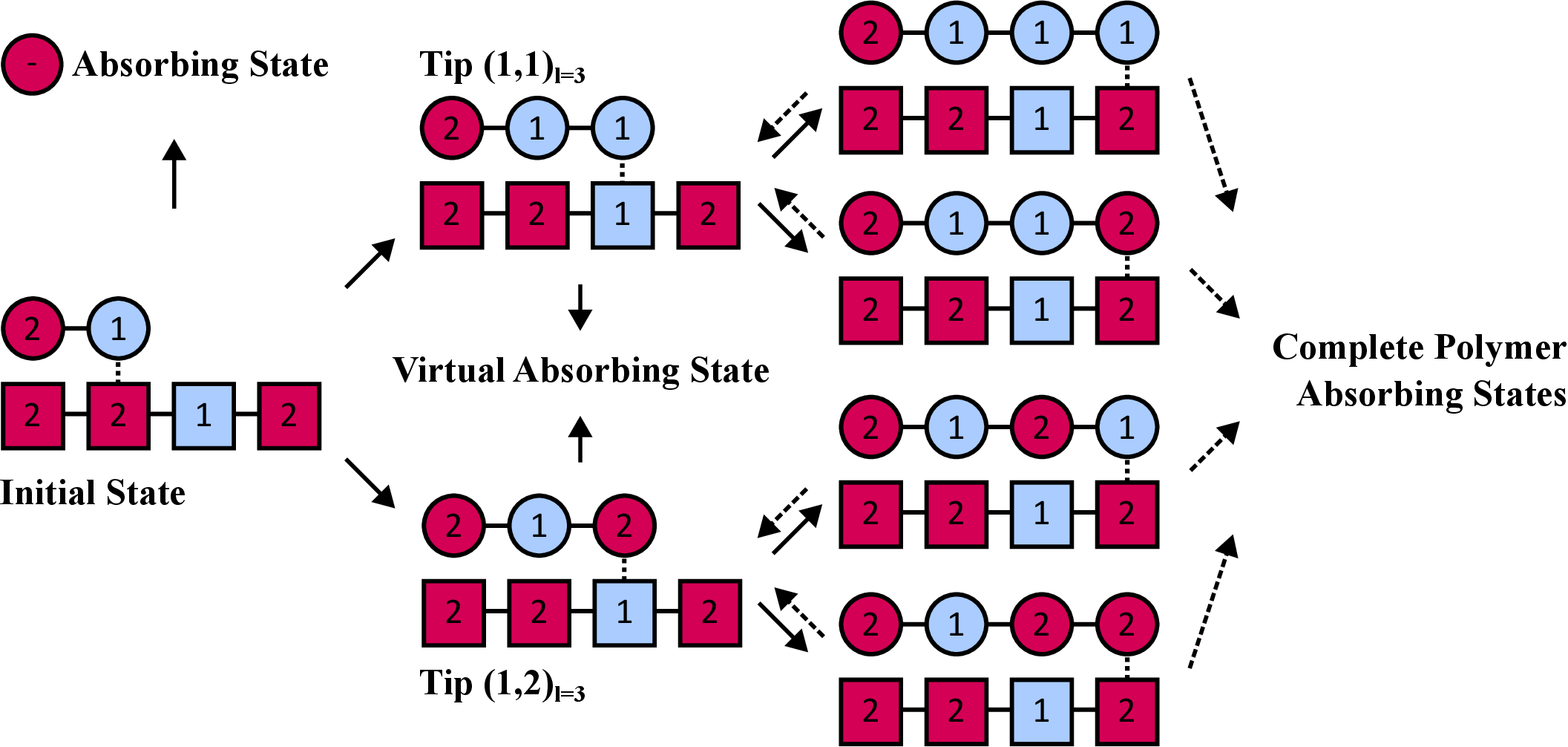}
\caption{ Markov chain from which expected visitation counts per visit of the previous tip state $\mathbb{E}[V_{\mathbf{x},l}(m_l,m_{l+1})|m_{l-1},m_l]$ can be calculated. Solid arrows represent local transition probabilities that can be calculated from coarse-grained rates $\Phi^+$ or $\Phi^-$, while dotted arrows represent the $Q$ variables obtained via Gaspard's iteration \cite{Gaspard2017} (Figure \ref{fig:Markov}). To avoid over-counting, the tip states whose visitation counts are of interest are not permitted to move back into the initial state, rather they point to a virtual absorbing state at a rate equal to the rate of monomer removal. Complete polymers are again treated as absorbing states.  Starting from the initial state, we can now count, through standard methods, the average number of times either tip state of interest (denoted by ``$\textnormal{Tip}(m_l,m_l+1)_l$") is visited before either the initial tip state is revisited or polymerization terminates. \label{fig:VisitationMarkov}}
\end{figure*}

$\mathbb{E}[V_{\mathbf{x},l+1}(m_{l} m_{l+1})]$ can be obtained through the following iteration:
\begin{align}
\mathbb{E}[V_{\mathbf{x},l+1}(m_lm_{l+1})] = \sum_{m_{l-1}} \mathbb{E} [V_{\mathbf{x},l+1}(m_l,m_{l+1})|m_{l-1}m_l] \nonumber \\ \mathbb{E}[V_{\mathbf{x},l}(m_{l-1}m_l)]
\label{eq:forward_iter}
.\end{align}
Here, $\mathbb{E}[V_{\mathbf{x},l} (m_l,m_{l+1})|m_{l-1},m_l]$ is the expected number of visits to tip state $(\mathbf{x}, m_l m_{l+1},l+1)$ from each visit of $(\mathbf{x}, m_{l-1} m_l,l)$. $\mathbb{E}[V_{\mathbf{x},l} (m_l,m_{l+1})|m_{l-1},m_l]$ can be obtained from the Markov chain constructed in Figure \ref{fig:VisitationMarkov}. This Markov chain is similar to that in Figure \ref{fig:Markov} with two important changes. First, instead of considering absorption probabilities from $(\mathbf{x},\&m_{l-1} m_l m_{l+1})$, we further extend the chain to  $(\mathbf{x},\& m_{l-1} m_l m_{l+1} m_{l+2})$ before allowing absorption (these absorption probabilities are again the $Q$ probabilities from Gaspard's iteration in Section \ref{sec:GaspardMethod}), allowing visits to $(\mathbf{x},\&m_{l-1} m_l m_{l+1})$ from $(\mathbf{x},\&m_{l-1} m_l m_{l+1} m_{l+2})$ to be counted. Second, monomer removal from $(\mathbf{x},\&m_{l-1} m_l m_{l+1})$ results in an artificial absorbing state instead of returning to $(\mathbf{x},\&m_{l-1} m_l)$. This approach prevents the double counting of visits to $(\mathbf{x},\& m_{l-1} m_l m_{l+1})$ from distinct visits of $(\mathbf{x},\& m_{l-1} m_l)$. Note that we have not imposed any additional assumptions unique to our system for this iteration to work, and it is applicable to any polymerisation system whose coarse-grained form (applying Qureshi et al.'s procedure \cite{Qureshi2023}) obeys the assumptions in \cite{Gaspard2017}. The derivation of a general form for $\mathbb{E}[V_{\mathbf{x},l+1} (m_lm_{l+1})|m_{l-1}m_l]$ is provided in Appendix \ref{app:VisitIter}.

\subsection{Thermodynamically Consistent Parameterization for Heterogeneous Separating Copiers \label{sec: Parameterization}}

Refer back to Figure \ref{fig:FineGrained} and Table \ref{table:summary_fine_grain} for the rates of our fine-grained system. In this section, we will assign parameters to these fine-grained rates such that thermodynamic consistency is maintained. We first assume that monomer concentrations $[M]$ are equal and chemostatted (unchanging over the copy process). We assume that binding of a monomer to a template obeys mass-action, so $K^+_{\textnormal{bind}}(n_{l-1}n_l,m_{l-1}m_l) = k_{\textnormal{bind}}(n_{l-1}n_l,m_{l-1})[M]$. The rate constants for monomer binding are set to be invariant between the different tip states, as they are chemically difficult to tune, so $K^+_{\textnormal{bind}}(n_{l-1}n_l,m_{l-1}m_l) = k_{\textnormal{bind}}[M]$. Tail binding is similarly difficult to tune and obeys a pseudo-mass action rule where rates are proportional to an effective local concentration of the tail monomer around the tip of the growing polymer, so $K^+_{\textnormal{tail}}(n_{l-1}n_l,m_{l-1}m_l) = k_{\textnormal{tail}}[M]_{\textnormal{eff}}$. Monomer polymerization rates, on the other hand, can conceivably be engineered, and here we envision a parameterization $K^+_{\textnormal{pol}}(n_{l-1}n_l,m_{l-1}m_l) = k_{\textnormal{pol}}(n_{l-1}n_l,m_{l-1}m_l)$.  

Since copying terminates on detachment, there is a net flow of molecules producing complete polymers, so the system does not satisfy a global detailed balance and is thus out of equilibrium. However, each fine-grained reaction step constitutes an elementary reaction step, and hence a generalized local detailed balance condition applies as follows \cite{Peliti2021}.

\begin{equation}
\frac{k^+}{k^-} = \exp\left({\frac{-\Delta G + \delta} {k_{\rm B} T}} \right)
.\end{equation}
Here, $k^+$ is the rate of some elementary forward reaction, $k^-$ is the rate of its corresponding backward reaction, $\Delta G$ is the free-energy change of the forward reaction, $\delta$ is an external driving, $k_{\rm B}$ is Boltzmann's constant and $T$ is the reservoir temperature. This generalized local detailed balance condition, along with the need for the polymer to separate as it copies, enforces constraints on the rates of the fine-grained reactions based on the free-energy changes incurred during transitions. Note that we use negative free energies when defining free-energy parameters (i.e. more positive parameter values correspond to greater stability) and then include the minus signs in the rate expressions. Further, there is a single global heat reservoir, and all free-energy terms are measured in units of $k_{\rm B} T$.

As in reference \cite{Juritz2021}, we assume the binding free energy between monomers $n_l$ and $m_l$ can be divided into a specific monomer-dependent free energy $\Delta G_{TT}(n_l,m_l)$ and a generic monomer-independent free energy $\Delta G_{\textnormal{gen}}$ (Figure \ref{fig:FineGrained}). Coupling the breaking of this generic bond to backbone formation was found to reduce product inhibition and allow detachment \cite{Juritz2021}. Monomer binding incurs a free-energy change of $-\ln{[M]}-\Delta G_{TT}(n_l,m_l)-\Delta G_{\textnormal{gen}}$, polymerization incurs a free-energy change of $-\Delta G_{BB}+\Delta G_{\textnormal{gen}}$ and tail unbinding results in a net free-energy change of $\Delta G_{TT}(n_{l-1},m_{l-1})+\ln{[M]_{\textnormal{eff}}}$. 

Due to the need to separate, copy-template interactions must be transient. Observe that $\Delta G_{\textnormal{gen}}$ cancels during the incorporation of a single monomer. Similarly, the $-\Delta G_{TT}(n_l,m_l)$ free-energy change from monomer binding is offset by the $ \Delta G_{TT}(n_l,m_l)$ change from tail unbinding during the incorporation of the {\it next} monomer. The persistent net polymerization free-energy change per incorporated monomer is then $-\Delta G_{\textnormal{pol}} = -\Delta G_{BB} - \ln{\frac{[M]}{[M]_{\textnormal{eff}}}}$, free of any $\Delta G_{TT}$ or $\Delta G_{\textnormal{gen}}$ terms and hence consistent with transient copy-template interactions. The resulting forward and backward rates, with ratio determined by generalized local detailed balance, are summarized in Table \ref{table:param_fine_grain}. 

\begin{table*}[]
\begin{tabular}{@{}|l|l|l|l|l|@{}}
\toprule
\textbf{Forward Reaction   } & 
\textbf{Forward Rate}                                               & \textbf{Parameterized Form}  & \textbf{Backward Rate} & \textbf{Parameterized Form} \\ \midrule
        Binding & $K^+_{\textnormal{bind}}(n_{l-1}n_l,m_{l-1}m_l)$              
    & $k_{\textnormal{bind}}[M]$ 
    & $K^-_{\textnormal{bind}}(n_{l-1}n_l,m_{l-1}m_l)$                 & $k_{\textnormal{bind}}e^{-\Delta G_{TT}(n_l,m_l)-\Delta G_{\textnormal{gen}}}$ \\
      Polymerization &  $K^+_{\textnormal{pol}}(n_{l-1}n_l,m_{l-1}m_l)$ 
    & $k_{\textnormal{pol}}(n_{l-1}n_l,m_{l-1}m_l)$                
    & $K^-_{\textnormal{pol}}(n_{l-1}n_l,m_{l-1}m_l)$                  & $k_{\textnormal{pol}}(n_{l-1}n_l,m_{l-1}m_l)e^{- \Delta G_{BB}+\Delta G_{\textnormal{gen}}}$ \\
        Tail Unbinding & $K^-_{\textnormal{tail}}(n_{l-1}n_l,m_{l-1}m_l)$   
    & $k_{\textnormal{tail}}e^{-\Delta G_{TT}(n_{l-1},m_{l-1})}$             
    & $K^+_{\textnormal{tail}}(n_{l-1}n_l,m_{l-1}m_l)$                 & $k_{\textnormal{tail}}[M]_{\textnormal{eff}}$                       \\ \bottomrule
\end{tabular}
\caption{Parameterized forms of each of the fine-grained reaction steps after local detailed balance is imposed and further simplifcations are made. \label{table:param_fine_grain}}
\end{table*}

Now that the fine-grained system has been parameterized, we can proceed to define transition propensities \cite{Qureshi2023}. These propensities are derived in full in Appendix \ref{app:Ben}, and the results are given in equations \ref{eq:phiplus} and \ref{eq:phimin}. Due to the forms taken by our fine-grained steps, we can impose a more restrictive condition on locality than required by Gaspard's method: that the propensities are also dependent only on local template monomers, as well as copy monomers, $\Phi^+(m_{l+1},\mathbf{x},\&m_{l-1}m_l) = \Phi^+(n_ln_{l+1},m_lm_{l+1})$ and $\Phi^-(n_{l-1}n_{l},\&m_{l-1}m_l) = \Phi^-(n_{l-1}n_{l},m_{l-1}m_l)$. We find
\begin{widetext}
\begin{align}
\Phi^+(n_{l-1}n_l, &m_{l-1}m_l) \nonumber \\ &= \frac{k_{\textnormal{bind}} k_{\textnormal{pol}}(n_{l-1}n_l,m_{l-1}m_l) k_{\textnormal{tail}} e^{-\Delta G_{TT}(n_l,m_l)} [M]}{ \begin{matrix} e^{-\Delta G_{BB} -\Delta G_{TT}(n_l,m_l)} k_{\textnormal{bind}} k_{\textnormal{pol}}(n_{l-1}n_l,m_{l-1}m_l) +  e^{-\Delta G_{TT}(n_{l-1},m_{l-1}) -\Delta G_{TT}(n_l,m_l) - \Delta G_{\textnormal{gen}}} k_{\textnormal{bind}} k_{\textnormal{tail}} &\\+ e^{-\Delta G_{TT}(n_{l-1},m_{l-1}) }k_{\textnormal{pol}}(n_{l-1}n_l,m_{l-1}m_l) k_{\textnormal{tail}} \end{matrix}} \label{eq:phiplus},\\
\Phi^-(n_{l-1}n_l, &m_{l-1}m_l) \nonumber \\ &= \frac{k_{\textnormal{bind}} k_{\textnormal{pol}}(n_{l-1}n_l,m_{l-1}m_l) k_{\textnormal{tail}} e^{-\Delta G_{BB} -\Delta G_{TT}(n_l,m_l)} [M]_{\textnormal{eff}}}{\begin{matrix} e^{-\Delta G_{BB} -\Delta G_{TT}(n_l,m_l)} k_{\textnormal{bind}} k_{\textnormal{pol}}(n_{l-1}n_l,m_{l-1}m_l) +  e^{-\Delta G_{TT}(n_{l-1},m_{l-1}) -\Delta G_{TT}(n_l,m_l) - \Delta G_{\textnormal{gen}}} k_{\textnormal{bind}} k_{\textnormal{tail}} &\\+ e^{-\Delta G_{TT}(n_{l-1},m_{l-1}) }k_{\textnormal{pol}}(n_{l-1}n_l,m_{l-1}m_l) k_{\textnormal{tail}} \end{matrix} \label{eq:phimin}}
.\end{align}
\end{widetext}

\noindent  The binding rate constants $k_{\textnormal{bind}}$ and $k_{\textnormal{tail}}$ are unlikely to be very different from each other (although binding and unbinding rates can be tuned by the concentration terms), and hence we can set $k_{\textnormal{bind}} = k_{\textnormal{tail}}$. To start with, we assume a sequence-independent $k_{\textnormal{pol}}$, although we later relax this assumption. We begin by investigating the following limits: 

\begin{enumerate}

\item Slow binding and unbinding of the free monomers, $[M] \rightarrow 0$ and $\Delta G_{\textnormal{gen}}, \Delta G_{BB} \rightarrow \infty$. For simplicity, we let $[M] = e^{-\Delta G_{\textnormal{gen}}}$ and we normalize rates such that $k_{\textnormal{bind}} [M] = 1$. Then,
\begin{align}
\Phi^+(&n_{l-1}n_l, m_{l-1}m_l) \nonumber \\ &= 1, \label{eq:phipslowbind}\\
\Phi^-(&n_{l-1}n_l, m_{l-1}m_l) \nonumber \\ &= e^{-\Delta G_{\textnormal{pol}} +\Delta G_{TT}(n_{l-1},m_{l-1})-\Delta G_{TT}(n_l,m_l)} \label{eq:phimslowbind}. 
\intertext{ \item Slow polymerization, $k_{\textnormal{pol}} \ll k_{\textnormal{bind}}$. Here, rates are normalized so that $k_{\textnormal{pol}}[M] = 1$. Then,} 
\Phi^+(&n_{l-1}n_l, m_{l-1}m_l) \nonumber \\ &=  e^{\Delta G_{TT}(n_l,m_l)}, \label{eq:phipslowpol}\\
\Phi^-(&n_{l-1}n_l, m_{l-1}m_l) \nonumber \\ &=  e^{-\Delta G_{\textnormal{pol}} +\Delta G_{TT}(n_{l-1},m_{l-1})} 
.\label{eq:phimslowpol}\end{align}
\end{enumerate}

Case 1 corresponds to a discrimination on backward propensities (consistent with the ``temporary thermodynamic discrimination'' described in \cite{Poulton2019}). Case 2, on the other hand, corresponds to a thermodynamically-driven discrimination on forward propensities (consistent with ``combined kinetic and thermodynamic discrimination'' as described in \cite{Poulton2019}). 

\section{Results}

\subsection{Consequences of Separation for the Velocity and Energy Landscapes of Heterogeneous Copying \label{sec:Velocity}}

\begin{figure*}
  \begin{subfigure}[t]{\textwidth}
    \includegraphics[width=0.65\linewidth]{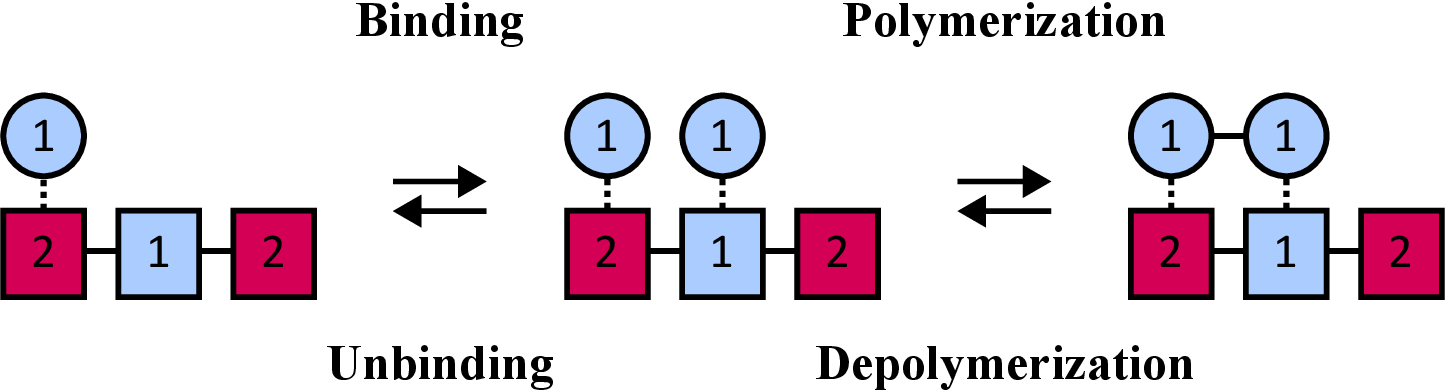}
    \caption{\label{fig:FineGrainPerm}}
  \end{subfigure}
  \begin{subfigure}[t]{\textwidth}
    \includegraphics[width=0.65\linewidth]{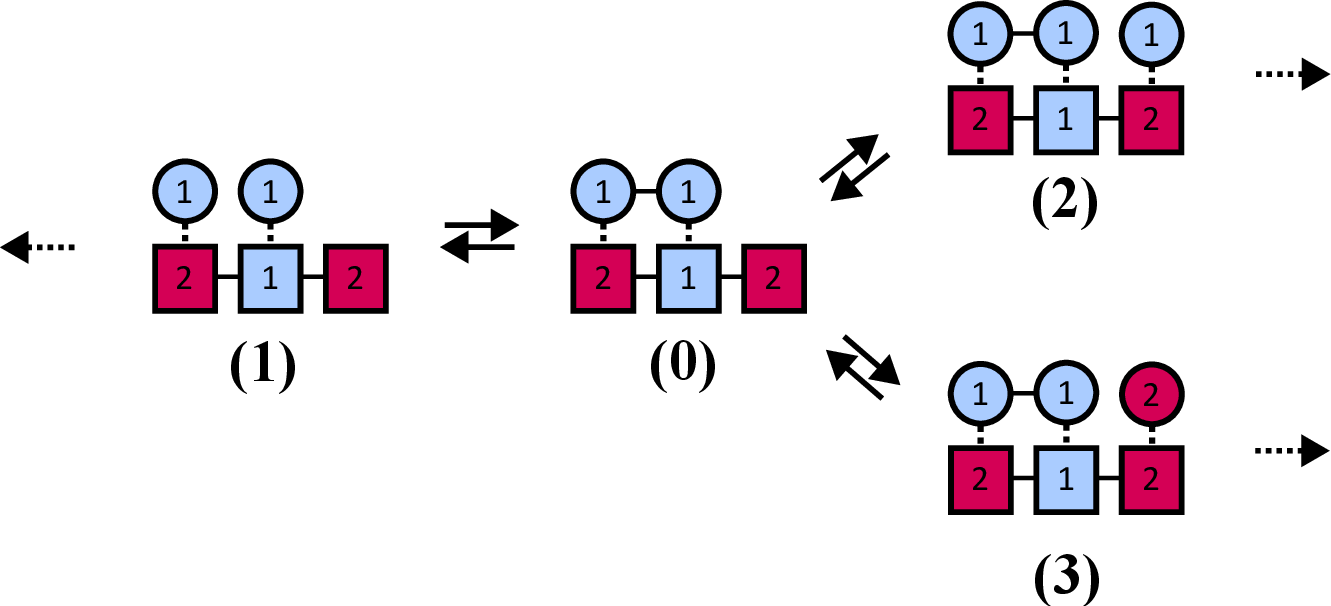}
\caption{\label{fig:FineGrainPermIndices}}
  \end{subfigure}
    \caption{ Fine-grained steps for a model of copying where the copy does not separate from the template. Only monomers in the vicinity of the copy tip are depicted. (a) New monomers bind and polymerize in separate steps, but there is no tail unbinding of the previously added monomer. (b) Labelling of fine-grained states associated with a given coarse-grained state. Indices $f$ are denoted by bold bracketed numbers. Transitions to other states of index $f = 0$ (denoted here by the dotted arrows) change the coarse-grained state $(\mathbf{x},\mathbf{y})$.
    \label{fig:FineGrainPermAll}}
\end{figure*}

We wish to compare the velocity profile of separating templated copolymerization (STC) and non-separating templated copolymerization (NTC, note the phrase `templated self-assembly' has been used in some literature \cite{Poulton2019}) as studied by Gaspard \cite{Gaspard2017}. In figure \ref{fig:FineGrainPerm}, we illustrate a fine-grained model of NTC which we will compare against our model of STC. The fine-grained dynamics is similar to that in Section \ref{sec: Parameterization}, with the exception that the tail binding step has been removed. We assume here that the binding free energy of any given monomer pair is independent of any other monomer pair, giving rise to the parameterization in Table \ref{table:param_fine_grain_perm}. We consider here the slow binding limit, leading to the following coarse-grained propensities:

\begin{align}
\Phi^+(&n_{l-1}n_l, m_{l-1}m_l) \nonumber \\ &= 1. \\
\Phi^-(&n_{l-1}n_l, m_{l-1}m_l) \nonumber \\ &= e^{-\Delta G_{\textnormal{pol}} -\Delta G_{PT}(n_l,m_l)} 
.\end{align}

\noindent This choice again corresponds to discrimination on backward propensities. We denote the sequence-independent non-equilibrium drive provided by monomer binding and backbone formation by $\Delta G_{\textnormal{pol}} = \Delta G_{BB} + \ln{[M]}$ (similar to copying with separation, but with the  $[M]_{\textnormal{eff}}$ term absent). However, observe that in the absence of separation, the sequence-dependent free-energy change $\Delta G_{PT}$ incurred on the addition of each monomer persists as the copy is produced (note $\Delta G_{TT}$ was used in \cite{Poulton2019} to refer to a `temporary thermodynamic' discrimination factor, that is no longer present after separation. To maintain consistency with the intended meaning in \cite{Poulton2019}, we have introduced a `permanent thermodynamic' discrimination factor $\Delta G_{PT}$ for NTC). 

\begin{figure*}
\begin{tabular}[t]{cc}

    \begin{subfigure}[t]{0.42\textwidth}
      \includegraphics[width=0.8\linewidth]{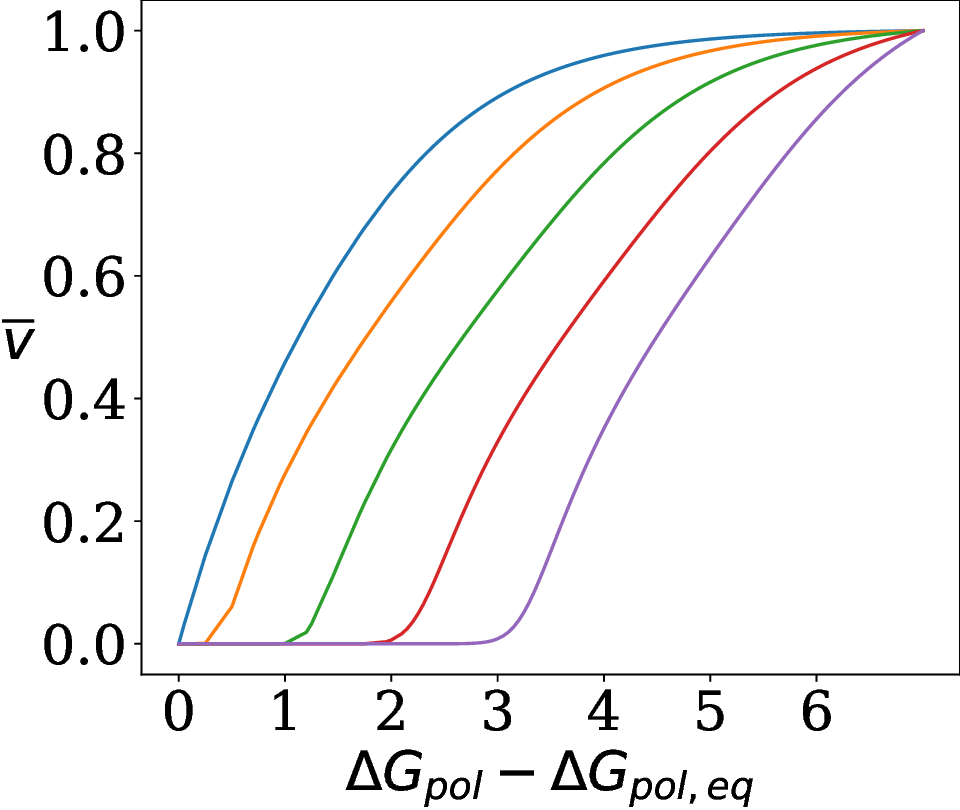}
      \caption{\label{fig:velPerm}}
    \end{subfigure}
    \begin{subfigure}[t]{0.42\textwidth}
        \centering
      \includegraphics[width=0.8\linewidth]{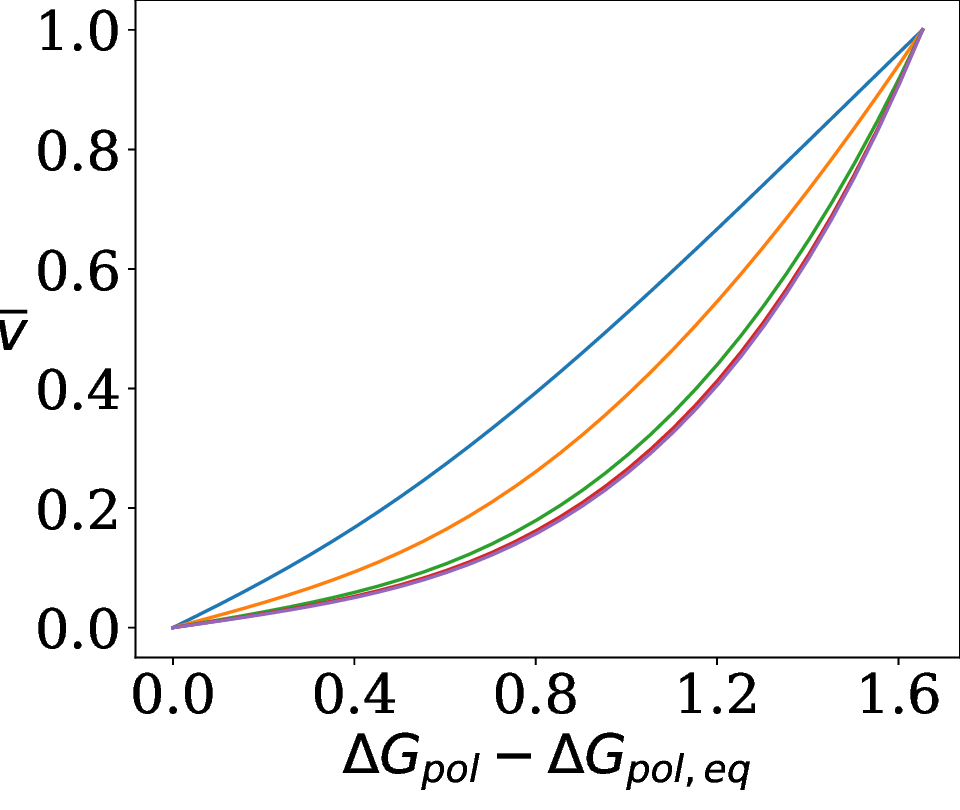}
      \caption{\label{fig:velTemp}}
    \end{subfigure} & 
        \begin{subfigure}[t]{0.14\textwidth}
          \includegraphics[width=0.8\linewidth]{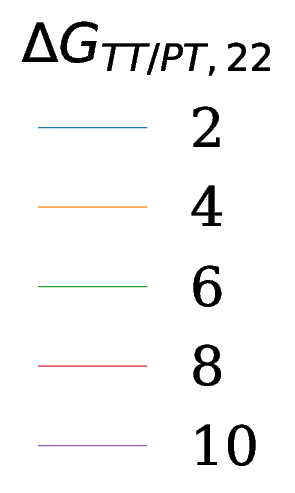}

        \end{subfigure}
        
    \\
    \begin{subfigure}[t]{0.42\textwidth}
      \includegraphics[width=1.0\linewidth]{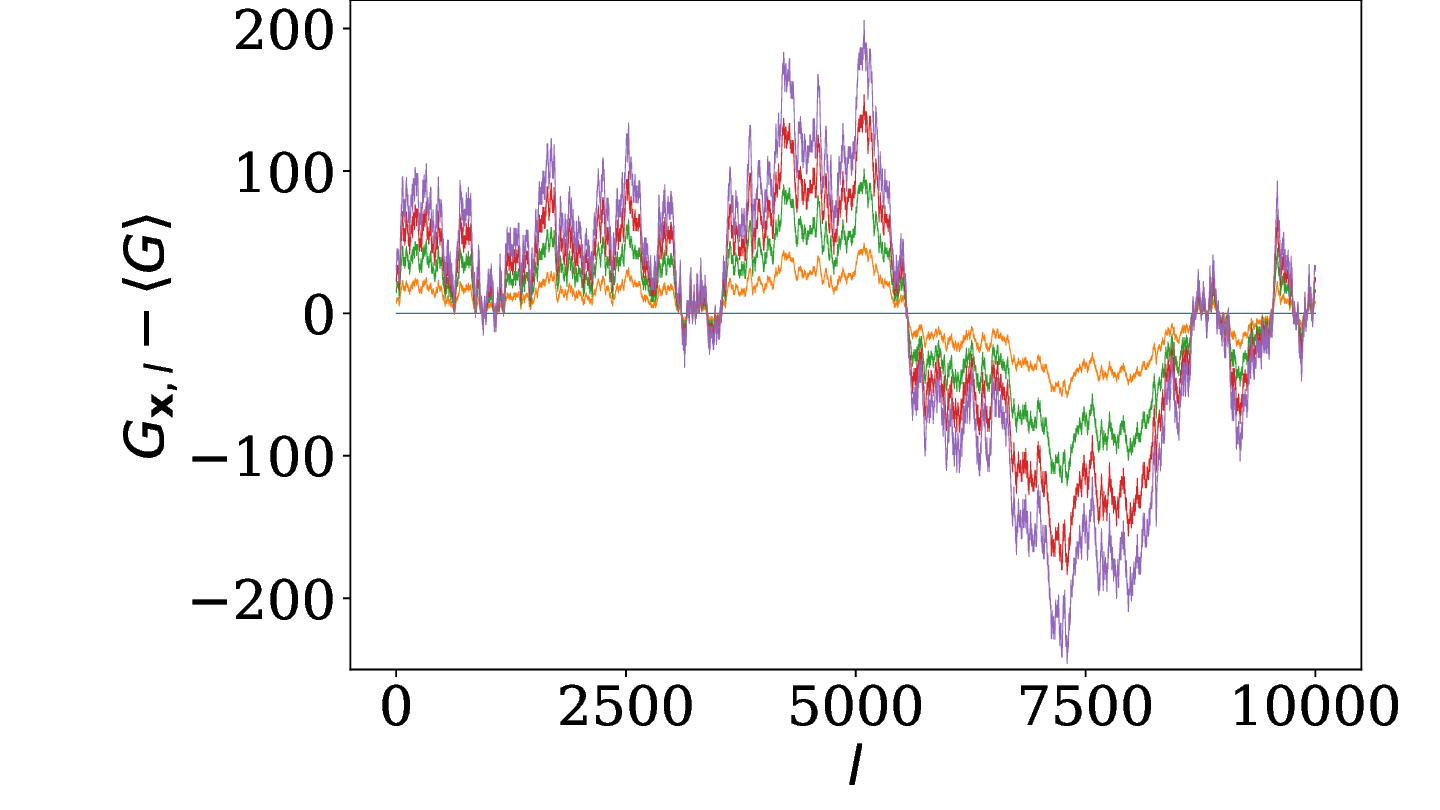}
      \caption{\label{fig:FEPerm}}
    \end{subfigure}
    \begin{subfigure}[t]{0.42\textwidth}
      \includegraphics[width=1.0\linewidth]{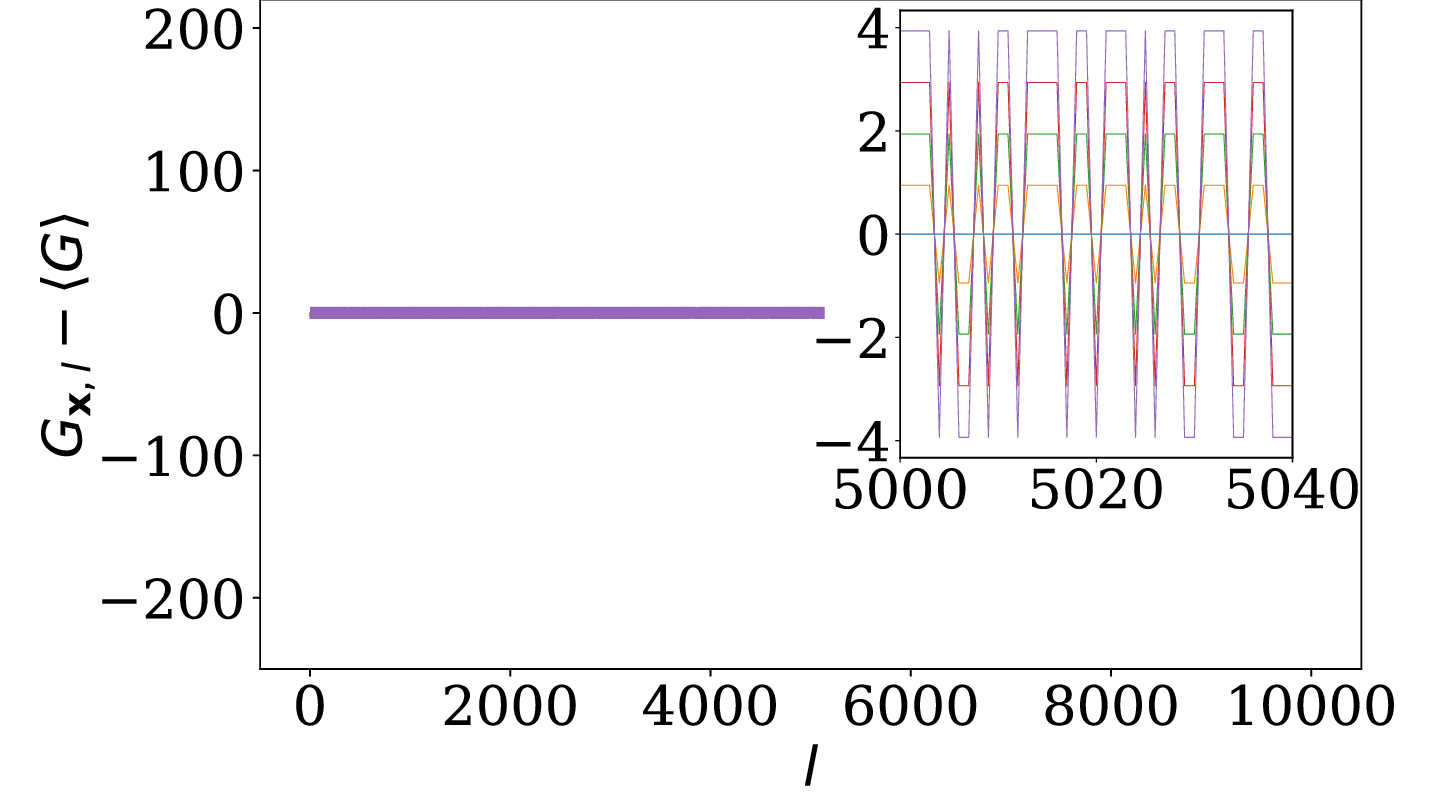}
      \caption{\label{fig:FETemp}}
    \end{subfigure} & 
        \begin{subfigure}[t]{0.14\textwidth}

        \end{subfigure}

\end{tabular}
\caption{Comparison of velocity profiles and free energy landscapes for STC and NTC. Graphs of normalized average velocity as a function of $\Delta G_{\textnormal{pol}} - \Delta G_{\textnormal{pol,eq}}$ are shown for NTC in (a) and STC in (b). Equilibrium $ G_{\mathbf{x},l} - \langle G \rangle$ free-energy landscapes (at $\Delta G_{pol} = \Delta G_{pol,eq}$) are shown for NTC in (c) and STC (d). A zoom inset is provided in (d). The long tails of zero velocity for copying without separation, caused by a rougher free-energy profile, do not occur for copying with simultaneous separation, where the velocity profile is linear close to $\Delta G_{\textnormal{pol}} - \Delta G_{\textnormal{pol,eq}} = 0$.   \label{fig:VelGraphs}}
\end{figure*}

We begin by considering how the velocity profiles for STC change with increasing nonequilibrium drive $\Delta G_{\textnormal{pol}}$. As with NTC, we consider backward propensity discrimination (that is, the slow binding limit). The binding free energies of the incorrect monomer (we remind readers that a correct pair is defined by $m_l = n_l$) pairs are kept constant, such that $\Delta G_{TT}(1,2) = \Delta G_{TT}(2,1) = 0$. The free energy of a correct monomer $1$ pair is held at $\Delta G_{TT}(1,1) = 2$, while $\Delta G_{TT}(2,2)$ is varied from $2$ to $10$. Gaspard's iterated function system \cite{Gaspard2017} was applied to obtain the absorption probabilities $Q$ for the copying of a polymer sequence of length $L = 10000$. Then, the methods in Section \ref{sec:VisitIter} and reference \cite{Qureshi2023} were used to find $\tau_{\mathbf{x},l}$, defined as the average amount of total time spent at position $l$ for a template sequence $\mathbf{x}$ before a complete polymer forms and detaches. $\tau_{\mathbf{x},l}$ is calculated by multiplying visits to each state by the waiting time for each visit to said state and then summing over all states of copy length $l$. The average completion times $t_c = \Sigma_l \tau_{\mathbf{x},l}$ can then be calculated, and the average copying velocity was calculated as $v = \frac{L}{t_c}$. As in reference \cite{Poulton2019}, the value of $\Delta G_{\textnormal{pol}}$ at equilibrium is given by $\Delta G_{\textnormal{pol,eq}} = -\ln{2}$, and we report copy velocities as a function of the difference in $\Delta G_{\textnormal{pol}}$ from this equilibrium value up to $\Delta G_{\textnormal{pol}} = 2$ in figure \ref{fig:velTemp}.

\begin{table*}[]
\begin{tabular}{@{}lllll@{}}
\toprule
\textbf{Forward Reaction   } &
\textbf{Forward Rate}                                               &   & \textbf{Backward Rate} & \\ \midrule
        Binding & $K^+_{\textnormal{bind}}(n_{l-1}n_l,m_{l-1}m_l)$              
    & $k_{\textnormal{bind}}[M]$ 
    & $K^-_{\textnormal{bind}}(n_{l-1}n_l,m_{l-1}m_l)$                 & $k_{\textnormal{bind}}e^{-\Delta G_{PT}(n_l,m_l)}$ \\
        Polymerization & $K^+_{\textnormal{pol}}(n_{l-1}n_l,m_{l-1}m_l)$ 
    & $k_{\textnormal{pol}}(n_{l-1}n_l,m_{l-1}m_l)$                
    & $K^-_{\textnormal{pol}}(n_{l-1}n_l,m_{l-1}m_l)$                  & $k_{\textnormal{pol}}(n_{l-1}n_l,m_{l-1}m_l)e^{- \Delta G_{BB}}$                      \\ \bottomrule
\end{tabular}
\caption{Parameterized forms of each of the fine-grained reaction steps for a model of copying where the copy does not separate from the template. \label{table:param_fine_grain_perm}}
\end{table*}

We now consider velocity profiles for NTC. We must first calculate the equilibrium  point $\Delta G_{\textnormal{pol,eq}}$, which follows from the partition function  of the  product. As the monomer binding free energies are uncorrelated in our NTC system, the (negative of) free energy $G_{\mathbf{x},l}$ of the ensemble of length of length $l$ product polymers can be calculated as follows:

\begin{align}
z_1 &= e^{\Delta G_{PT}(1,1)}+e^{\Delta G_{PT}(2,1)}. \\
z_2 &= e^{\Delta G_{PT}(2,2)}+e^{\Delta G_{PT}(1,2)}. \\
Z_{\mathbf{x},l} &= Z_{\mathbf{x},1} e^{(l_1+l_2)\Delta G_{\textnormal{pol}}}z_1^{l_1}z_2^{l_2}. \\
G_{\mathbf{x},l} &= \ln[Z_l]
.\end{align}

\noindent Here, $z_i$ is the partition function for a template monomer of type $i$ over its copies, %{\color{purple} [introduced explicit $\mathbf{x},l$ dependence to make it consistent with ohter variables]} 
$Z_{\mathbf{x},l}$ is the partition function of length $l$ products for a template $\mathbf{x}$, $Z_{\mathbf{x},1}$ is the partition function of the very first copy monomer, $l_1$ and $l_2$ are the numbers of monomers of types 1 and 2, respectively, in the template $\mathbf{x}$ up to index $l$ excluding the first monomer, and $l = l_1 + l_2 + 1$. We now introduce a new expectation operator $\mathbb{E}_{\mathbf{x}}$ averaging over all templates $\mathbf{x}$ for a fixed template distribution. Then, for equilibrium between polymer growth and shrinking, the  system free energy averaged over templates $\mathbb{E}_{\mathbf{x}}[G_{\mathbf{x},l}]$ should not change with length index $l$. Furthermore, the average over positional indices in the $L \rightarrow \infty$ limit $\langle G \rangle$ should be the same as the per-site expectation $\mathbb{E}_{\mathbf{x}}[G_{\mathbf{x},l}]$,
\begin{equation}
\mathbb{E}_{\mathbf{x}}[G_{\mathbf{x},l}] = \mathbb{E}_{\mathbf{x}}[G_{\mathbf{x},l}] = \langle G \rangle
.\end{equation}

\noindent As the system free energy is self-averaging, the equilibrium drive $\Delta G_{\textnormal{pol,eq}}$ can be calculated as follows:

\begin{align}
\Delta G_{\textnormal{pol,eq}} = \lim_{l,L \rightarrow \infty}\frac{-\ln{z_1^{l_1}z_2^{l_2}}}{(l_1+l_2)}
.\end{align}

\noindent We plot the normalized copy velocity (obtained by  dividing velocity by its maximum value over the range of $\Delta G_{\textnormal{pol}}$ considered; in practice this maximum value always occurs at our maximum considered drive $\Delta G_{\textnormal{pol}}= 0$) $\overline{v} = v/(\max_{\Delta G_{\textnormal{pol}}} v)$ as a function of $\Delta G_{\textnormal{pol}} - \Delta G_{\textnormal{pol,eq}}$ for NTC in Figure \ref{fig:velPerm}. Comparing Figures \ref{fig:velPerm} and \ref{fig:velTemp}, we observe a few common features. First, velocity tends to zero as the equilibrium $\Delta G_{\textnormal{pol}}$ is reached. At equilibrium there should be no net flux, and hence this behaviour is expected. Copy velocity  increases with increasing $\Delta G_{\textnormal{pol}}$, as the rate of monomer removal decreases. 

However, the models approach zero velocity in different ways. As noted by Gaspard \cite{Gaspard2017}, the heterogeneous model of NTC displays a long tail of near-zero velocity as it approaches equilibrium. Such behaviour is not observed in our model of heterogeneous copying (Figure \ref{fig:velTemp}). This difference can be explained by considering sample free-energy landscapes in the heterogeneous NTC model and the STC model. We plot free-energy deviation $G_{\mathbf{x},l}-\langle G \rangle$ for a fixed sample template at various copy lengths $l$ for both heterogeneous NTC (Figure \ref{fig:FEPerm}) and heterogeneous STC (Figure \ref{fig:FETemp}). 

For heterogeneous STC, the free-energy deviations switch stochastically between two levels, determined by the template monomer at position $l$, and  $G_{\mathbf{x},l}-\langle G \rangle$ is therefore constrained to be close to zero. By contrast, the free-energy deviations for heterogeneous NTC are large. On scales of length $l_d$ we expect to see barriers of height $ \sim\sqrt{l_d}$ \cite{Sinai1983} that trap the system, resulting in slow sub-diffusive motion close to equilibrium driving \cite{Sinai1983,Gaspard2017}. Hence we observe a region of zero velocity in Figure \ref{fig:velPerm}. These tall barriers do not occur for separating copiers as the length of product interacting with the template at any given time is bounded. In our system, only two monomers may be bound at a given time, but this intuition generalizes to other parameter sets and even completely different models that include separation as the copy grows (even realistic models of transcription and translation). If only a finite number of copy monomers bind to the template simultaneously, the roughness of the free-energy landscape associated with sequence heterogeneity is inherently limited.  

\begin{figure*}
\begin{tabular}[t]{cccc}
        \begin{subfigure}[b]{0.28\textwidth}
          \includegraphics[width=1.0\linewidth]{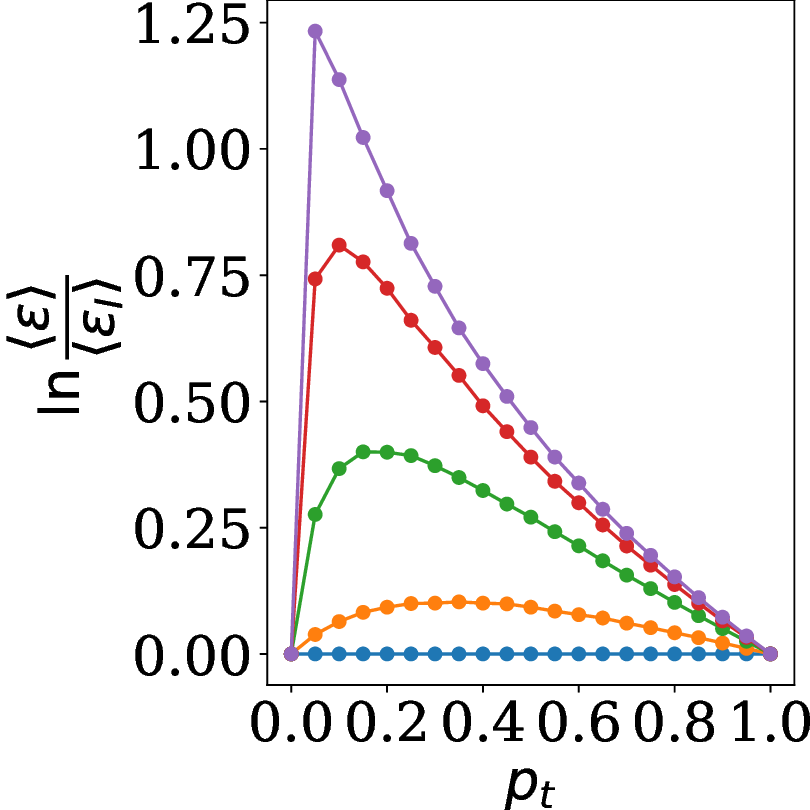}
          \caption{\label{fig:ThermoCorrect.Error}}
        \end{subfigure} &
        \begin{subfigure}[b]{0.28\textwidth}
            \centering
          \includegraphics[width=1.0\linewidth]{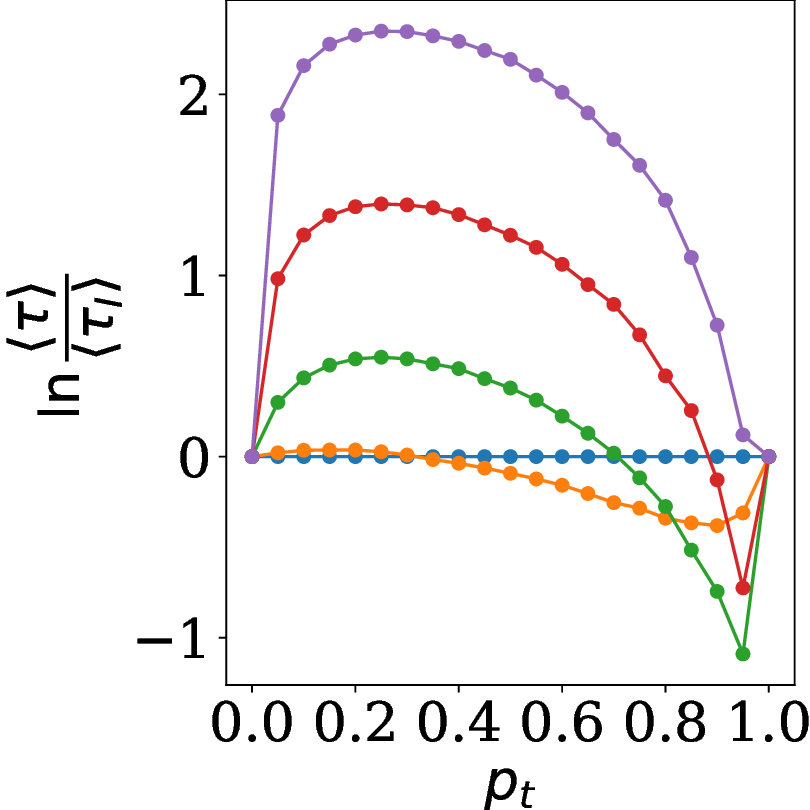}
          \caption{\label{fig:ThermoCorrect.Times}}
        \end{subfigure} &
        \begin{subfigure}[b]{0.28\textwidth}
              \includegraphics[width=1.0\linewidth]{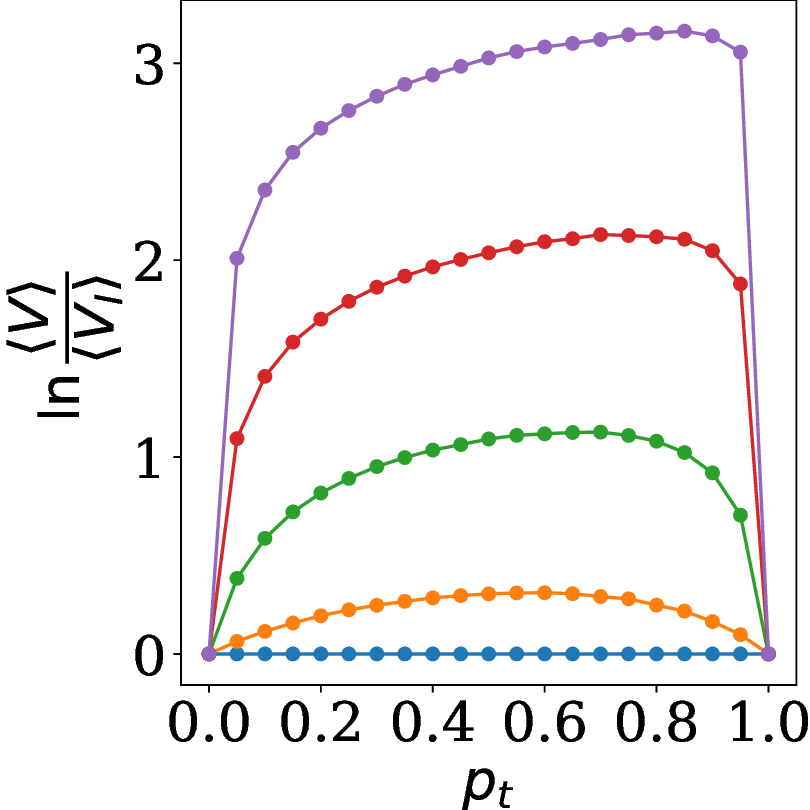}
            \caption{\label{fig:ThermoCorrect.Visits}}
            \end{subfigure} &
            \begin{subfigure}[b]{0.14\textwidth}
              \includegraphics[width=0.8\linewidth]{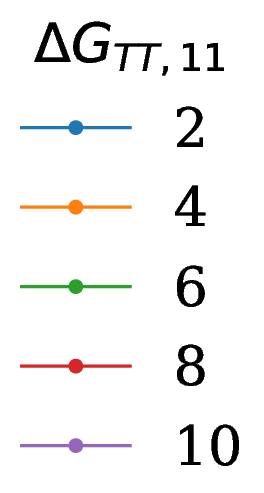}
    
            \end{subfigure} \\
            
        \begin{subfigure}[b]{0.28\textwidth}
          \includegraphics[width=1.0\linewidth]{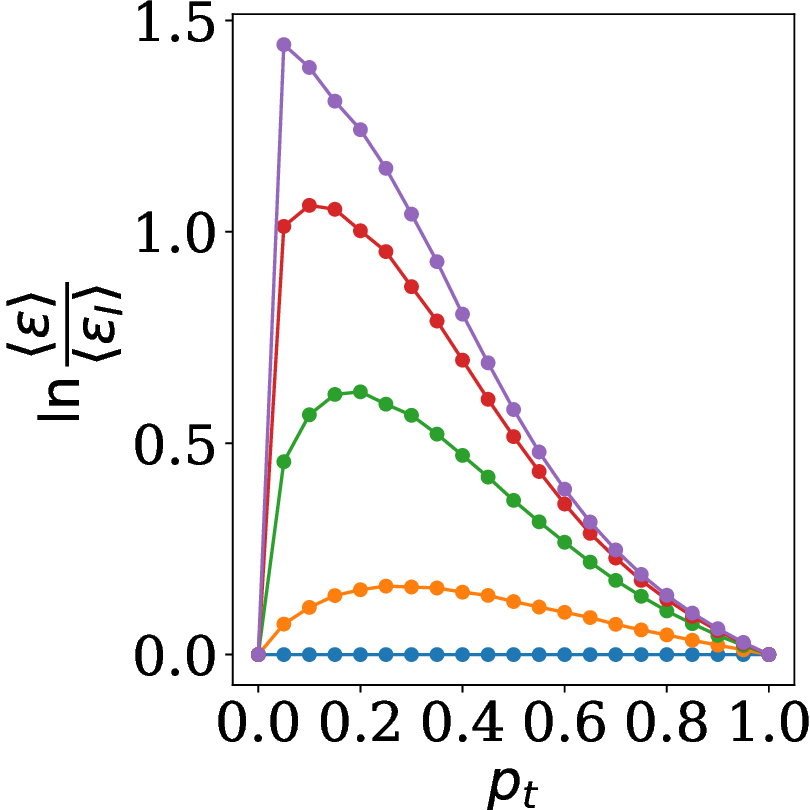}
          \caption{\label{fig:CombCorrect.Error}}
        \end{subfigure} &
        \begin{subfigure}[b]{0.28\textwidth}
            \centering
          \includegraphics[width=1.0\linewidth]{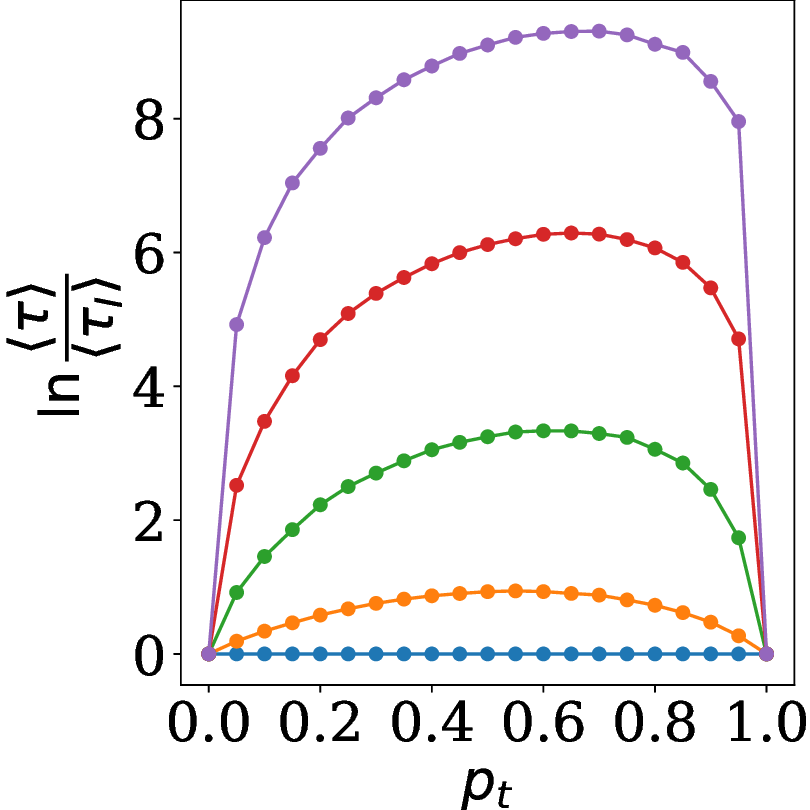}
          \caption{\label{fig:CombCorrect.Times}}
        \end{subfigure} &
        \begin{subfigure}[b]{0.28\textwidth}
              \includegraphics[width=1.0\linewidth]{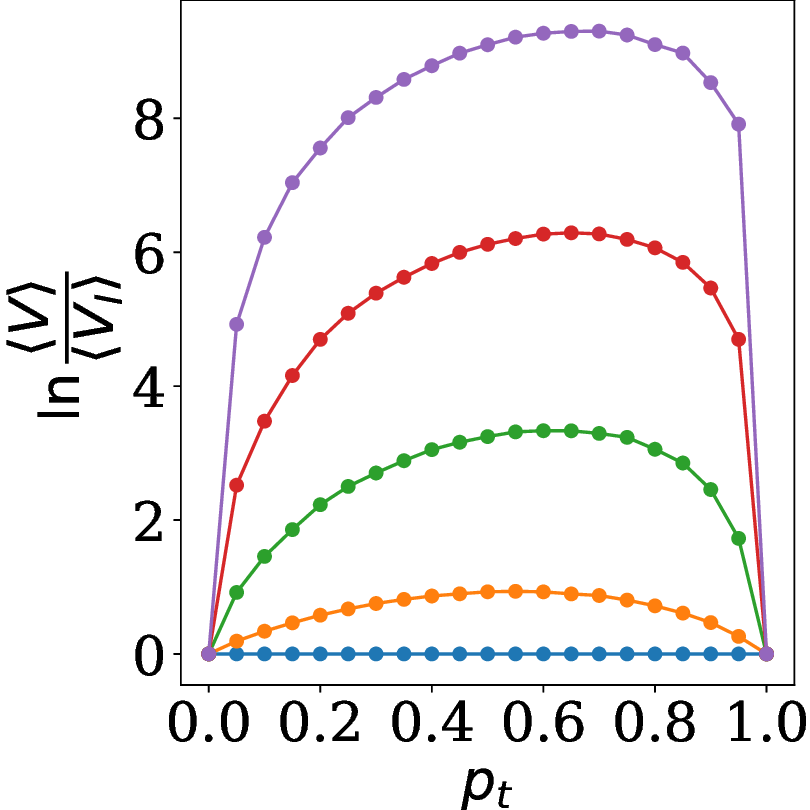}
            \caption{\label{fig:CombCorrect.Visits}}
            \end{subfigure}
\end{tabular}
\caption{Deviations in $\epsilon$, $\tau$ and $V$ due to heterogeneity in correct monomer interactions. as a function of the monomer $2$ content $p_t$, relative to homogeneous copying. Log-ratio-means $\ln{\frac{\langle \epsilon \rangle}{\langle \epsilon_I \rangle}}$, $\ln{\frac{\langle \tau \rangle}{\langle \tau_I \rangle}}$ and $\ln{\frac{\langle V \rangle}{\langle V_I \rangle}}$ are plotted for backward propensity discrimination in (a, b, c) and forward propensity discrimination in (d, e, f).
\label{fig:Correct}}
\end{figure*}

\begin{figure*}
\begin{tabular}[t]{cccc}
        \begin{subfigure}[b]{0.28\textwidth}
          \includegraphics[width=1.0\linewidth]{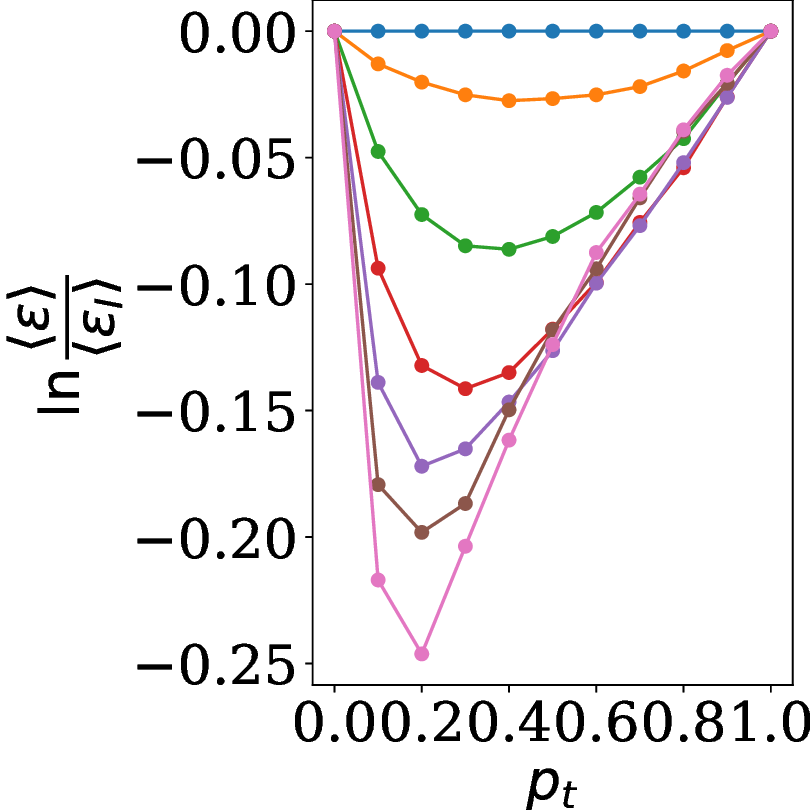}
          \caption{\label{fig:ThermoIncorrect.Error}}
        \end{subfigure} &
        \begin{subfigure}[b]{0.28\textwidth}
            \centering
          \includegraphics[width=1.0\linewidth]{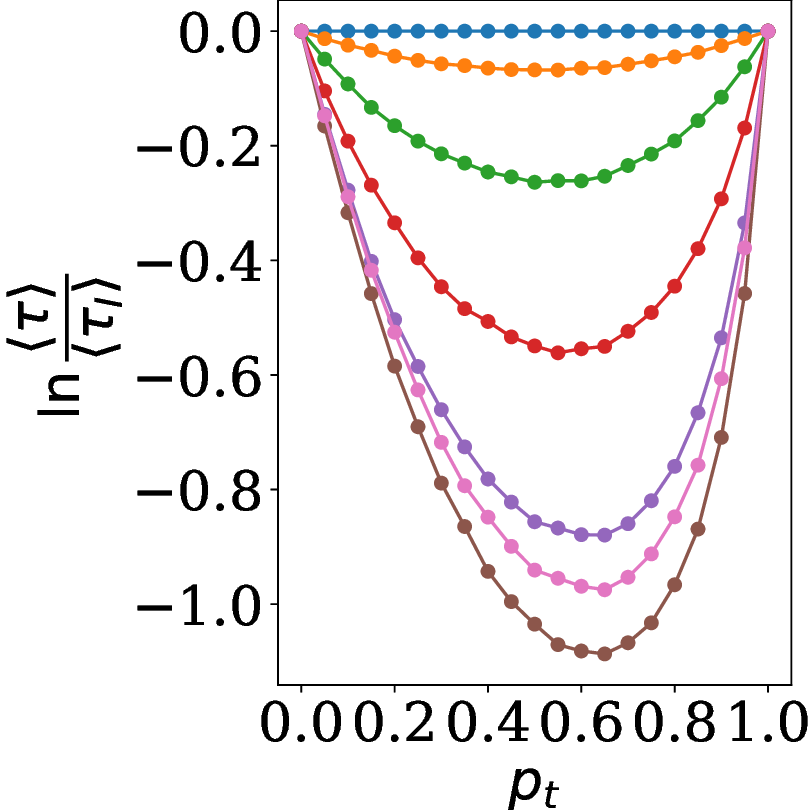}
          \caption{\label{fig:ThermoIncorrect.Times}}
        \end{subfigure} &
        \begin{subfigure}[b]{0.28\textwidth}
              \includegraphics[width=1.0\linewidth]{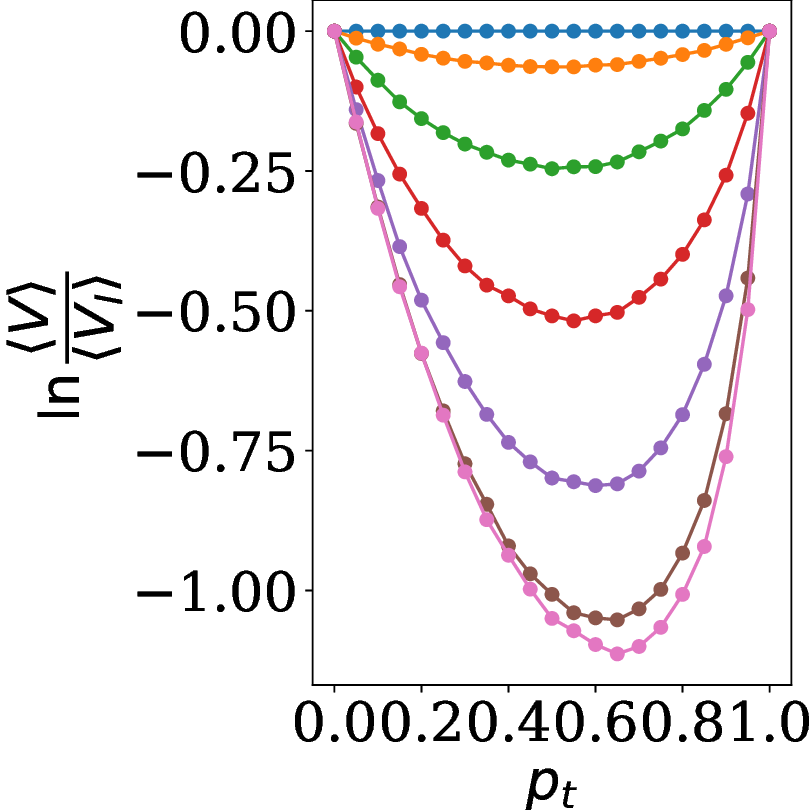}
            \caption{\label{fig:ThermoIncorrect.Visits}}
            \end{subfigure} &
            \begin{subfigure}[b]{0.14\textwidth}
              \includegraphics[width=0.74\linewidth]{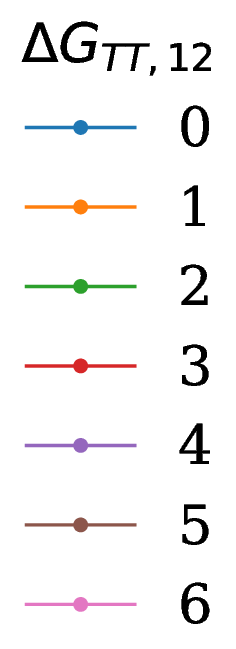}
    
            \end{subfigure} \\
        \begin{subfigure}[b]{0.28\textwidth}
          \includegraphics[width=1.0\linewidth]{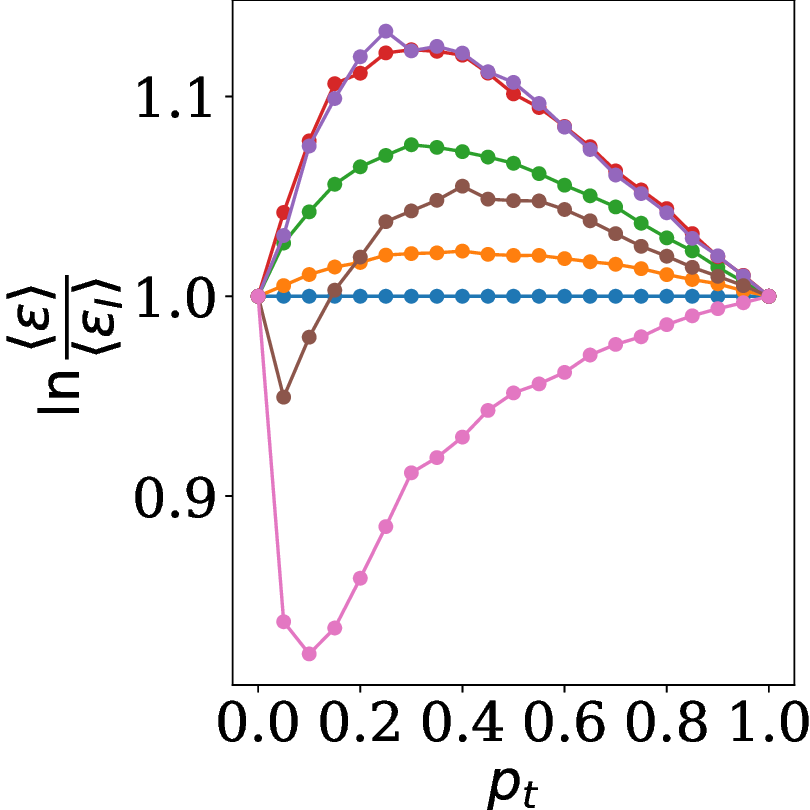}
          \caption{\label{fig:CombIncorrect.Error}}
        \end{subfigure} &
        \begin{subfigure}[b]{0.28\textwidth}
            \centering
          \includegraphics[width=1.0\linewidth]{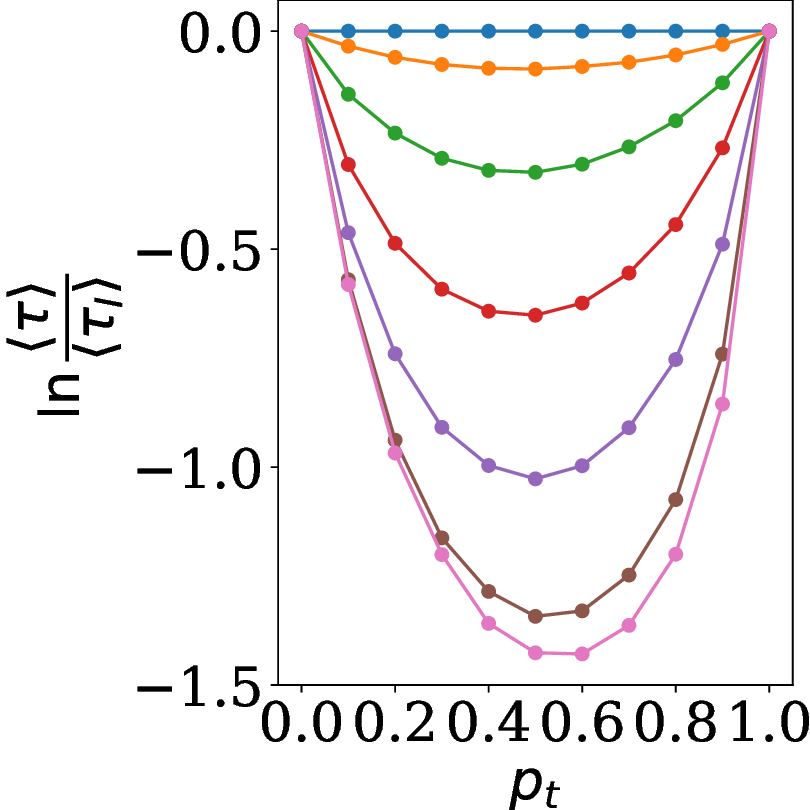}
          \caption{\label{fig:CombIncorrect.Times}}
        \end{subfigure} &
        \begin{subfigure}[b]{0.28\textwidth}
              \includegraphics[width=1.0\linewidth]{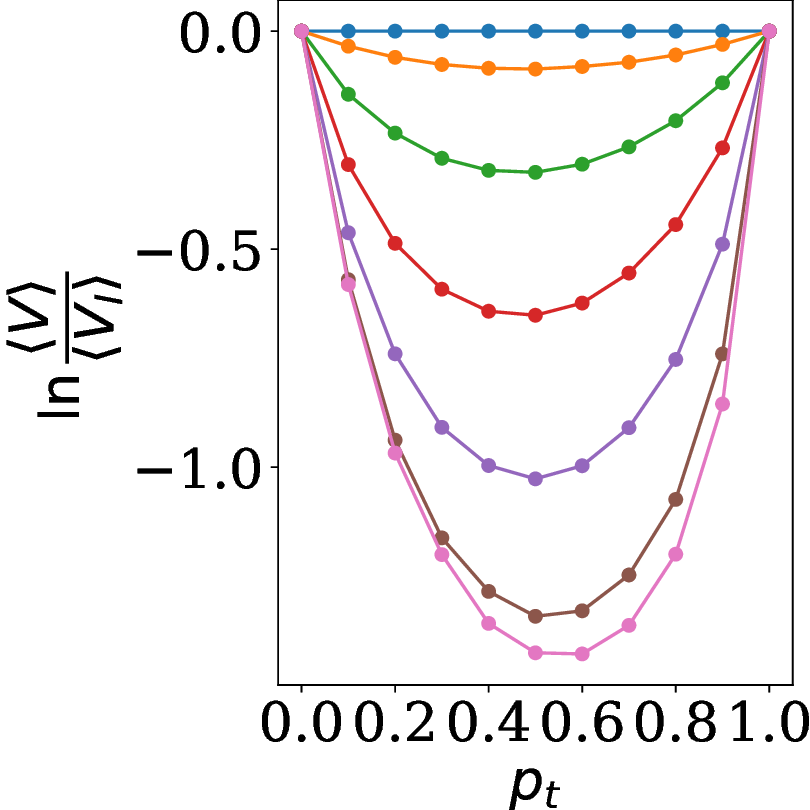}
            \caption{\label{fig:CombIncorrect.Visits}}
            \end{subfigure} &
\end{tabular}
\caption{ Deviations in $\epsilon$, $\tau$ and $V$ due to heterogeneity in incorrect monomer interactions, as a function of the monomer $2$ content $p_t$, relative to homogeneous copying. Log-ratio-means $\ln{\frac{\langle \epsilon \rangle}{\langle \epsilon_I \rangle}}$, $\ln{\frac{\langle \tau \rangle}{\langle \tau_I \rangle}}$ and $\ln{\frac{\langle V \rangle}{\langle V_I \rangle}}$ are plotted for backward propensity discrimination in (a, b, c) and forward propensity discrimination in (d, e, f). \label{fig:Incorrect}}
\end{figure*}

\subsection{Effects of Separation and Heterogeneity on Error Probability and Average Time Spent Per Site \label{sec:ErrorTime}}

We now investigate the effect of heterogeneity on two key indicators of a copy system: the error probability, $\epsilon_{\mathbf{x},l}$, and the total time spent at a site $l$ before the monomer is permanently incorporated, $\tau_{\mathbf{x},l}$ (contrast with $\theta_{\mathbf{x},l}$ the waiting time for each visit of a tip state). We also consider visitations to a given position $V_{\mathbf{x},l}$, for reasons that will become apparent. As indicated by the subscripts, these three quantities depend on the positional index $l$ underlying template $\mathbf{x}$.  We assume these properties are self-averaging if the underlying template distribution is stationary \cite{Gaspard2017}, such that in the limit $L \rightarrow \infty$, $\langle \epsilon \rangle = \frac{\Sigma_{l=1}^L \epsilon_{\mathbf{x},l}}{L}$, $\langle V \rangle = \frac{\Sigma_{l=1}^L V_{\mathbf{x},l}}{L}$ and $\langle \tau \rangle = \frac{\Sigma_{l=1}^L \tau_{\mathbf{x},l}}{L}$ are well-defined. Throughout this section, we will calculate these averages for a single long template $\mathbf{x}$ of length $L = 10^5$ whose monomers are drawn from a Bernoulli distribution $B(L,p_t)$ with $p_t$ representing the average proportion of monomers of type $2$, similar to \cite{Gaspard2017}.

The quantities $\langle \epsilon \rangle$, $\langle V \rangle$ and $\langle \tau \rangle$ are then dependent on $p_t$, the proportion of monomers of type $2$. 
Consider now random variables $\epsilon_1$ and $\epsilon_2$. $\epsilon_1$ is the error rate (in the large $L$ limit) for copying a homogeneous template with monomers of type $1$, while $\epsilon_2$ is the error rate  for copying a homogeneous template with monomers of type $2$. Let $\epsilon_{I,\mathbf{x},l}$ be a random variable that takes on a value $\epsilon_1$ if the monomer at position $l$ for template $x$ is of type $1$, and $\epsilon_2$ if the monomer at position $l$ for template $x$ is of type $1$. In the long $L$ limit, for a given probability $p_t$ of monomer 2, then $\langle \epsilon_I \rangle = (1-p_t)\epsilon_1 + p_t \epsilon_2$.  Similarly, we can apply analogous definitions to $\tau$ and $V$
such that 
$\langle V_I \rangle = (1-p_t)V_1 + p_t V_2$ and $\langle \tau_I \rangle = (1-p_t)\tau_1 + p_t \tau_2$.
If copying of subsequent monomers were independent of each other, we would expect $\langle \epsilon \rangle = \langle \epsilon_I \rangle$, $\langle V \rangle = \langle V_I \rangle$ and $\langle \tau \rangle = \langle \tau_I \rangle$. However, these equalities generally do not hold due to inter-monomer correlations in the product \cite{Gaspard2014,Gaspard2017,Poulton2019}. To infer whether heterogeneity tends to improve or worsen copying performance for a particular parameter set, we now consider, for specific parameter regimes, the sign of log-ratio-averages $\ln{\frac{\langle \epsilon \rangle}{\langle \epsilon_I \rangle}}$, $\ln{\frac{\langle V \rangle}{\langle V_I \rangle}}$ and $\ln{\frac{\langle \tau \rangle}{\langle \tau_I \rangle}}$. As an example, $\ln{\frac{\langle \epsilon \rangle}{\langle \epsilon_I \rangle}} > 0$ would imply that for a given parameter set, heterogeneity in monomer interactions tend to increase average errors. %On the other hand $\ln{\frac{\langle \epsilon \rangle}{\langle \epsilon_I \rangle}} \langle 0$ implies a parameter set where heterogeneity in monomer interactions tends to decrease average errors.

We will separately consider both limits mentioned in Section \ref{sec: Parameterization}, that is the slow binding limit with backward propensity discrimination and the slow $k_{\textnormal{pol}}$ limit with forward propensity discrimination. Where parameters tend to $\infty$, a multiplier of $e^{12}$ is used for the numerics. For the backward propensity discrimination case, $[M] = e^{-12}$ and $k_{\textnormal{bind}} = e^{12}$. In Section \ref{section:CorrectDisc}, we consider the case where the interaction between correct monomer pairs $(1,1)$ and $(2,2)$ are made heterogeneous, while in Section \ref{section:IncorrectDisc}, we consider the case where interactions between incorrect pairs $(2,1)$ and $(1,2)$ are made heterogeneous. For each section, we will be plotting $\ln{\frac{\langle \epsilon \rangle}{\langle \epsilon_I \rangle}}$,$\ln{\frac{\langle \tau \rangle}{\langle \tau_I \rangle}}$ and $\ln{\frac{\langle V \rangle}{\langle V_I \rangle}}$ as a function of $p_t$, and we provide a framing for our findings in Section \ref{section:DiscExplain}.

\subsubsection{Heterogeneity in Correct Monomer Interactions}
\label{section:CorrectDisc}

Consider the coarse-grained dynamics given in equations \ref{eq:phipslowbind}-\ref{eq:phimslowpol}. To investigate the effects of heterogeneous correct monomer interactions, we fix $\Delta G_{TT}(2,2) = 2$ and $\Delta G_{TT}(1,2) = \Delta G_{TT}(2,1) = 0$, while varying $\Delta G_{TT}(1,1)$ from $2$ to $10$. Keep in mind that with backward propensity discrimination, $\Delta G_{TT}$ modifies the rates of monomer unbinding, while in the case of forward propensity discrimination, $\Delta G_{TT}$ modifies the rate of monomer binding (Section \ref{sec:  Parameterization}). Throughout, $\Delta G_{\textnormal{pol}}$ is arbitrarily held at $0$, which corresponds to relatively weak but non-zero driving.

\textbf{Backward Propensity Discrimination: }  Plots are shown in Figures \ref{fig:ThermoCorrect.Error}, \ref{fig:ThermoCorrect.Times} and  \ref{fig:ThermoCorrect.Visits}. We see that $\ln{\frac{\langle \epsilon \rangle}{\langle \epsilon_I \rangle}} > 0$, implying that in this regime errors tend to be increased by heterogeneity. The trend in $\ln{\frac{\langle \tau \rangle}{\langle \tau_I \rangle}}$ is more complex, as there appears to be a small dip below $0$ for high $p_t$. Note that the time spent per site is obtained by modulating the total number of state visits with the average waiting time at each tip state (Section \ref{sec:VisitIter}). If we instead turn our attention to the visitation counts, we once again obtain $\ln{\frac{\langle V \rangle}{\langle V_I \rangle}} > 0$, implying that heterogeneity in this regime tends to increase the number of tip states visited before completion.  

\textbf{Forward Propensity Discrimination: }    Plots are shown in Figures \ref{fig:CombCorrect.Error}, \ref{fig:CombCorrect.Times} and  \ref{fig:CombCorrect.Visits}. Again $\ln{\frac{\langle \epsilon \rangle}{\langle \epsilon_I \rangle}} > 0$, so errors tend to increase in this regime due to heterogeneity. Here, both time spent per site $\tau$ and the number of state visits tend to increase as a result of heterogeneity.

\subsubsection{Heterogeneity in Incorrect Monomer Interactions}
\label{section:IncorrectDisc}

We now consider the case where the binding strengths of incorrect monomers are heterogeneous, $\Delta G_{TT}(1,2) \neq \Delta G_{TT}(2,1)$ while $\Delta G_{TT}(1,1) = \Delta G_{TT}(2,2) = 6$. $\Delta G_{\textnormal{pol}}$ is again arbitrarily held at $0$. $\Delta G_{TT}(2,1) = 2$ is kept constant and $\Delta G_{TT}(1,2)$ is varied from $0$ to $6$.

\textbf{Backward Propensity Discrimination: } Plots are shown in Figures \ref{fig:ThermoIncorrect.Error}, \ref{fig:ThermoIncorrect.Times} and  \ref{fig:ThermoIncorrect.Visits}. We see that $\ln{\frac{\langle \epsilon \rangle}{\langle \epsilon_I \rangle}} < 0$, implying that in this regime errors tend to be decreased by heterogeneity. Unlike in the case where heterogeneity is applied to the binding free energies of correct pairs, both $\ln{\frac{\langle \tau \rangle}{\langle \tau_I \rangle}} < 0$ and $\ln{\frac{\langle V \rangle}{\langle V_I \rangle}} < 0$. Hence, heterogeneity here tends to decrease visitation counts. The graphs have similar shapes, and hence waiting time modulation does not result in qualitative differences in the total time spent per site. 

\begin{figure}
\includegraphics[width=1.0\linewidth]{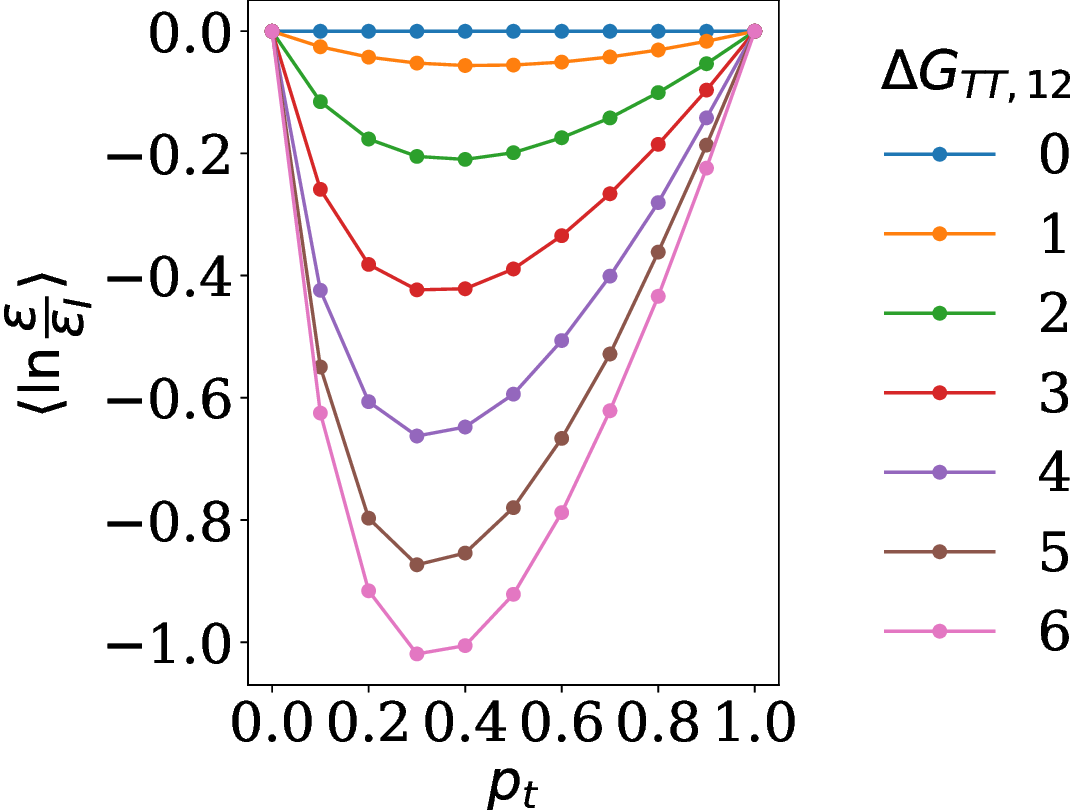}
    \caption{ Mean-log-ratios of error $\langle\ln{\frac{\epsilon}{\epsilon_{I}}}\rangle$ for forward propensity discrimination with heterogeneity on incorrect monomers. $\langle\ln{\frac{\epsilon}{\epsilon_{I}}}\rangle$ is negative for all values of $p_t$ and $\Delta G_{TT,12}$ with heterogeneous interactions.      
    \label{fig:CombIncorrect.RelativeError}}
\end{figure}

\textbf{Forwards Propensity Discrimination: }  Plots are shown in Figures \ref{fig:CombIncorrect.Error}, \ref{fig:CombIncorrect.Times} and  \ref{fig:CombIncorrect.Visits}. We see that there is no clear trend in
$\ln{\frac{\langle \epsilon \rangle}{\langle \epsilon_I \rangle}}$. Let us turn to a different measure, $\langle\ln{\frac{\epsilon}{\epsilon_{I}}}\rangle$. This average-log-ratio of errors is a measure of changes in {\em relative error at each site} that occur due to heterogeneity in monomer interactions. For example, $\langle\ln{\frac{\epsilon}{\epsilon_{I}}}\rangle > 0$ if the average factor of error increase for one monomer is greater than the average factor of error decrease for the other monomer. We will argue the significance of this error measure in the following subsection. For now, observe that $\langle\ln{\frac{\epsilon}{\epsilon_{I}}}\rangle < 0$ in Figure \ref{fig:CombIncorrect.RelativeError}, implying that relative errors on each site are, on average, reduced. We continue to observe $\ln{\frac{\langle \tau \rangle}{\langle \tau_I \rangle}} < 0$ and $\ln{\frac{\langle V \rangle}{\langle V_I \rangle}} < 0$.

\subsubsection{Discussion}
\label{section:DiscExplain}

\begin{figure*}

    \begin{subfigure}[t]{0.25\textwidth}
      \includegraphics[width=1.0\linewidth]{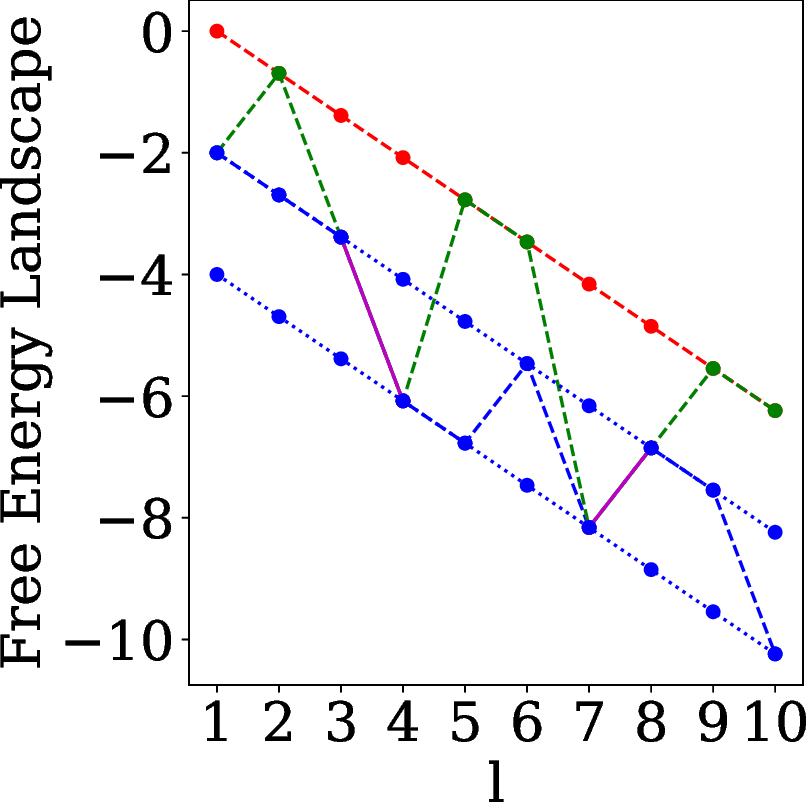}
      \caption{\label{fig:lchoice.chet}}
    \end{subfigure}
    \begin{subfigure}[t]{0.25\textwidth}
        \centering
      \includegraphics[width=1.0\linewidth]{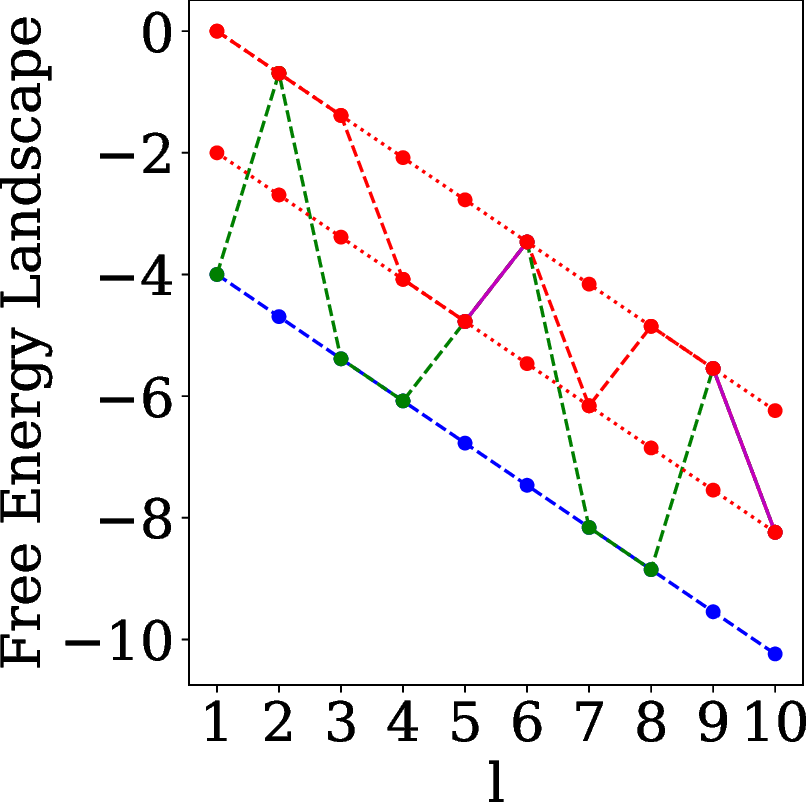}
      \caption{\label{fig:lchoice.ihet}}
    \end{subfigure} 
    \begin{subfigure}[t]{0.25\textwidth}
        \centering
      \includegraphics[width=1.0\linewidth]{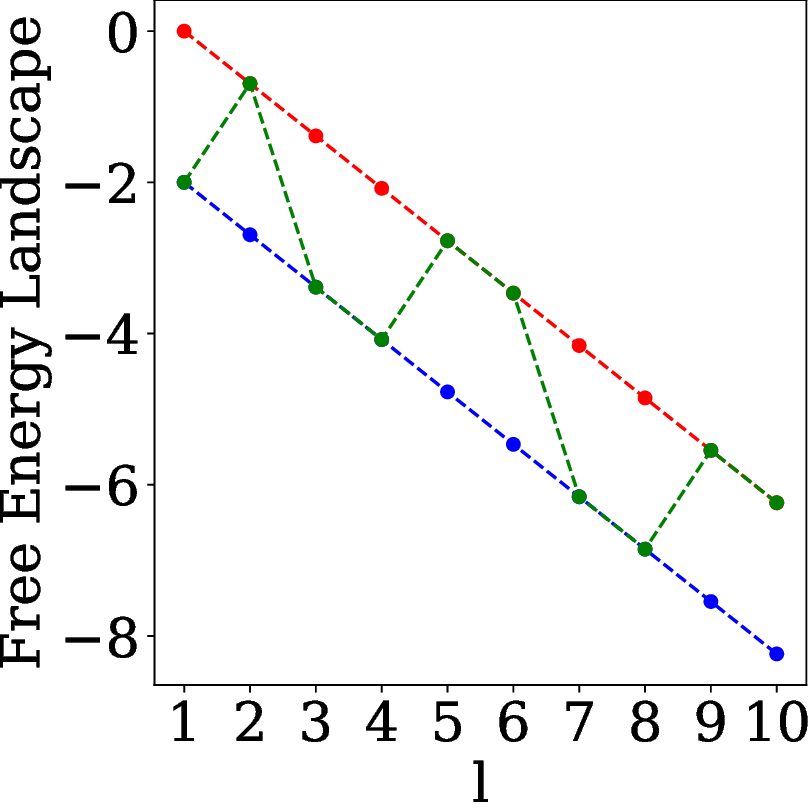}
      \caption{\label{fig:lchoice.ihet}}
    \end{subfigure}
        \begin{subfigure}[t]{0.23\textwidth}
          \includegraphics[width=1.0\linewidth]{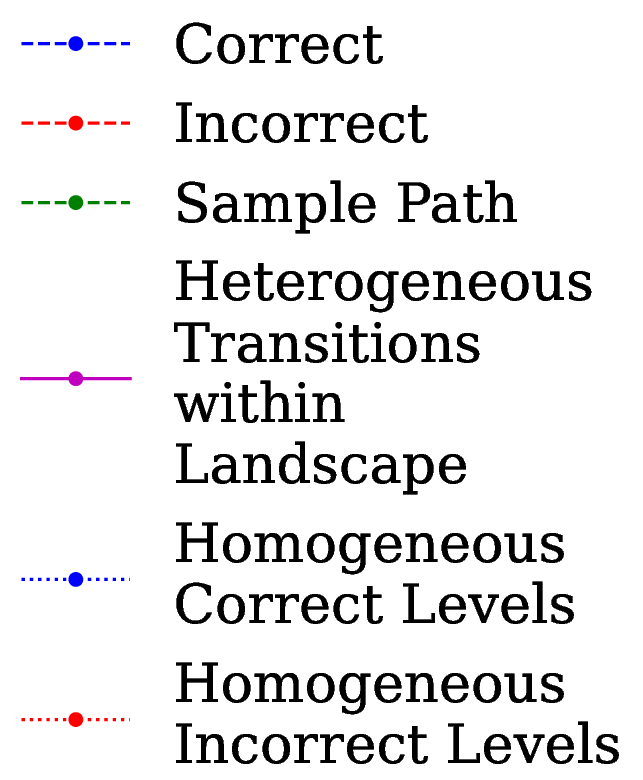}

        \end{subfigure}
\caption{ 
%{\color{purple} Re:sign of G, I was concerned that G could be interpreted as a free energy parameter rather than free energy itself. I changed all y labels to explicitly say 'Free energy landscape' to do away with this ambiguity. Thought since I include the template I could also include gtt matrices - do these add anything?} 
Conceptual diagram of the ``choice of landscape'' model. Three cases are considered: (a) Heterogeneous correct monomer interactions $\Delta G_{TT} = \begin{pmatrix}
    2 & 0,& 0 & 4
\end{pmatrix}$. (b)  Heterogeneous incorrect monomer interactions $\Delta G_{TT} = \begin{pmatrix}
    4 & 0,& 2 & 4
\end{pmatrix}$. (c) Homogeneous copying $\Delta G_{TT} = \begin{pmatrix}
    2 & 0,& 0 & 2
\end{pmatrix}$. The dashed red line represents the landscape that must be traversed to form incorrect pairings, dashed blue represents the landscape that must be traversed to form correct pairings. The dashed green line is a sample path including both correct and incorrect monomers. This sample path is coloured purple instead when transitions occur within a landscape with free-energy changes altered as a result of heterogeneity.  Dotted lines in graphs (a) and (b) represent landscapes of homogeneous analogs (i.e. free energy landscapes of the two homogeneous systems that involve a template with only monomer $1$ and only monomer $2$). 
A template $\mathbf{x} = 1112212112$ is used throughout, and the sample path taken by the dashed green line corresponds to $\mathbf{y} = 1212122121$.  \label{fig:LandscapeChoice}}
\end{figure*}

Heterogeneity on the correct monomers tends to make copying both slower (up to modifications due to waiting time) and more error-prone, while heterogeneity on incorrect monomers tends to make copying faster and more accurate (in some cases, only the average relative error is made better). We now attempt to explain these results. We can divide the overall coarse-grained state-space into two: one in which the correct monomer is bound at the tip, and one in which the incorrect monomer is bound at the tip. Due to detachment of the copy behind the tip, this division is enough to specify the chemical free energy at each step (the same does not apply to models of NTC). Copying can then be thought of as a special example of 1D diffusion with a choice of free-energy landscape at each step (Figure \ref{fig:LandscapeChoice}). The addition of a correct monomer corresponds to moving through a more favourable landscape (the `correct landscape'), while the addition of an incorrect monomer involves moving through a less favourable landscape (the `incorrect landscape'). To understand the effects of heterogeneity, we only need to consider transitions with free-energy changes modified by heterogeneity (purple lines in Figure \ref{fig:LandscapeChoice}), as all other transitions would be found in equivalent forms in corresponding homogeneous copying systems. Observe that the introduction of heterogeneity in the interaction of correct monomers results in an increased roughness of the correct landscape, while the introduction of heterogeneity in the interaction of incorrect monomers results in an increased roughness of the incorrect landscape. This relative roughness is particularly evident as the overall variability in $\Delta G_{\mathbf{x},l}$ is constrained within finite bounds (Figure \ref{fig:FETemp}).

Upward slopes tend to slow down motion more than downward slopes tend to speed up motion, and this asymmetry drives the well-documented observation that rough landscapes tend to be harder to traverse than smoother ones \cite{Sinai1983}. A ``choice of landscape'' model therefore explains why heterogeneity in the interaction of correct monomers tends to increase errors - traversing the incorrect landscape becomes more favourable relative to the correct one (compared to a copying system with homogeneous interactions). Similarly,  incorrect monomer interaction heterogeneity tends to reduce errors, as the incorrect landscape becomes harder to traverse. This effect is most robust when viewed in terms of averaging over relative errors at each site rather than average absolute errors. A sweep over copying systems with various heterogeneous discrimination factors for both backward and forward discrimination at $\Delta G_{\textnormal{pol}} = 0$ (we expect the effect to be stronger with weaker driving) uniformly shows $\langle\ln{\frac{ \epsilon }{\epsilon_I}} \rangle > 0$ when correct monomer interactions are heterogeneous and  $\langle\ln{\frac{\epsilon}{\epsilon_I}} \rangle < 0$ when incorrect monomer interactions are heterogeneous (figures are presented in Appendix \ref{app:ParamSweep}).

 The sign of $\langle\ln{\frac{\epsilon}{\epsilon_I}}\rangle$ influences the sign of $\ln{\frac{\langle \epsilon \rangle}{\langle \epsilon_I \rangle}}$, but the latter may experience a sign reversal, for instance if monomer $2$ is the monomer that experiences an error reduction and $\epsilon_2 \ll \epsilon_1$ already.
This argument suggests that, generally, the potential benefits of heterogeneity on copying error rates are more limited than the potential drawbacks. 
Heterogeneity will not tend to increase the probability of finding a correct monomer after another correct monomer, but instead can reduce the probability of finding an incorrect monomer placed after another incorrect monomer by making the landscape of incorrect monomers more difficult to traverse. However, finding an incorrect monomer after another incorrect monomer is usually a rare event for a good copying system, and reducing the occurrences of such events would have a smaller impact on the average error rates of a copying system. Conversely, deleterious effects would be expected to be larger as correct monomer pairs would be dominant in a good copying system, and there is more room to reduce the probability of consecutive correct monomers. This reasoning is consistent with the fact that the deterioration factors in Figure \ref{fig:Correct} tend to be larger than the improvement factors in Figure \ref{fig:Incorrect}. 

The observed trends in $\ln{\frac{\langle V \rangle}{\langle V_I \rangle}}$ also follow from this ``choice of landscape'' model after a few additional considerations. When correct monomer interactions are made heterogeneous, moving through the correct landscape becomes more difficult, and hence more state visits are required. This trend is again very robust (Appendix \ref{app:ParamSweep}). On the other hand, when incorrect monomer interactions are made heterogeneous, barriers to movement in the incorrect landscape are increased. Here, we need to consider two separate effects. First, traversal of the incorrect landscape gets harder, increasing the number of visits to states on average. Second, the system preferentially traverses the correct landscape. The expected number of steps of the coarse-grained model along the correct landscape will be less than that along the incorrect landscape since the correct tip will be harder to remove on average, thus decreasing state visits. The second effect is usually dominant (Appendix \ref{app:ParamSweep}), but exceptions occur when the correct and incorrect landscapes have a small free-energy gap (implying a small benefit to visitation counts when traversing the correct landscape).  The sign of $\ln{\frac{\langle V \rangle}{\langle V_I \rangle}}$ feeds forward into the sign of $\ln{\frac{\langle \tau \rangle}{\langle \tau_I \rangle}}$, but the latter may experience a sign reversal depending on the waiting times at each template monomer. 

\begin{figure}
\includegraphics[width=1.0\linewidth]{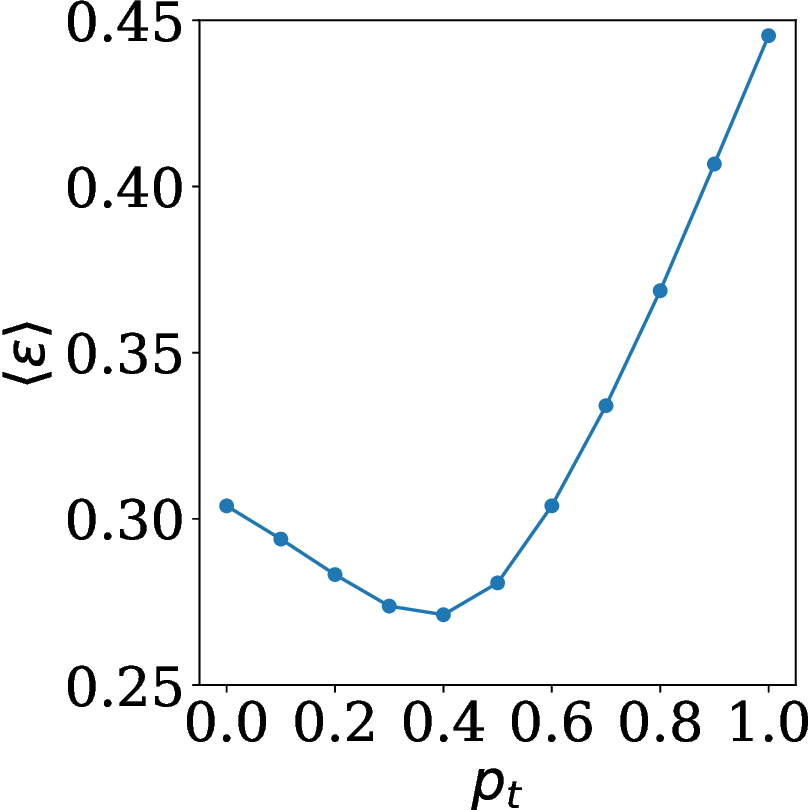}
    \caption{ Graph of $\langle \epsilon \rangle$ against $p_t$ showing $\langle \epsilon \rangle$ dipping below the values of $\langle \epsilon \rangle$ at both $p_t = 0$ and $p_t = 1$. Parameters are $\Delta G_{\textnormal{pol}} = -0.3$, $\Delta G_{TT,11} = \Delta G_{TT,22} = 6.0$, $\Delta G_{TT,12} = 5.0$ and $\Delta G_{TT,21} = 0.0$.     
    \label{fig:veryconvex}}
\end{figure}

We perform a sweep over parameters to identify  regimes where heterogeneity results in the greatest improvement in error rate (results in Appendix \ref{app:ParamSweep2}). Interestingly, our parameter sweep revealed parameter sets where a heterogeneous system has a lower mean error than either of its constituent monomers used in isolation. Consider for instance the plot in Figure \ref{fig:veryconvex}. There is a clear minimum in $\langle \epsilon \rangle$ at $p_t = 0.4$, implying that some combination of the two considered monomers performs better than either one on its own. Thus, it is generally untrue that the error performance of a heterogeneous copying system is bounded by the error performance of its constituent monomers.

\subsection{Entropy, Mutual Information and Efficiency in the Low $\Delta G_{\textnormal{pol}}$ Regime \label{sec:Entropy}}

Recall that $Y$ is the random variable representing the product sequence, and $X$ is the random variable representing the template sequence. As the sequence of copy polymers $Y$ conditioned on a template $\mathbf{x}$ is Markov, the template-conditioned entropy rate of complete polymers can be calculated as follows \cite{Gaspard2014,Gaspard2017,Poulton2019}.
\begin{align}
     h&(Y|X=\mathbf{x})= \lim_{L \rightarrow \infty}\frac{1}{L} \Sigma_{l=1}^LH(Y_l|Y_{1:l-1},X=\mathbf{x}) \nonumber\\ &= \lim_{L \rightarrow \infty}\frac{1}{L} \Sigma_{l=1}^L H(Y_l|Y_{l-1},X=\mathbf{x}) \nonumber \\&= -\lim_{L \rightarrow \infty}\frac{1}{L} \Sigma_{l=1}^L\Sigma_{ij}  P_{\mathbf{x},l-1}(m_{l-1}) \nonumber \\&\;\;\;\;\;\;\;\;P_{\mathbf{x},l}(m_l|m_{l-1})\ln{P_{\mathbf{x},l}(m_l|m_{l-1})} 
.\end{align}
Assuming this entropy rate is self-averaging for stationary distributions $p(\mathbf{x})$ of $X$ \cite{Gaspard2017}, $h(Y|X=\mathbf{x})$ is the same for any typical \cite{Shannon1948} instance $\mathbf{x}$ of $X$. Then,   
\begin{align}
    h(Y|X) &= \Sigma_{\mathbf{x}} p(\mathbf{x})h(Y|X=\mathbf{x}) \nonumber\\&=  h(Y|X=\mathbf{x})
.\end{align}

For our investigations here, we average entropy over the middle $80\%$ of a long template to mitigate edge effects. Similarly to the error rate, the entropy rate captures uncertainty in the identity of successive monomers in the copy. It is, however, the more thermodynamically relevant parameter as it directly bounded by the drive $\Delta G_{\textnormal{pol}}$. For $\Delta G_{\textnormal{pol}} > 0$, copying can in principle be arbitrarily accurate. However, if $\Delta G_{\textnormal{pol}} < 0$ (as $\Delta G_{\textnormal{pol}} = \Delta G_{BB} + \ln{\frac{[M]}{[M]_{\textnormal{eff}}}}$ and generally $[M] < [M]_{\textnormal{eff}}$, this regime corresponds to one where the chemical free-energy drop due to backbone formation cannot compensate for the entropic drop as a result of attaching a monomer to the polymer tail), then there is a fundamental limit on how accurate a copying system can be when operating in the limit of long polymer length \cite{Poulton2019}.

At equilibrium, copying has no accuracy as every possible polymer product has the same chemical free energy given by $(L-1) \Delta G_{\textnormal{pol}}$. For our model of copying, this equilibrium point occurs when $\Delta G_{\textnormal{pol}} = -\ln{2}$, as this is the point where the forward drive provided by the sequence entropy of the copied polymer is exactly cancelled by the polymerization drive $\Delta G_{\textnormal{pol}}$, resulting in no net free-energy change with increasing length of the copy polymer \cite{Poulton2019}. Furthermore, if  $-\ln 2 < \Delta G_{\textnormal{pol}} < 0$, then the entropy rate decrease relative to the equilibrium entropy rate of $\ln{2}$ per monomer, $\ln{2} - h(Y|X)$, is bounded by $\Delta G_{\textnormal{pol}}+\ln{2}$, leading to a thermodynamic measure of efficiency
\begin{equation}
    \eta_{\textnormal{Therm}} = \frac{\ln{2}-h(Y|X)}{\Delta G_{\textnormal{pol}}+\ln{2}}
.\end{equation}

\noindent $\eta_{\textnormal{Therm}}$ is the efficiency with which free energy is converted into the low entropy of the product state. Poulton observed in \cite{Poulton2019} that the entropy drop during (homogeneous) copying tends to be quite far from the fundamental bound at most values of $\Delta G_{\textnormal{pol}}$, and hence it would be instructive to see if we can get closer to this fundamental bound by permitting heterogeneity. One important caveat in this discussion is that this bound is only valid in the infinite length limit \cite{Poulton2021}. At the polymer length we consider, $L = 10^4$, the behaviour of most monomers far from polymer boundaries approaches the behaviour for the infinite length limit, and hence while methodological limitations force us to work with finite length polymers, we are endeavouring to make statements on the efficiency of copying of infinite length polymers.  

\begin{figure*}
\begin{tabular}[t]{ccc}

    \begin{subfigure}[b]{0.43\textwidth}
      \includegraphics[width=1.0\linewidth]{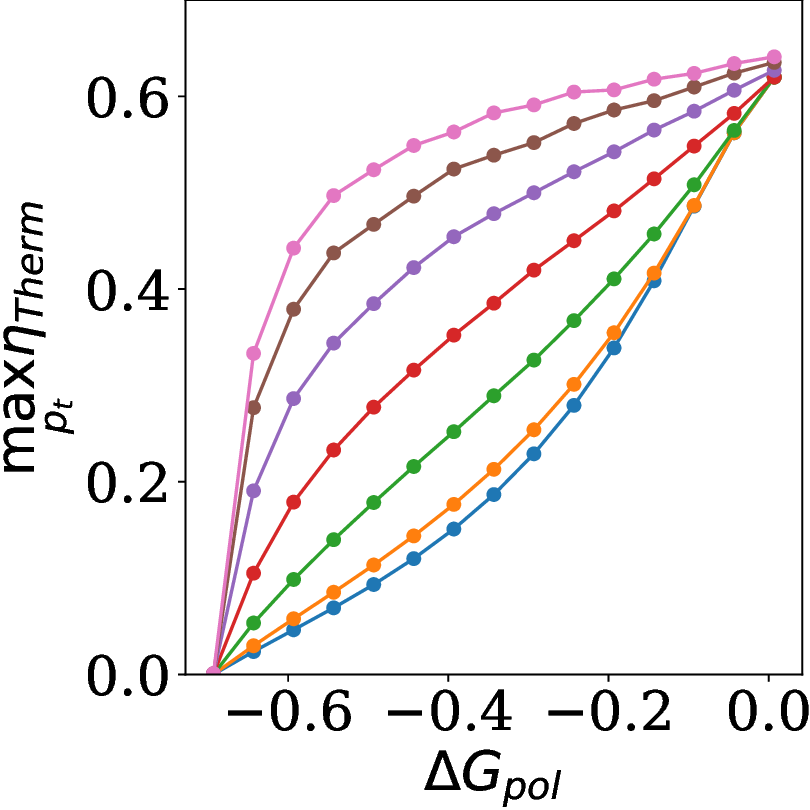}
      \caption{\label{fig:etaTemp}}
    \end{subfigure}
    \begin{subfigure}[b]{0.43\textwidth}
        \centering
      \includegraphics[width=1.0\linewidth]{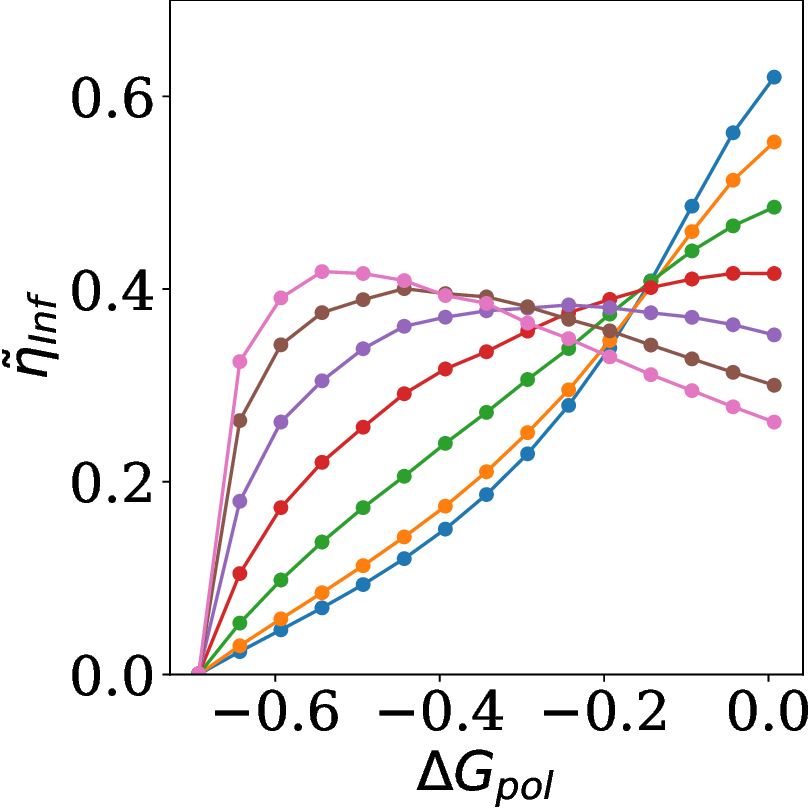}
      \caption{\label{fig:etaTempInf}}
    \end{subfigure} & 
        \begin{subfigure}[b]{0.14\textwidth}
          \includegraphics[width=0.8\linewidth]{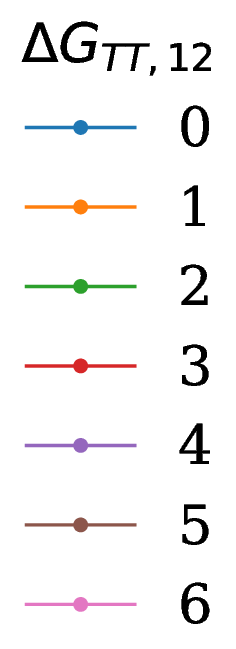}
        \end{subfigure}
    \\
    \begin{subfigure}[b]{0.43\textwidth}
      \includegraphics[width=1.0\linewidth]{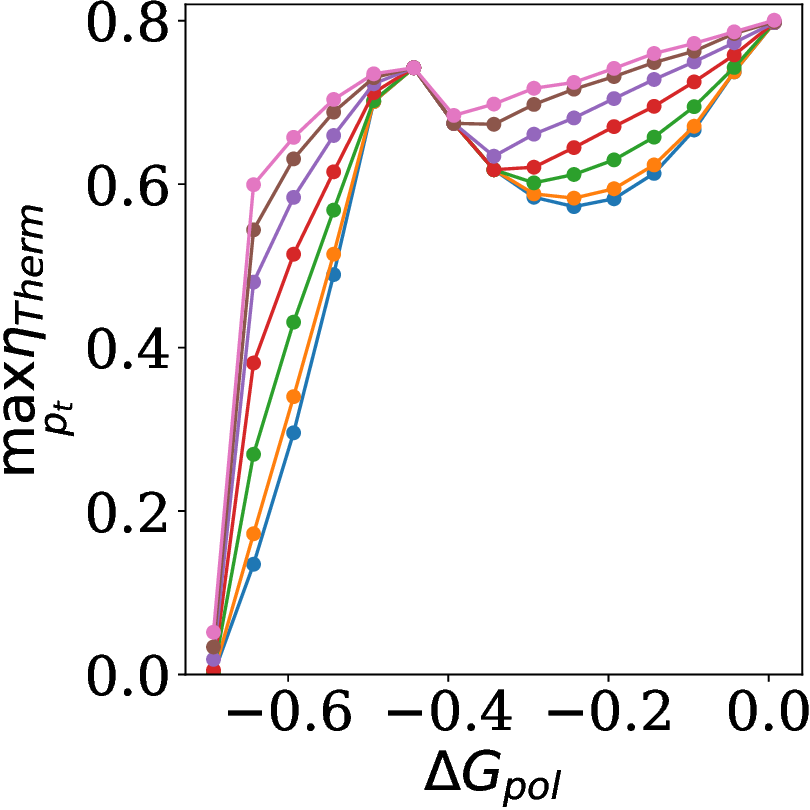}
      \caption{\label{fig:etaComb}}
    \end{subfigure}
    \begin{subfigure}[b]{0.43\textwidth}
        \centering
      \includegraphics[width=1.0\linewidth]{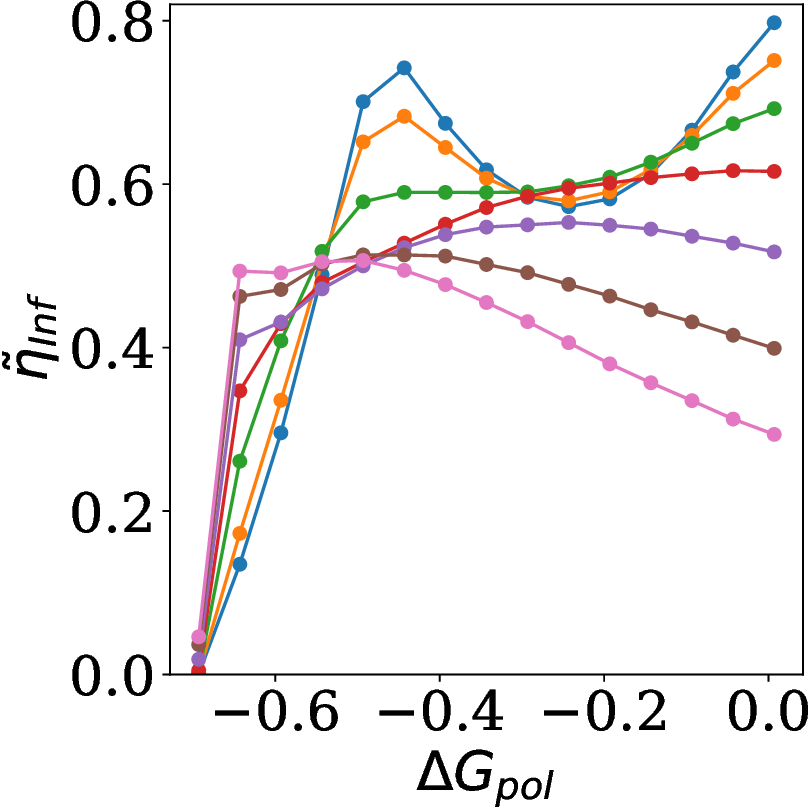}
      \caption{\label{fig:etaCombInf}}
    \end{subfigure} 

\end{tabular}
\caption{Thermodynamic ($\eta_{\textnormal{Therm}}$) and (estimated) information ($\tilde{\eta}_{\textnormal{Inf}}$) efficiencies as a function of $\Delta G_{\textnormal{pol}}$ when incorrect monomer interactions are made heterogeneous by varying $\Delta G_{TT}(1,2)$ from $0$. Plots for backward propensity discrimination are shown in (a, b) and plots for forward propensity discrimination are shown in (c, d). For both types of discrimination, entropy rate is reduced by heterogeneity, and hence  $\eta_{\textnormal{Therm}}$ is increased (a,c). $\tilde{\eta}_{\textnormal{Inf}}$ for backward propensity discrimination is increased by heterogeneity up to about $\Delta G_{\textnormal{pol}} = -0.2$ to $-0.3$, after which it starts decreasing. For forward propensity discrimination, heterogeneity tends to reduce $\eta_{\textnormal{Inf}}$ except for very low values of $\Delta G_{\textnormal{pol}} < -0.55$.}
\end{figure*}

Setting $L = 10^4$, and using the coarse-grained model defined by the propensities in equations \ref{eq:phipslowbind}-\ref{eq:phimslowpol} with $\Delta G_{TT}(1,1) = \Delta G_{TT}(2,2) = 6$, $\Delta G_{TT}(2,1) = 0$ and varying $\Delta G_{TT}(1,2)$ from $0$ (the homogeneous case) to $6$, we plot $\max_{p_t} \eta_{\textnormal{Therm}}$ in the low $\Delta G_{\textnormal{pol}}$ region as a function of $\Delta G_{\textnormal{pol}}$ for copying with backward propensity and forward propensity discrimination in Figures \ref{fig:etaTemp} and \ref{fig:etaComb}, respectively. We see quite significant increases in efficiency relative to the homogeneous case. The source of this increased efficiency is the mechanism identified in Section \ref{sec:ErrorTime}: roughness in the free-energy landscape of incorrect monomers makes correct monomers more favourable. Interestingly, these increases in efficiency occur despite combining a monomer with another monomer having worse discrimination than itself, hence (congruent with our observations of error performance in Section \ref{sec:ErrorTime}) the thermodynamic performance of a heterogeneous copying system is not bounded by the performance of its constituent monomers. 
%In the forward discrimination case, $\max_{p_t} \eta_{\textnormal{Therm}}$ converges at the secondary peak, identified in \cite{Poulton2019}, at around $\Delta G_{\textnormal{pol}} = -0.44$ (given our sampling rate). 
Note that because of finite size effects and sampling, $h(Y|X)$ does not quite achieve $\ln{2}$ at $\Delta G_{\textnormal{pol}} = -\ln{2}$, so the leftmost data points are calculated at $\Delta G_{\textnormal{pol}} = -\ln{2} + 0.0001$ and $L = 10^5$ to prevent numerical instabilities.

Instead of the entropy rate drop $\ln{2} - h(Y|X)$, we  may instead wish to consider the mutual information rate $\dot{I}(X,Y) = h(Y) - h(Y|X)$ where $h(Y)$ is the product entropy rate unconditioned on the template, 
\begin{align}
h(Y) = \lim_{L \rightarrow \infty}\frac{1}{L} \Sigma_{l=1}^LH(Y_l|Y_{1:l-1})
\end{align}
\noindent Per Shannon's Coding Theorem \cite{Shannon1948}, the mutual information quantifies our ability to uniquely distinguish a message encoded in our original template from our obtained copy or, equivalently, tell which template a copy came from. The mutual information maximized over input distributions defines a channel capacity $C = \max_{p(X)} \dot{I}(X,Y)$. We shall now explore this channel capacity, although for practical purposes we will consider optimizing over Bernoulli input distributions. In the case of homogeneous copying, the channel capacity is exactly equal to the entropy rate drop $\ln{2} - h(Y|X)$, as the entropy rate $h(Y|X)$ is independent of the template distribution, and due to symmetry, $h(Y) = h(X) = \ln{2}$ when the template is drawn from a Bernoulli distribution with $p_t = \frac{1}{2}$. To understand how $C$ differs from the entropy drop $\ln{2} - h(Y|X)$ in the case of heterogeneous copying, consider a hypothetical copying system that always maps template monomers to a copy monomer of type $1$. This copying system would have a maximum entropy rate drop of $\ln{2}$, but a mutual information rate of $0$. For a given template distribution,  $\dot{I}(X,Y) = h(Y) - h(Y|X) \leq \ln{2} - h(Y|X)$ and $\ln{2} - h(Y)$ can be regarded as the proportion of the entropy drop from the equilibrium distribution that does not contribute towards information transfer. Maximizing over $p(X)$ should maintain our inequality such that $C = \max_{p(X)} h(Y) -  h(Y|X) \leq \max_{p(X)} \ln{2} - h(Y|X)$. We may thus define an information efficiency $\eta_{\textnormal{Inf}} \leq \max_{p(X)} \eta_{\textnormal{Therm}}$ as follows.  

\begin{equation}
    \eta_{\textnormal{Inf}} = \frac{C}{\Delta G_{\textnormal{pol}}+\ln{2}}
.\end{equation}

The entropy rate $h(Y)$ is difficult to calculate exactly for heterogeneous polymer copying. It can be expressed as follows:
\begin{align}
    h(Y) &= \lim_{L \rightarrow \infty}\frac{1}{L}\Sigma_{l=1}^LH(Y_l|Y_{1:l-1}) \nonumber\\&= -\lim_{L \rightarrow \infty}\frac{1}{L}\Sigma_{l=1}^L\Sigma_{m_{1:l}}p(m_1,m_2,..., m_l) \nonumber\\  &\;\;\;\;\ln{p(m_l|m_1,m_2,..,m_{l-1})},
\end{align}
 These probabilities are marginalised over the template $X$. The joint and conditional probabilities can be expanded as follows:
\begin{align}
    p(m_l|&m_1,m_2,...m_{l-1}) \nonumber\\
    &= \Sigma_{\mathbf{x}} p(m_l|m_1,m_2,..,m_{l-1},\mathbf{x})p(\mathbf{x}|m_1,m_2,...m_{l-1}) \nonumber\\
    &= \Sigma_{\mathbf{x}} p(m_l|m_{l-1},\mathbf{x})p(\mathbf{x}|m_1,m_2,...m_{l-1}), \label{eq:conditionalY}\\
    p(m_1,&m_2,...m_{l-1},m_l) \nonumber\\
    &= \Sigma_{\mathbf{x}} p(m_1,m_2,..,m_{l-1},m_l,\mathbf{x}) \nonumber\\
    &= p(m_1,m_2,...m_{l-1}) \Sigma_{\mathbf{x}} p(m_l,\mathbf{x}|m_1,m_2,...m_{l-1}) \nonumber\\
    &= p(m_1,m_2,...m_{l-1}) p(m_l|m_1,m_2,...m_{l-1}) 
.\end{align}

In Equation \ref{eq:conditionalY}, the $p(\mathbf{x},m_1,m_2,...m_{l-1})$ term means that we cannot assume $p(m_l|m_1,m_2,...m_{l-1}) = p(m_l|m_{l-1})$. In order to proceed, we first make the assumption that $Y$ having marginalised over $X$, has finite-length templated-mediated correlations and can be approximated as an $i^{th}$-order Markov process (Appendix \ref{app:EntropyInfo}). Remarkably, numerical evaluations show that the change in estimated entropy going from $i = 1$ to $i = 8$ is not significant for the parameters we consider when sampling the middle $80\%$ of a length $L = 10^4$ template. Furthermore, by sampling length $L = 10^5$ and $L = 10^6$ templates at select parameter values, we found that a significant proportion of this error (at least for some regimes) is likely attributable to sampling issues instead of genuine long-range correlations (Appendix \ref{app:EntropyInfo}), and hence we are justified in estimating entropies by treating $Y$ as a $1^{st}$-order Markov chain. Formally, this assumption means that the information efficiency we calculate $\tilde{\eta}_{\textnormal{Inf}}$ is an upper bound of some true information efficiency $\eta_{\textnormal{Inf}}$ for Bernoulli templates. However, appendix \ref{app:EntropyInfo} suggests that $\tilde{\eta}_{\textnormal{Inf}}$ is a very tight upper bound. On the other hand, $\eta_{\textnormal{Inf}}$ may be higher when considering non-Bernoulli templates.

In Figures \ref{fig:etaTempInf} and \ref{fig:etaCombInf}, we plot our estimated $\tilde{\eta}_{\textnormal{Inf}}$ as a function of $\Delta G_{\textnormal{pol}}$ for the backward propensity and forward propensity discrimination regimes, respectively. As expected, we observe $\tilde{\eta}_{\textnormal{Inf}}\leq \max_{p_t} \eta_{\textnormal{Therm}}$. Increases in efficiency relative to the homogeneous case $\Delta G_{TT}(1,2)=0$ are better preserved with backward propensity discrimination compared to forward propensity discrimination. $P(Y)$ is, in essence, more skewed in the forward propensity discrimination case, which is detrimental for mutual information.

\section{Conclusion}

Using minimal models with fine-grained steps \cite{Juritz2021}, we have investigated how heterogeneity interacts with separation to affect copying error, velocity and thermodynamic efficiency in STC. We have thus far investigated heterogeneity in monomer binding energies and kinetics; the effects of heterogeneous monomer concentrations,
which would naturally manifest as heterogeneity in binding rates, will be an interesting topic for future research. Our first contribution is an approach for extending Qureshi et al's coarse-graining method \cite{Qureshi2023} to generalize to heterogeneous systems. We have thus far used this method to find polymer completion times for our coarse-grained model. A natural extension would be to use the method to investigate a heterogeneous version of kinetic proofreading \cite{Hopfield1974,Qureshi2023}, where it can be used to find the average expected free-energy consumption per polymer copy. 

Using this method, we were able to characterize velocity profiles of a heterogeneous separating templated copolymerization system. In contrast to non-separating templated copolymerization systems \cite{Gaspard2017}, we do not observe a long tail of zero velocity as equilibrium is approached. This absence makes sense in light of the template-dependent free-energy profiles of partial copies, which have significantly higher barriers (scaling as $\sqrt{l_D}$ for a length scale $l_D$) in the case of NTC. Put another way, each monomer type in heterogeneous NTC has a different pseudo-equilibrium point, and copying with a drive lower than a monomer's pseudo-equilibrium point makes traversing a long stretch of said monomer more difficult. This effect is totally absent in the case of STC, since the sequence-specific interactions with the template are transient. We expect that this observation would generalise to more realistic models of transcription or translation, as long as the copy continuously separates from the template and the number of copy monomers interacting with the template at a given time is effectively bounded. 
 
Our results on the effect of heterogeneity on error rates are more surprising. It was not initially obvious that discriminating on correct monomer interactions, arguably the more natural form of heterogeneous discrimination, would tend to increase errors. There is evidence that in protein translation, the ribosome grips on tRNA have identical strengths \cite{Grosjean2016}. This grip is analogous to our transient copy-template bond as it is not persistent, and so it is plausible that the homogeneity of this grip was selected for due to similar mechanisms that increase heterogeneous error rates (consider that in our case, error increase factors of up to $e^{1.25} = 3.5$-fold were observed). In the context of artificial systems, our results would suggest that it would be wise to minimize interaction heterogeneity on correct pairs of monomers. 

The relative error reduction observed when discriminating on incorrect monomers is equally surprising. Applying heterogeneity to incorrect monomer interactions could be a useful design motif for the design of accurate copiers, and we have made some attempt towards this goal by scanning over parameter space. We find that having backward and forward propensity discrimination on separate monomers, with roughly equal effect sizes, tends to produce the best results. However, we emphasize that the usefulness of this design motif depends on the space of accessible parameter sets. Under some chemical restrictions (in particular, if we had to operate in the low $\Delta G_{\textnormal{pol}}$ regime, or if our copying mechanism only permits backward propensity discrimination) it may make sense to apply heterogeneity to incorrect monomer interactions for the roughly 10-20\% improvement in error rates. Note that operating under low $\Delta G_{\textnormal{pol}}$ may be necessary for synthetic systems.   Reassuringly, there does not appear to be a trade-off between using heterogeneity to reduce error or using it to speed up copying, as regimes that tend to reduce error tend to reduce copying time as well. Both performance measures prefer heterogeneity on incorrect monomer interactions as opposed to correct monomer interactions, as heterogeneity on incorrect pairs makes it harder to move through the landscape of incorrect monomers.  

Our final result relates to the entropy drop and mutual information in heterogeneous copying systems.  In homogeneous copying, the entropy drop from equilibrium is exactly the mutual information when template monomers are equally distributed. In contrast, we showed that there is a meaningful difference between this entropy drop and mutual information in the case of heterogeneous copying due to the skewing of the copy polymer distributions. For a given template distribution, we can evaluate a channel capacity, the mutual information maximized over input distributions. Channel capacity can in turn be used to define an information efficiency measure. We showed that both thermodynamic and information efficiencies can be improved by heterogeneity in the incorrect monomer interactions at low $\Delta G_{\textnormal{pol}}$, but that information efficiency is always less than or equal to thermodynamic efficiency. Per Shannon's channel coding theory, channel capacity represents the minimal bitrate above which arbitrarily accurate decoding is possible, and hence some of the thermodynamic entropy drop in heterogeneous copying systems does not contribute to reducing this bitrate. On the other hand, for homogeneous copying systems there is always a template distribution ($p_t = \frac{1}{2}$) where the thermodynamic entropy drop in the copy (relative to equilibrium) fully contributes to the reduction of this bitrate. 

\section*{Funding}
JEBG was supported by an Imperial College President’s PhD Scholarship, BJQ by the European Research Council under the European Union’s Horizon 2020 Research and Innovation Program (Grant Agreement No. 851910) and TEO by a Royal Society University Research Fellowship.

\section*{Data Availability}
Code and data are available at doi.org/10.5281/zenodo.14003309

\section*{Conflict of Interest}

All authors certify that they have no affiliations with or involvement in any organization or entity with any financial interest or non-financial
interest in the subject matter or materials discussed in this manuscript.

\section*{Author Contributions}

JEBG, BJQ and TEO planned the research. JEBG performed the research. JEBG, BJQ and TEO wrote the manuscript.

\appendix

\section{Gaspard's Method via Absorbing Probabilities of Markov Chains \label{app:Gaspard}}
Consider again Figure \ref{fig:Markov} in the main text. We wish to calculate the probability $R_{\mathbf{x},l}(m_{l+1}|m_{l-1} m_l)$ of being absorbed into a complete polymer with $m_{l+1}$ after $m_l$ starting from the initial state $(\mathbf{x},\&m_{l-1}m_l)$, without going back a step. To aid in our calculations, we introduce the absorption probabilities $R_{\mathbf{x},l}(m_{l+1}| m_{l-1} m_l m_r)$, the probability of completing polymerization with $m_{l+1}$ after $m_l$ starting from a system state $(\mathbf{x},\&m_{l-1}m_lm_r)$ in the context of the Markov chain given in Figure \ref{fig:Markov} ($m_{l+1}$ may be the same or different from $m_r$). To clarify, this probability includes the probability that we move backward from $(\mathbf{x},\&m_{l-1}m_lm_r)$ to $(\mathbf{x},\&m_{l-1}m_l)$ and then after an arbitrary non-absorbing set of moves eventually absorb into $(\mathbf{x},\&m_{l-1}m_lm_{l+1}...0)$, where $0$ indicates a detached, complete polymer. Note moving back two consecutive times from $m_r$ always results in the backward absorbing state. Also, if $m_{l+1} = m_r$, then absorption without moving back is included in the absorption probability $R_{\mathbf{x},l}(m_{l+1}| m_{l-1} m_l m_r)$.   

We can now use the fact that for an arbitrary Markov chain, the absorbing probability to an absorbing state $A$ from a transient state $x$ can be calculated as $p_{\textnormal{abs}}(A|x) = \Sigma_i p(x_i|x)p_{\textnormal{abs}}(A|x_i) + p(A|x)$ where $x_i$ are all states with transitions from $x$ (setting $p(A|x) = 0$ if $x$ cannot directly transition into $A$).  Reminding readers that $p_{\mathbf{x},l}(m_{l+1}|m_{l-1} m_l)$ are the local transition probabilities for the addition of monomer $m_{l+1}$ in a single coarse-grained step, the system of equations \ref{eq:AbsSys1} and \ref{eq:AbsSys2} are obtained. 
\begin{widetext}
\begin{align}
    R_{\mathbf{x},l}(m_{l+1}|m_{l-1} m_l) &= \Sigma_{m_r} p_{\mathbf{x},l}(m_r|m_{l-1} m_l) R_{\mathbf{x},l}(m_{l+1}|m_{l-1} m_l m_r) \label{eq:AbsSys1},\\
    R_{\mathbf{x},l}(m_{l+1}|m_{l-1} m_l m_r) &= \begin{cases}
    ( 1 - Q_{\mathbf{x},l+1}(m_lm_r) )R_{\mathbf{x},l} (m_{l+1}|m_{l-1} m_l) &m_{l+1} \neq m_r\\
    ( 1 - Q_{\mathbf{x},l+1}(m_lm_r) )R_{\mathbf{x},l}(m_{l+1}|m_{l-1} m_l) +  Q_{\mathbf{x},l+1}(m_lm_{l+1})\;\; &m_{l+1} = m_r
    \label{eq:AbsSys2}\end{cases}
.\end{align}

Substitution of \ref{eq:AbsSys2} into \ref{eq:AbsSys1} yields 
\begin{align}
     R_{\mathbf{x},l}(m_{l+1}|m_{l-1} m_l) = p_{\mathbf{x},l}(m_{l+1}|m_{l-1} m_l) &Q_{\mathbf{x},l+1}(m_lm_{l+1}) \nonumber\\ &+ \Sigma_{m_r} p_{\mathbf{x},l}(m_r|m_{l-1} m_l)( 1 - Q_{\mathbf{x},l+1}(m_lm_r) )R_{\mathbf{x},l}(m_{l+1}| m_{l-1} m_l)\label{eq:AbsSys3}.
\end{align}
Equation \ref{eq:AbsSys3} can be rearranged with all $R_{\mathbf{x},l}(m_{l+1}|m_{l-1} m_l)$ terms to the left, yielding
\begin{align}
    R_{\mathbf{x},l}(m_{l+1}|m_{l-1} m_l) &= \frac{p_{\mathbf{x},l}(m_{l+1}|m_{l-1} m_l)Q_{\mathbf{x},l+1}(m_lm_{l+1})}{1 - \Sigma_{m_r} p_{\mathbf{x},l}(m_r|m_{l-1} m_l) + \Sigma_{m_r} p_{\mathbf{x},l}(m_r|m_{l-1} m_l) Q_{\mathbf{x},l+1}(m_lm_r)} \label{eq:AbsSyssol}
.\end{align}
We can perform the substitutions $p_{\mathbf{x},l}(m_{l+1}|m_{l-1} m_l) = \frac{\Phi^+_l(\mathbf{x},m_lm_{l+1})}{\Phi^-_l(\mathbf{x},m_{l-1}m_l) + \Sigma_{m_r}\Phi^+_l(\mathbf{x},m_lm_r)}$ to obtain equation \ref{eq:AbsSyssub}, and then perform the summation in equation \ref{eq:oneIter} of the main text 
to obtain equation \ref{eq:AbsSyssub2}
\begin{align}
    R_{\mathbf{x},l}(m_{l+1}|m_{l-1} m_l) &= \frac{\Phi^+_l(\mathbf{x},m_lm_{l+1})Q_{\mathbf{x},l+1}(m_lm_{l+1})}{\Phi^-_l(\mathbf{x},m_{l-1}m_l) + \Sigma_{m_r} \Phi^+_l(\mathbf{x},m_lm_r) Q_{\mathbf{x},l+1}(m_lm_r) },\label{eq:AbsSyssub}\\
    Q_{\mathbf{x},l}(m_{l-1}m_l) &= \frac{\Sigma_{m_r} \Phi^+_l(\mathbf{x},m_lm_r)Q_{\mathbf{x},l+1}(m_lm_r)}{\Phi^-_l(\mathbf{x},m_{l-1}m_l) + \Sigma_{m_r} \Phi^+_l(\mathbf{x},m_lm_r) Q_{\mathbf{x},l+1}(m_lm_r) } \label{eq:AbsSyssub2}
.\end{align} 
\end{widetext}
 
\noindent A slight difference from the derivation in \cite{Gaspard2017} is that we have opted to express our iteration in terms of $Q$ instead of local velocities $v_{\mathbf{x},l}$ as our derivation of expected visits in Appendix \ref{app:VisitIter} uses $Q_{\mathbf{x},l}$ rather than $v_{\mathbf{x},l}$. However, we emphasize that the underlying mathematics is the same, and the local velocities $v_{\mathbf{x},l}$ in \cite{Gaspard2017} can be obtained using $v_{\mathbf{x},l} = \Sigma_{m_{l+1}} \Phi^+_l(\mathbf{x}, m_lm_{l+1})Q_{\mathbf{x},l+1}(m_lm_{l+1})$.

The monomeric conditional probabilities $P_{\mathbf{x},l+1}(m_{l+1}|m_l)$ can be calculated by considering the Markov chain in figure \ref{fig:Markov} and then omitting the possibility of removal of monomer $m_l$. To understand why, note that for complete polymers, the distribution of $m_{l+1}$ is only affected by the copying trajectory after $m_l$ is added for the final time and then never removed. $P_{\mathbf{x},l+1}(m_{l+1}|m_l)$ is then simply the absorption probability into $(\mathbf{x},\&m_{l-1}m_lm_{l+1}...0)$ for this modified Markov chain. The calculation of absorbing probabilities is identical to our previous derivation up to equation \ref{eq:AbsSyssol}. Our final substitution now must omit the backward rate such that $p_{\mathbf{x},l}(m_{l+1}|m_{l-1} m_l) = \frac{\Phi^+_l(\mathbf{x},m_lm_{l+1})}{ \Sigma_{m_r}\Phi^+_l(\mathbf{x},m_lm_r)}$, resulting in:
\begin{equation}
P_{\mathbf{x},l+1}(m_{l+1}|m_l) = \frac{\Phi^+_l(\mathbf{x},m_lm_{l+1}) Q_{\mathbf{x},l+1}(m_lm_{l+1})} {\Sigma_{m_r} \Phi^+_l(\mathbf{x},m_lm_r) Q_{\mathbf{x},l+1}(m_lm_r)}  
.\end{equation}

\section{Derivation of Coarse-Grained Propensities \label{app:Ben}}

We use the method in \cite{Qureshi2023} to derive our coarse-grained propensities for the model with detachment. Refer to equation \ref{eq:Ben} and assume a starting completed state (as a reminder, completed states are states with $f = 0$) $(\mathbf{x},\&m_{l-1},0)$
and a second completed state $(\mathbf{x},\&m_{l-1}m_l,0)$. In Figure \ref{fig:FineGrained}, there is only one spanning tree rooted at either of these completed states (the leftmost and rightmost states), and a straightforward multiplication of forward rates results in $\Lambda^+_{\mathbf{x}}(n_{l-1}n_l,m_{l-1}m_l)$ (Equation \ref{eq:lambdaplus}), while a multiplication of backward rates results in $\Lambda^-_{\mathbf{x}}(n_{l-1}n_l,m_{l-1}m_l)$ (Equation \ref{eq:lambdaminus}).  

\begin{figure*}
\begin{subfigure}[t]{0.8\textwidth}
  \includegraphics[width=1.0\linewidth]{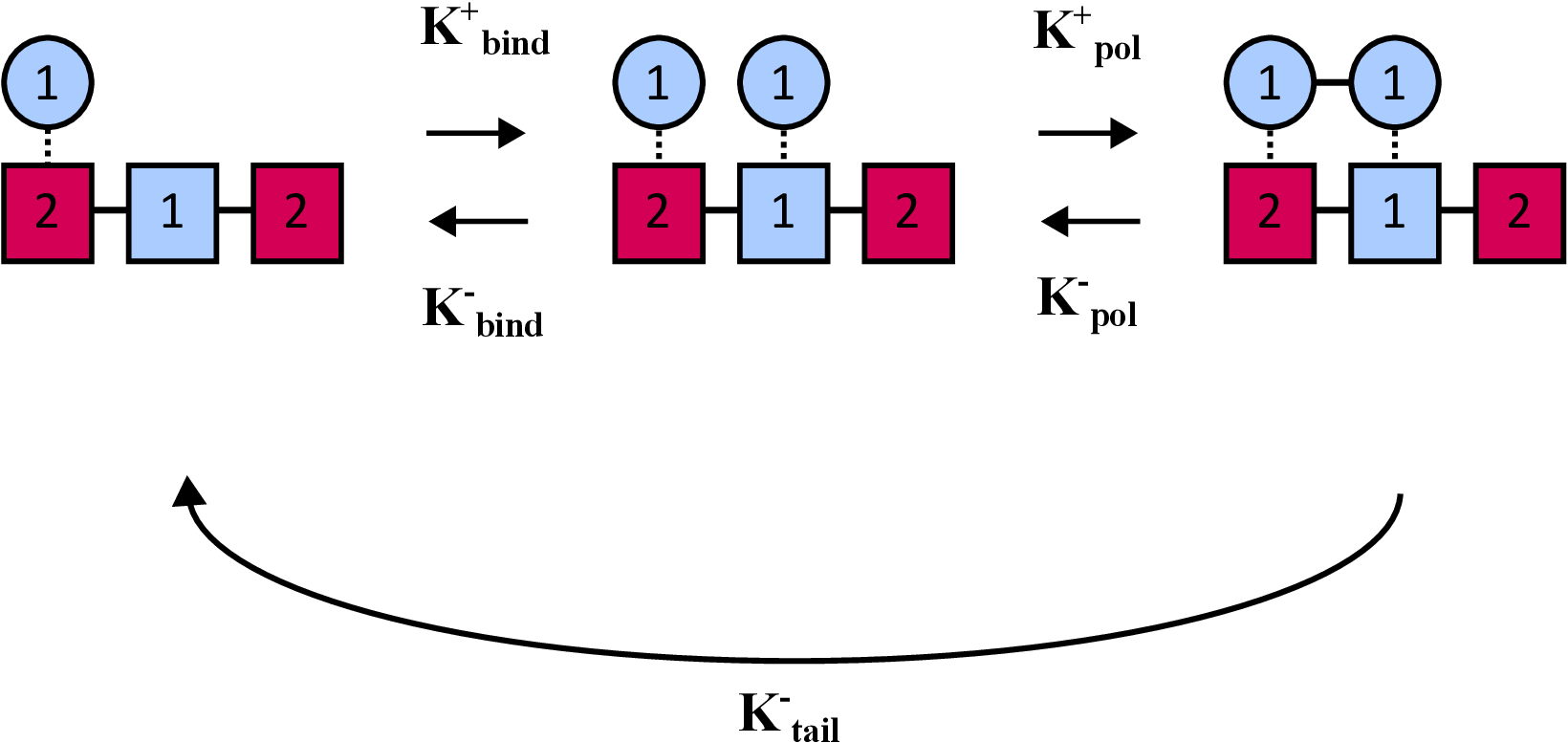}
  \caption{\label{fig:CoarseBen}}
\end{subfigure}
\begin{subfigure}[t]{1.0\textwidth}
    \centering
  \includegraphics[width=1.0\linewidth]{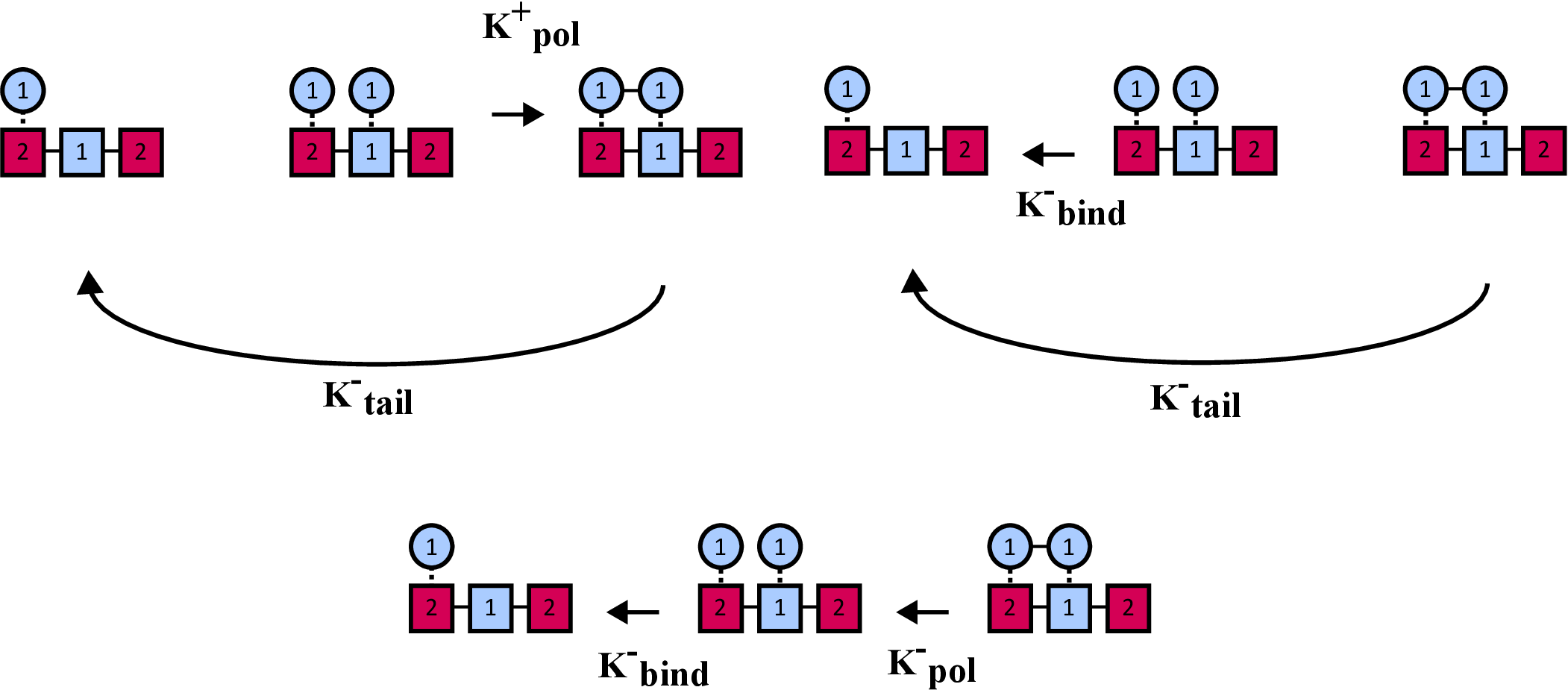}
  \caption{\label{fig:CoarseBenSpanning}}
  \end{subfigure}
\caption{ Enumeration of spanning trees for the calculation of coarse-grained propensities. a) The modified network for the calculation of $A$. b) Spanning trees of the network rooted at the initial coarse-grained state.}
\end{figure*}

\begin{widetext}
\begin{align}
\Lambda^+_{\mathbf{x}}(n_{l-1}n_l,m_{l-1}m_l) &= K^+_{\textnormal{bind}}(n_{l-1}n_l,m_{l-1}m_l)K^+_{\textnormal{pol}}(n_{l-1}n_l,m_{l-1}m_l)K^-_{\textnormal{tail}}(n_{l-1}n_l,m_{l-1}m_l) \nonumber \\
&= k_{\textnormal{bind}} k_{\textnormal{pol}}(n_{l-1}n_l,m_{l-1}m_l) k_{\textnormal{tail}} e^{-\Delta G_{TT}(n_l,m_l)} [M]. \label{eq:lambdaplus}\\
\Lambda^-_{\mathbf{x}}(n_{l-1}n_l,m_{l-1}m_l) &= K^-_{\textnormal{bind}}(n_{l-1}n_l,m_{l-1}m_l)K^-_{\textnormal{pol}}(n_{l-1}n_l,m_{l-1}m_l)K^+_{\textnormal{tail}}(n_{l-1}n_l,m_{l-1}m_l) \nonumber \\
&= k_{\textnormal{bind}} k_{\textnormal{pol}}(n_{l-1}n_l,m_{l-1}m_l) k_{\textnormal{tail}} e^{-\Delta G_{BB} -\Delta G_{TT}(n_l,m_l)} [M]_{\textnormal{eff}}. \label{eq:lambdaminus}\\
\intertext{The modified process used to obtain $A_{\mathbf{x}}(n_{l-1}n_l,m_{l-1}m_l)$ (with transitions to $(\mathbf{x},\&m_{l-1}m_l)$ redirected back to $(\mathbf{x},\&m_{l-1})$) and its spanning trees rooted at $(\mathbf{x},\&m_{l-1})$ are depicted in Figures \ref{fig:CoarseBen} and \ref{fig:CoarseBenSpanning} respectively. The resulting $A_{\mathbf{x}}(n_{l-1}n_l,m_{l-1}m_l)$ is given in Equation \ref{eq:modifiedA}. }
A_{\mathbf{x}}(n_{l-1}n_l,m_{l-1}m_l) &= K^-_{\textnormal{bind}}(n_{l-1}n_l,m_{l-1}m_l)K^-_{\textnormal{pol}}(n_{l-1}n_l,m_{l-1}m_l) \nonumber
+ K^-_{\textnormal{bind}}(n_{l-1}n_l,m_{l-1}m_l)K^-_{\textnormal{tail}}(n_{l-1}n_l,m_{l-1}m_l)
\\
&\;\;\;\;\;\;\;\;  + K^+_{\textnormal{pol}}(n_{l-1}n_l,m_{l-1}m_l)K^-_{\textnormal{tail}}(n_{l-1}n_l,m_{l-1}m_l)
\nonumber \\
 &= e^{-\Delta G_{BB} -\Delta G_{TT}(n_l,m_l)} k_{\textnormal{bind}} k_{\textnormal{pol}}(n_{l-1}n_l,m_{l-1}m_l) \nonumber \\ &\;\;\;\;\;\;\;\; +  e^{-\Delta G_{TT}(n_{l-1},m_{l-1}) -\Delta G_{TT}(n_l,m_l) - \Delta G_{\textnormal{gen}}} k_{\textnormal{bind}} k_{\textnormal{tail}}  \nonumber \\ &\;\;\;\;\;\;\;\;
+ e^{-\Delta G_{TT}(n_{l-1},m_{l-1}) }k_{\textnormal{pol}}(n_{l-1}n_l,m_{l-1}m_l) k_{\textnormal{tail}}. 
\label{eq:modifiedA}\\
\intertext{Equations \ref{eq:lambdaplus}, \ref{eq:lambdaminus} and \ref{eq:modifiedA} are consistent with the coarse-grained propensities in Equations \ref{eq:phiplus} and \ref{eq:phimin}. Now refer to Figure~\ref{fig:FineGrainPerm} for the case of non-separating templated copolymerization. $\Lambda^+_{\textnormal{NTC},\mathbf{x}}(n_{l-1}n_l,m_{l-1}m_l)$ (Equation \ref{eq:lambdaplusperm}) and $\Lambda^-_{\textnormal{NTC},\mathbf{x}}(n_{l-1}n_l,m_{l-1}m_l)$ (Equation \ref{eq:lambdaminusperm}) can again be obtained by a straightforward multiplication of forward and backward rates}
\Lambda^+_{\textnormal{NTC},\mathbf{x}}(n_{l-1}n_l,m_{l-1}m_l) &= K^+_{\textnormal{bind}}(n_{l-1}n_l,m_{l-1}m_l)K^+_{\textnormal{pol}}(n_{l-1}n_l,m_{l-1}m_l) \nonumber \\
&= k_{\textnormal{bind}} k_{\textnormal{pol}}(n_{l-1}n_l,m_{l-1}m_l) [M], \label{eq:lambdaplusperm}\\
\Lambda^-_{\textnormal{NTC},\mathbf{x}}(n_{l-1}n_l,m_{l-1}m_l) &= K^-_{\textnormal{bind}}(n_{l-1}n_l,m_{l-1}m_l)K^-_{\textnormal{pol}}(n_{l-1}n_l,m_{l-1}m_l) \nonumber \\
&= k_{\textnormal{bind}} k_{\textnormal{pol}}(n_{l-1}n_l,m_{l-1}m_l) e^{-\Delta G_{BB} -\Delta G_{TT}(n_l,m_l)}. \label{eq:lambdaminusperm}\\
\intertext{ For $A_{\textnormal{NTC},\mathbf{x}}(n_{l-1}n_l,m_{l-1}m_l)$, note that the only two spanning trees for a modified process with polymerization redirected back towards the initial state are the single-step processes of unbinding and polymerization, leading to the form in Equation \ref{eq:modifiedAperm}.}
A_{\textnormal{NTC},\mathbf{x}}(n_{l-1}n_l,m_{l-1}m_l) &= K^-_{\textnormal{bind}}(n_{l-1}n_l,m_{l-1}m_l) + K^+_{\textnormal{pol}}(n_{l-1}n_l,m_{l-1}m_l)
\nonumber \\
 &= e^{-\Delta G_{TT}(n_l,m_l)} k_{\textnormal{bind}} 
+ k_{\textnormal{pol}}(n_{l-1}n_l,m_{l-1}m_l).
\label{eq:modifiedAperm}
\end{align}
\end{widetext}

\section{Derivation of Average Tip Waiting Times \label{app:BenTimes}}

\begin{figure*}
\includegraphics[width=1.0\linewidth]{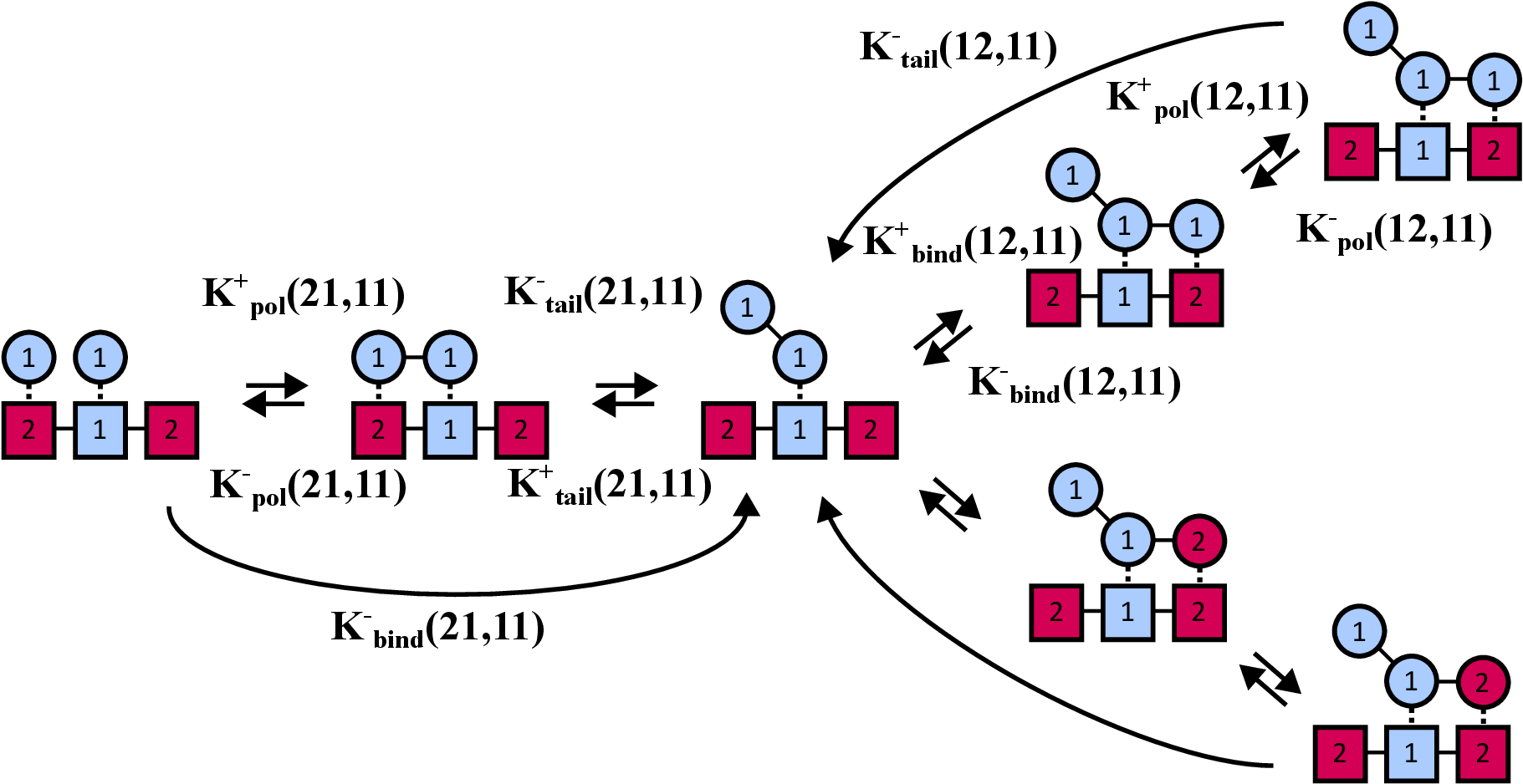}
\caption{\label{fig:allpetals} Fine-grained network for calculating first passage between coarse-grained states. Transitions leading to other coarse-grained tip states are redirected back to the initial coarse-grained state. Rates for the bottom arm are ommitted, but are analogous to those for the top arm with $(12,12)$ in place of $(12,11)$.}
\end{figure*}

We apply the method in \cite{Qureshi2023} to calculate the expected waiting times $\mathbb{E}_V[\theta_{\mathbf{x},l}]$ at a given tip state of the coarse-grained model. Note the waiting times will be dependent on template as well as copy monomers near the tip, hence in this section a tip state will refer to $(n_{l-1}n_ln_{l+1},m_{l-1}m_l)$. For each coarse-grained tip state, we wish to consider the network of reactions into and out of the completed state corresponding to $f = 0$ (Figure \ref{fig:FineGrainedStates}). Refer to Figure \ref{fig:allpetals} for the first passage network from a completed state, with all transitions to other completed states redirected back to the initial completed state. As our propensities are dependent only on local template and copy monomers, $\mathbb{E}_V[\theta_{\mathbf{x},l}] = \mathbb{E}_V[\theta(n_{l-1}n_ln_{l+1},m_{l-1}m_l)]$ for our case. As per \cite{Qureshi2023}, $\mathbb{E}_V[\theta(n_{l-1}n_ln_{l+1},m_{l-1}m_l)] = 1/J(n_{l-1}n_ln_{l+1},m_{l-1}m_l)$ where $J(n_{l-1}n_ln_{l+1},m_{l-1}m_l)$ is the flux into completed states corresponding to distinct coarse-grained states. To find this flux, we will need to calculate the stationary probability distribution $p_{ss}(\mathbf{x},\&m_{l-1}m_l,f)$ of fine-grained states in the network in Figure \ref{fig:allpetals}. 

Our discussion for the remainder of this section will take place in the context of a single network of the type in Figure \ref{fig:allpetals} (with state indices given by Figure \ref{fig:FineGrainedStates}). Hence, $\mathbf{x}$ and $\&m_{l-1}m_l$ are fixed, and we can omit them from variables for simplicity. Rates of the form $K^{\pm}(n_l n_{l+1},m_l m_{l+1})$ (with dependence on $m_{l+1}$) are abbreviated to $K^{\pm}(m_{l+1})$, rates of the form $K^{\pm}(n_{l-1} n_l,m_{l-1} m_l)$ (without dependence on $m_{l+1}$) are abbreviated to $K^{\pm}(-)$,
$J(n_{l-1}n_ln_{l+1},m_{l-1}m_l)$ is abbreviated to $J$ and $p_{ss}(\mathbf{x},\&m_{l-1}m_l,f)$ is abbreviated to $p_{ss}(f)$. Finally, we introduce a function $\chi(m_{l+1}) = 2+2 \times m_{l+1}$ that maps a monomer $m_{l+1}$ to the transitory state index $f_{\textnormal{exit}}$ such that  $(\mathbf{x},\&m_{l-1}m_l,f_{\textnormal{exit}})$ has a transition into $(\mathbf{x},\&m_{l-1}m_lm_{l+1},0)$. That is, $\chi$ returns the last fine-grained state encountered before the complete addition of a monomer $m_{l+1}$. Then, the flux $J$ can be calculated as follows \cite{Qureshi2023}:
\begin{align}
J =  K_{\textnormal{bind}}^-&(-) p_{ss}(2) \nonumber \\ & + \Sigma_{m_{l+1}} K^-_{\textnormal{tail}}(m_{l+1}) p_{ss}(\chi(m_{l+1})).
\end{align}
Here, the first term is the backward flux, and the second term is the sum of all forward fluxes. The steady state probabilities of states with transitions to other coarse-grained tip states can be found through standard methods, and they are expressed in Equations \ref{eq:pssAdd} and \ref{eq:pssRem} (and a normalization variable $\mathcal{N}$ expressed in Equation \ref{eq:pssnormalization}).

\begin{widetext}
\begin{equation}
p_{ss}(\chi(m_{l+1})) = \frac{K_{\textnormal{bind}}^+(m_{l+1}) K_{\textnormal{pol}}^+(m_{l+1})}{\mathcal{N} (K_{\textnormal{bind}}^-(m_{l+1})(K_{\textnormal{pol}}^-(m_{l+1})+K_{\textnormal{tail}}^-(m_{l+1}))+K_{\textnormal{tail}}^-(m_{l+1})K_{\textnormal{pol}}^+(m_{l+1})) },
\label{eq:pssAdd}
\end{equation}
\begin{equation}
p_{ss}(2) = \frac{K_{\textnormal{tail}}^+(-) K_{\textnormal{pol}}^-(-)}{\mathcal{N}(K^-_{\textnormal{tail}}(-)(K_{\textnormal{pol}}^+(-)+K^-_{\textnormal{bind}}(-))+K^-_{\textnormal{bind}}(-)K^-_{\textnormal{pol}}(-))},
\label{eq:pssRem}
\end{equation}
\begin{align}
\mathcal{N} &= 1 + \frac{K_{\textnormal{tail}}^+(-) K_{\textnormal{pol}}^-(-)}{K^-_{\textnormal{tail}}(-)(K_{\textnormal{pol}}^+(-)+K^-_{\textnormal{bind}}(-))+K^-_{\textnormal{bind}}(-)K^-_{\textnormal{pol}}(-)}(1+\frac{K^+_{\textnormal{pol}}(-)+K^-_{\textnormal{bind}}(-)}{K^-_{\textnormal{pol}}(-)}) \nonumber\\
&\;\;\;\; + \Sigma_{m_{l+1}}
\frac{K_{\textnormal{bind}}^+(m_{l+1}) K_{\textnormal{pol}}^+(m_{l+1})}{K_{\textnormal{bind}}^-(m_{l+1})(K_{\textnormal{pol}}^-(m_{l+1})+K_{\textnormal{tail}}^-(m_{l+1}))+K_{\textnormal{tail}}^-(m_{l+1})K_{\textnormal{pol}}^+(m_{l+1}) }
(1+\frac{K^-_{\textnormal{pol}}(m_{l+1})+K^-_{\textnormal{tail}}(m_{l+1})}{K^+_{\textnormal{pol}}(m_{l+1})}).
\label{eq:pssnormalization}
\end{align}

Analogous definitions and abbreviations can be applied to NTC. The function $\chi_{\textnormal{NTC}}(m_{l+1}) = 1+m_{l+1}$ now maps each $m_{l+1}$ to $f_{\textnormal{exit}}$, its final state before transition. The flux $J_{\textnormal{NTC}}$ is then calculated as follows (refer to Figure \ref{fig:FineGrainPermIndices} for state indices):
\begin{align}
J_{\textnormal{NTC}} =  K_{\textnormal{bind}}^-(-) p_{ss,\textnormal{NTC}}(1) \nonumber \\ & +\Sigma_{m_{l+1}} K^+_{\textnormal{pol}}(m_{l+1}) p_{ss}(\chi_{\textnormal{NTC}}(m_{l+1})).
\end{align}
Steady state probabilities and normalization variables for NTC without detachment, calculated through standard methods, are given in equations \ref{eq:pssAddNTC}, \ref{eq:pssRemNTC} and \ref{eq:pssnormalizationNTC}. 

\begin{equation}
p_{ss,\textnormal{NTC}}(\chi_{\textnormal{NTC}}(m_{l+1})) = \frac{K^+_{\textnormal{bind}}(m_{l+1})}{\mathcal{N}_{\textnormal{NTC}} (K_{\textnormal{bind}}^-(m_{l+1}) 
+ K_{\textnormal{pol}}^+(m_{l+1}))},
\label{eq:pssAddNTC}
\end{equation}
\begin{equation}
p_{ss,\textnormal{NTC}}(1) = \frac{K^-_{\textnormal{pol}}(-)}{\mathcal{N}_{\textnormal{NTC}} (K_{\textnormal{pol}}^+(-) 
+ K_{\textnormal{bind}}^-(-))},
\label{eq:pssRemNTC}
\end{equation}
\begin{align}
\mathcal{N}_{\textnormal{NTC}} &= 1 + \frac{K^-_{\textnormal{pol}}(-)}{ K_{\textnormal{pol}}^+(-) 
+ K_{\textnormal{bind}}^-(-)} + \Sigma_{m_{l+1}}
\frac{K^+_{\textnormal{bind}}(m_{l+1})}{ K_{\textnormal{bind}}^-(m_{l+1}) 
+ K_{\textnormal{pol}}^+(m_{l+1})}.
\label{eq:pssnormalizationNTC}
\end{align}
\end{widetext}

\section{Iteration for the Expected Number of Visits to a Tip State \label{app:VisitIter}}

Let a tip state $(m_l,m_{l+1})_{\mathbf{x},l+1}$ refer to a class of states with template $\mathbf{x}$ and a copy of length $l+1$ ending in $m_l,m_{l+1}$. This definition extends to longer tails; for example, $(m_{l-1},m_l,m_{l+1})_{\mathbf{x},l+1}$ would refer to states ending in $m_{l-1},m_l,m_{l+1}$. In this section, we aim to calculate  $\mathbb{E} [V_{\mathbf{x},l}(m_l,m_{l+1})|m_{l-1},m_l]$, the expected number of visits to $(m_l,m_{l+1})_{\mathbf{x},l+1}$ for each visit to $(m_{l-1},m_{l})_{\mathbf{x},l}$. We argue that this quantity can be calculated by considering visits to transient tip states in the Markov chain in Figure \ref{fig:VisitationMarkov}. We will use abbreviations of the form $V_{l+1}(m_{l+1}) = \mathbb{E} [V_{\mathbf{x},l}(m_l,m_{l+1})|m_{l-1},m_l]$ for convenience, noting that we always refer to the Markov process in Figure \ref{fig:VisitationMarkov}, and hence the conditioned tip $m_{l-1},m_l$ and template $\mathbf{x}$ are defined. 

Each tip state $(m_{l-1},m_l,m_{l+1})_{\mathbf{x},l+1}$ will be visited an expected $p_{\mathbf{x},l}(m_{l+1}| m_{l-1},m_l)$ number of times from each transition out of $(m_{l-1},m_l)_{l}$ (we call this visit, resulting directly from monomer addition, a `forward' visit). We can count `backward' visits to $(m_{l-1},m_l,m_{l+1})_{\mathbf{x},l+1}$ from $(m_{l-1},m_l,m_{l+1},m_{l+2})_{\mathbf{x},l+2}$ by noting that each visit of $(m_{l-1},m_l,m_{l+1},m_{l+2})_{\mathbf{x},l+2}$ has a 
probability $1-Q_{\mathbf{x},l+2}(m_{l+1}m_{l+2})$ of returning back to $(m_{l-1},m_l,m_{l+1})_{\mathbf{x},l+1}$. Hence, denoting forward visits to $(m_{l-1},m_l,m_{l+1},m_{l+2})_{\mathbf{x},l+2}$ in the Markov chain given in Figure \ref{fig:VisitationMarkov} by 
$V_{l+2}^f(m_{l+1}m_{l+2})$, we obtain an expression for $V_{l+1}(m_{l+1})$ in equation \ref{eq:visits1}.  
\begin{widetext}
\begin{align}
V_{l+1}(m_{l+1}) &= p_{\mathbf{x},l}(m_{l+1}|m_{l-1} m_l) + \Sigma_{m_{l+2}} (1-Q_{\mathbf{x},l+2}(m_{l+1}m_{l+2}))V^f_{l+2}(m_{l+1},m_{l+2})\label{eq:visits1}.
\end{align}
We further know that visits to $(m_{l-1},m_l,m_{l+1},m_{l+2})_{\mathbf{x},l+2}$ are just the visits to $(m_{l-1},m_l,m_{l+1})_{\mathbf{x},l+1}$ weighted by the probability of adding $m_{l+2}$ so that $V^f_{l+2}(m_{l+1},m_{l+2}) = p_{\mathbf{x},l+1}(m_{l+2}|m_l m_{l+1}) V_{l+1}(m_{l+1})$, leading to equation \ref{eq:visits2}.
\begin{align}
V_{l+1}(m_{l+1}) &= p_{\mathbf{x},l}(m_{l+1}|m_{l-1} m_l) + \Sigma_{m_{l+2}} (1-Q_{\mathbf{x},l+2}(m_{l+1}m_{l+2})) p_{\mathbf{x},l+1}(m_{l+2}|m_l m_{l+1}) V_{l+1}(m_{l+1})\label{eq:visits2}.
\end{align}
Rearranging for $V_{l+1}(m_{l+1})$, we obtain \ref{eq:visits3}
\begin{align}
\mathbb{E} [V_{\mathbf{x},l}(m_l,m_{l+1})|m_{l-1},m_l] = V_{l+1}(m_{l+1}) = \frac{p_{\mathbf{x},l}(m_{l+1}|m_{l-1} m_l)}{1-\Sigma_{m_{l+2}} (1-Q_{\mathbf{x},l+2}(m_{l+1}m_{l+2})) p_{\mathbf{x},l+1}(m_{l+2}|m_l m_{l+1})}\label{eq:visits3}
.\end{align}
$\mathbb{E} [V_{\mathbf{x},l}(m_l,m_{l+1})|m_{l-1},m_l]$ is then calculable from previously calculated $Q$ variables (and local rates). Then, the absolute visitations to any tip state $(m_{l-1}m_l)_{\mathbf{x},l}$ can be calculated using equation \ref{eq:forward_iter} of the main text.
\end{widetext}

\section{Parameter Sweep of Relative Error and Total Visitations \label{app:ParamSweep}}

To test the robustness of our qualitative predictions on the effect of heterogeneity on relative error and state visits, we perform a sweep of parameter space. Letting $\Delta G_{\textnormal{pol}} = 0$, $L = 10^4$ and considering a high discrimination level $\Delta G_{TT,H}$ and a low discrimination level $\Delta G_{TT,L}$ with $\Delta G_{TT,H} \geq \Delta G_{TT,L}$ we consider forward/backward discrimination matrices of the following form for heterogeneity on correct monomers:
\begin{equation}
    \Delta G_{TT} = \begin{pmatrix}
       \Delta G_{TT,H} & 0\\
       0 & \Delta G_{TT,L}
    \end{pmatrix}
.\end{equation}

\begin{figure*}
\begin{tabular}[t]{cc}

    \begin{subfigure}[t]{0.39\textwidth}
      \includegraphics[width=1.0\linewidth]{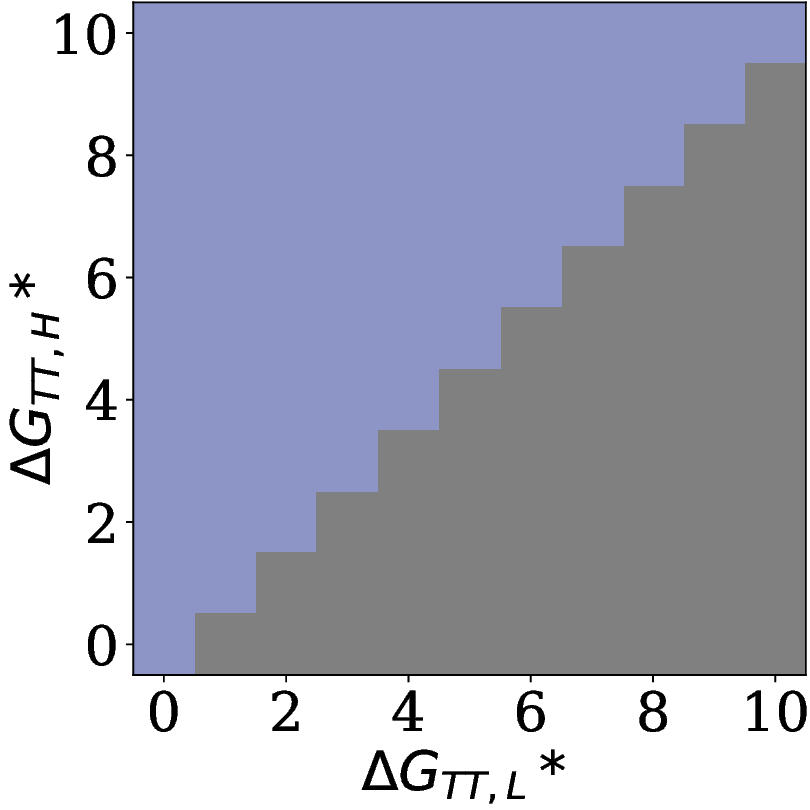}
      \caption{\label{fig:heatmaperror.thermcorr}}
    \end{subfigure}
    \begin{subfigure}[t]{0.39\textwidth}
        \centering
      \includegraphics[width=1.0\linewidth]{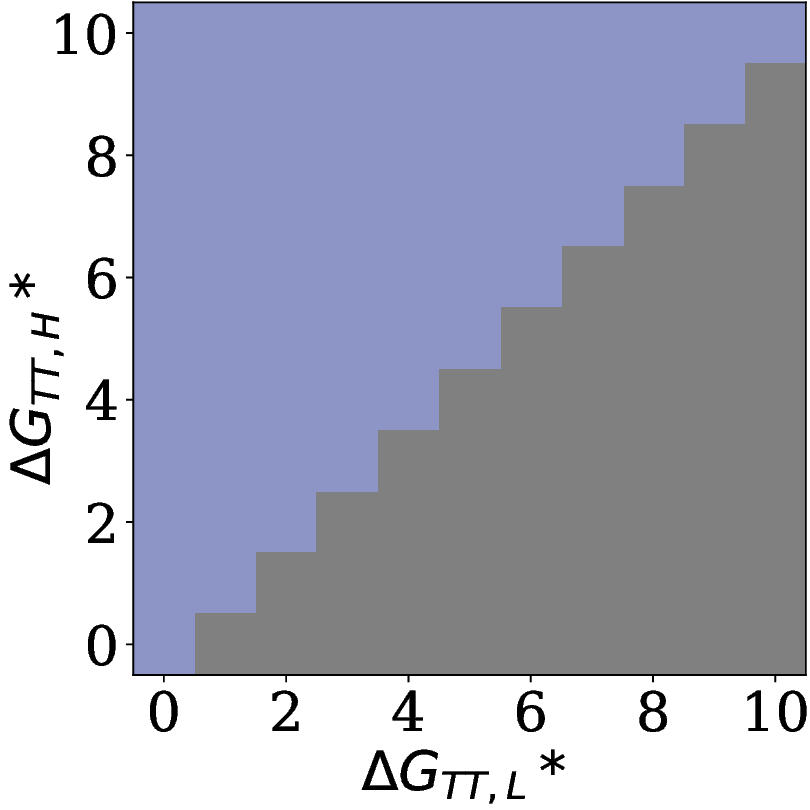}
      \caption{\label{fig:heatmaperror.combcorr}}
    \end{subfigure} & 
        \begin{subfigure}[t]{0.21\textwidth}
          \includegraphics[width=1.0\linewidth]{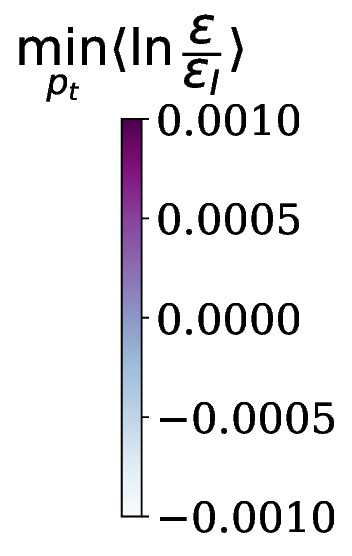}

        \end{subfigure}
        
    \\
    \begin{subfigure}[t]{0.39\textwidth}
      \includegraphics[width=1.0\linewidth]{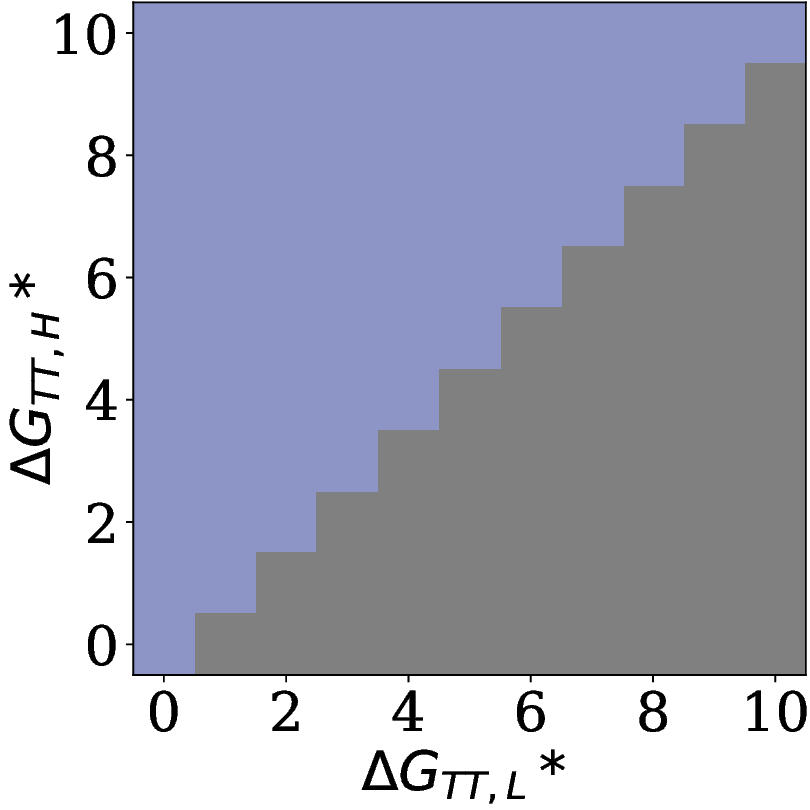}
      \caption{\label{fig:heatmaperror.thermincorr}}
    \end{subfigure}
    \begin{subfigure}[t]{0.39\textwidth}
      \includegraphics[width=1.0\linewidth]{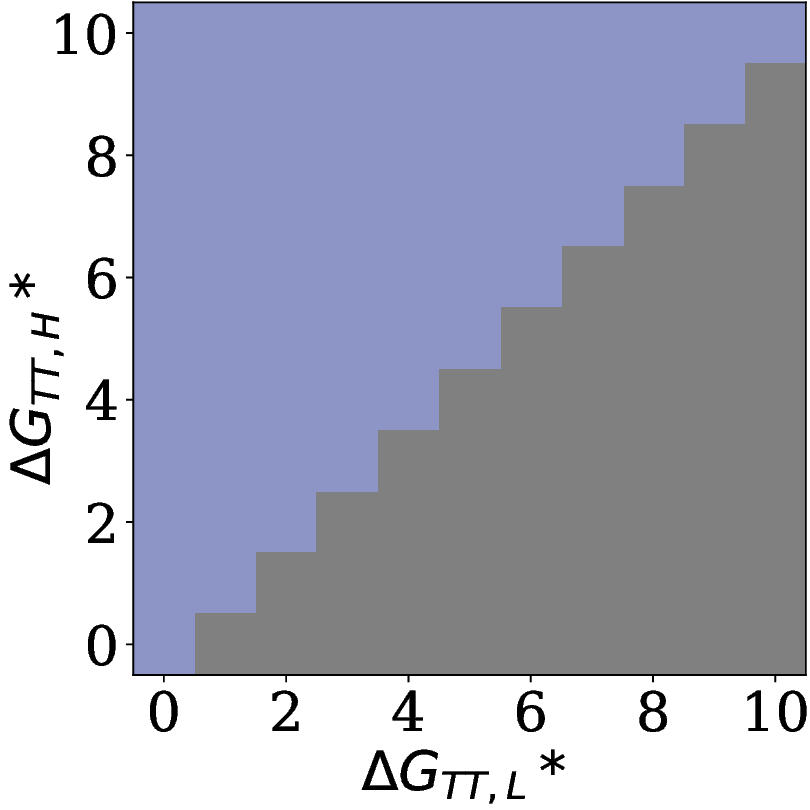}
      \caption{\label{fig:heatmaperror.combincorr}}
    \end{subfigure} &
         \begin{subfigure}[t]{0.21\textwidth}
          \includegraphics[width=1.0\linewidth]{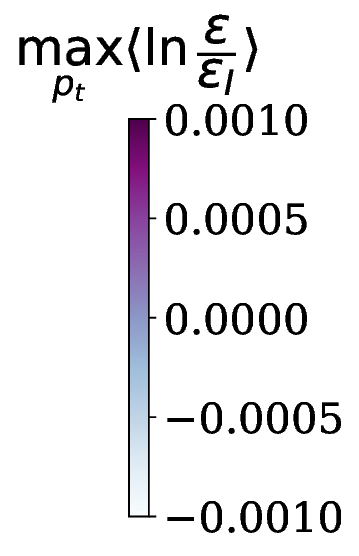}
        \end{subfigure}

\end{tabular}
\caption{Heat map illustrating the resulting relative error change $\langle\ln{\frac{\epsilon}{\epsilon_I}}\rangle$ from parameter sweeps for heterogeneity, having extremised over $p_t$. We consider heterogeneity on correct (backward propensity discrimination: (a); forward  discrimination propensity discrimination: (b)) and incorrect (backward propensity discrimination: (c); forward propensity discrimination: (d)) monomers. In the case of heterogeneity on correct monomer interactions, this relative error change is minimised over $p_t$; for heterogeneity on incorrect monomer interactions it is maximised over $p_t$. Only the upper triangle $\Delta G_{TT,H} \geq \Delta G_{TT,L}$ is plotted. For heterogeneity on correct monomers, $\min_{p_t} \langle\ln{\frac{\epsilon}{\epsilon_I}}\rangle = 0$ so relative error decreases do not occur, and for heterogeneity on incorrect monomers, $\max_{p_t} \langle\ln{\frac{\epsilon}{\epsilon_I}}\rangle = 0$ so relative error increases do not occur.}
\end{figure*}

\begin{figure*}
\begin{tabular}[t]{cc}

    \begin{subfigure}[t]{0.39\textwidth}
      \includegraphics[width=1.0\linewidth]{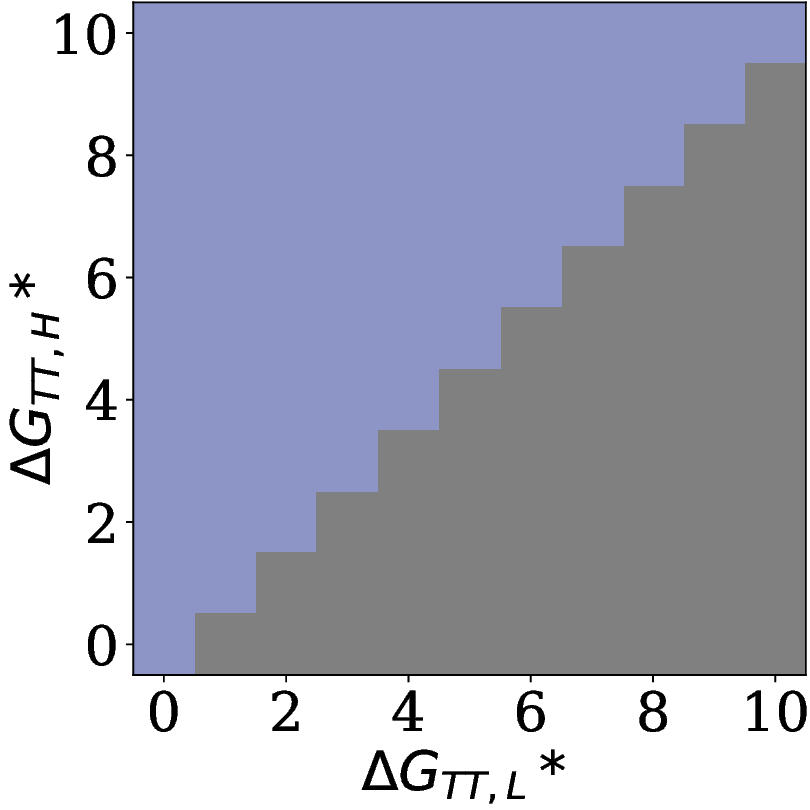}
      \caption{\label{fig:heatmapvisit.thermcorr}}
    \end{subfigure}
    \begin{subfigure}[t]{0.39\textwidth}
        \centering
      \includegraphics[width=1.0\linewidth]{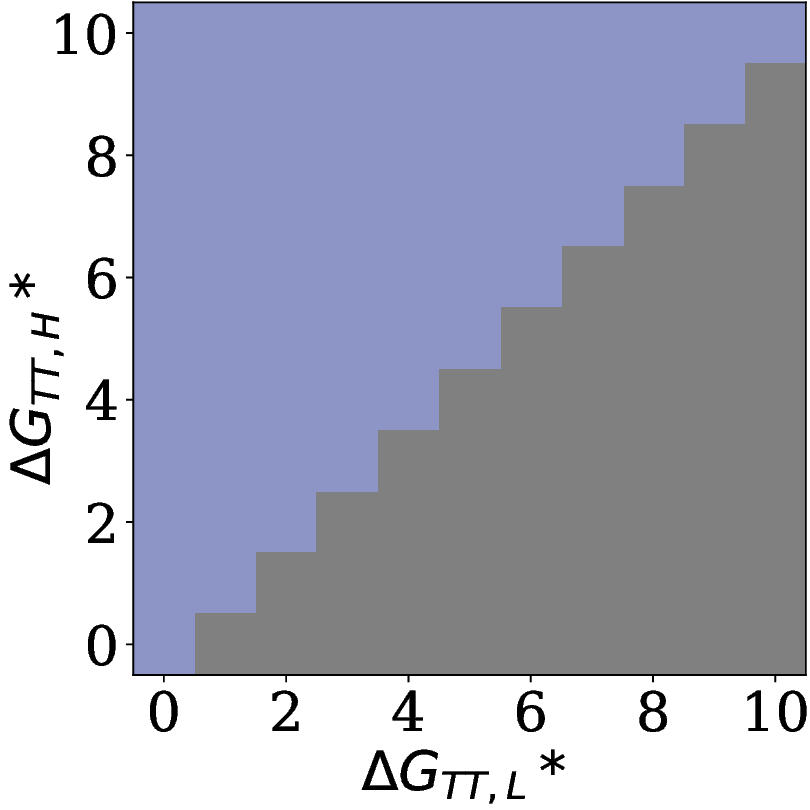}
      \caption{\label{fig:heatmapvisit.combcorr}}
    \end{subfigure} & 
        \begin{subfigure}[t]{0.21\textwidth}
          \includegraphics[width=1.0\linewidth]{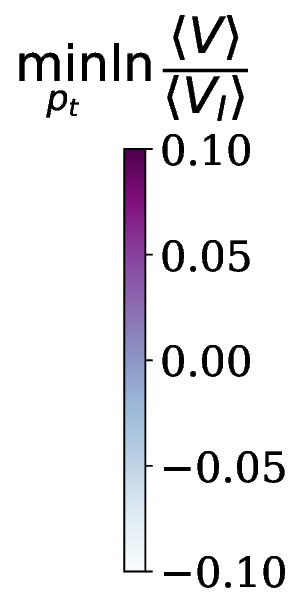}

        \end{subfigure}
        
    \\
    \begin{subfigure}[t]{0.39\textwidth}
      \includegraphics[width=1.0\linewidth]{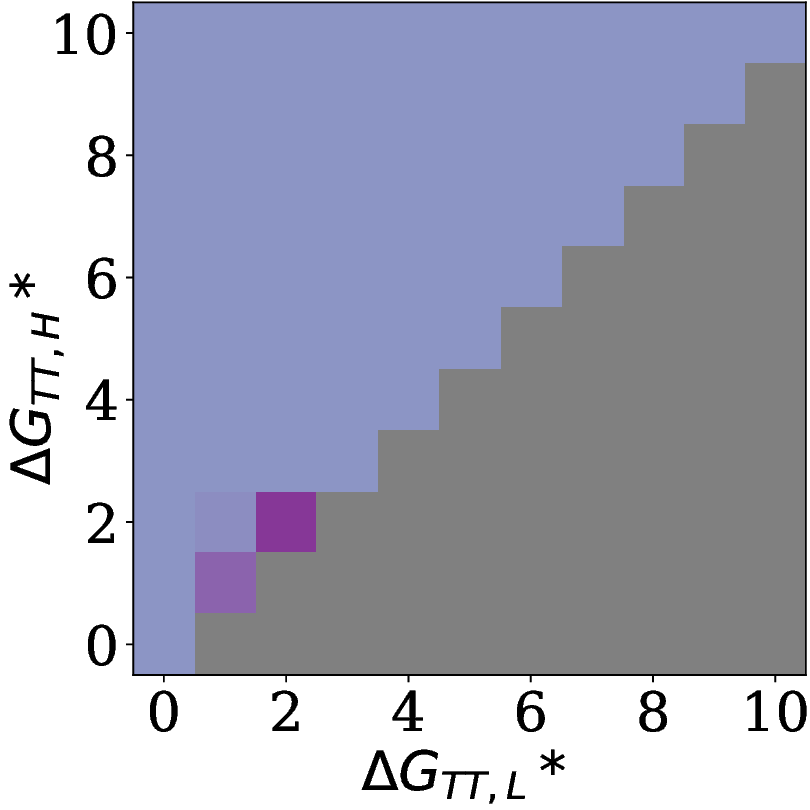}
      \caption{\label{fig:heatmapvisit.thermincorr}}
    \end{subfigure}
    \begin{subfigure}[t]{0.39\textwidth}
      \includegraphics[width=1.0\linewidth]{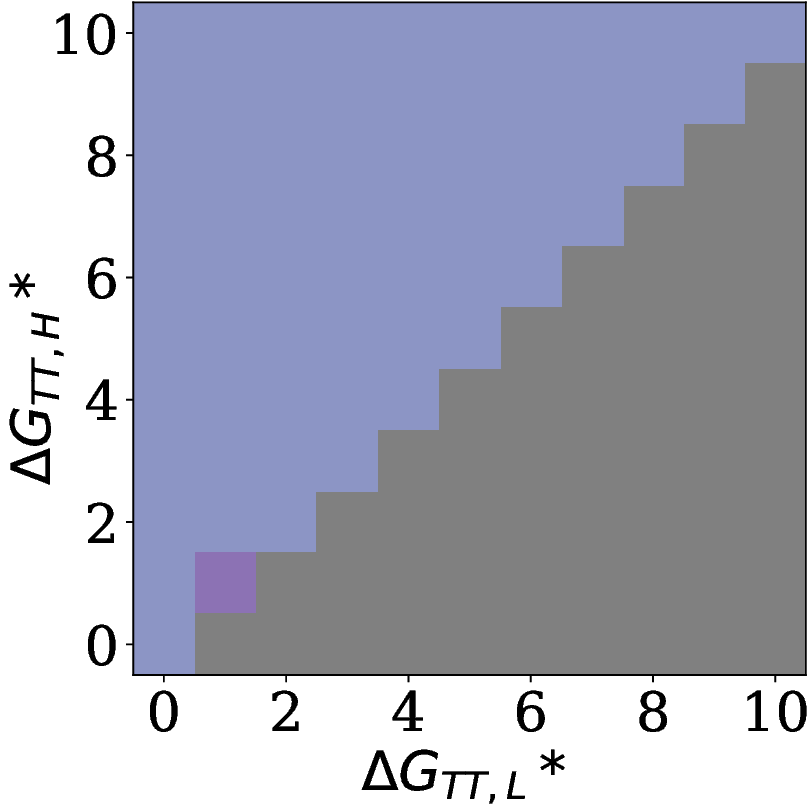}
      \caption{\label{fig:heatmapvisit.combincorr}}
    \end{subfigure} &
         \begin{subfigure}[t]{0.21\textwidth}
          \includegraphics[width=1.0\linewidth]{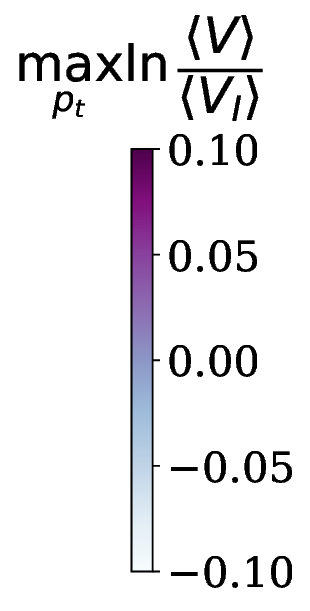}
        \end{subfigure}

\end{tabular}
\caption{Heat map illustrating the resulting log heterogeneous change in state visits $\ln{\frac{\langle V \rangle}{\langle V_I \rangle}}$ from parameter sweeps for heterogeneity, having extremised over $p_t$. We consider heterogeneity on correct (backward propensity discrimination: (a); forward  discrimination propensity discrimination: (b)) and incorrect (backward propensity discrimination: (c); forward propensity discrimination: (d)) monomers. In the case of heterogeneity on correct monomer interactions, $\ln{\frac{\langle V \rangle}{\langle V_I \rangle}}$ is minmised over $p_t$; for heterogeneity on incorrect monomer interactions it is maximised over $p_t$. Only the upper triangle $\Delta G_{TT,H} \geq \Delta G_{TT,L}$ is plotted. For heterogeneity on correct monomers,$\min_{p_t} \ln{\frac{\langle V \rangle}{\langle V_I \rangle}} = 0$ so decreases in state visits do not occur. For heterogeneity on incorrect monomers, $\max_{p_t} \ln{\frac{\langle V \rangle}{\langle V_I \rangle}} = 0$ for the vast majority of parameters, so increases in state visits do not occur. However, deviations $\max_{p_t} \ln{\frac{\langle V \rangle}{\langle V_I \rangle}} > 0$ are observed in the small discrimination regime for heterogeneity on incorrect monomer interactions. }
\end{figure*}

On the other hand, for heterogeneity on incorrect monomers, we consider matrices of the following form:
\begin{equation}
    \Delta G_{TT} = \begin{pmatrix}
       \Delta G_{TT,H} & \Delta G_{TT,L}\\
       0 & \Delta G_{TT,H}
    \end{pmatrix}
.\end{equation}
We then sweep over these parameters for different $p_t$. In the case of heterogeneity on correct monomers, we want to show that relative error and state visits are always increased, and we thus expect $\min_{p_t} \langle\ln{\frac{\epsilon}{\epsilon_I}}\rangle \geq 0$ (Figure \ref{fig:heatmaperror.thermcorr} and \ref{fig:heatmaperror.combcorr}) and $\min_{p_t} \ln{\frac{\langle V \rangle}{\langle V_I \rangle}} \geq 0$ (Figure \ref{fig:heatmapvisit.thermcorr} and \ref{fig:heatmapvisit.combcorr}). On the other hand, for heterogeneity on incorrect monomers $\max_{p_t} \langle\ln{\frac{\epsilon}{\epsilon_I}}\rangle \leq 0$ (Figure \ref{fig:heatmaperror.thermincorr} and \ref{fig:heatmaperror.combincorr}) and $\max_{p_t} \ln{\frac{\langle V \rangle}{\langle V_I \rangle}} \leq 0$ (Figure \ref{fig:heatmapvisit.thermincorr} and \ref{fig:heatmapvisit.combincorr}) are expected. In this $L=10^4$, regime the noise due to edge effects becomes significant and so we extract values from the central $80\%$ of the polymer. Only the upper right triangle of all of the heat maps are plotted. Despite allowing the system to optimise over $p_t$, the maximum and the minimum in each case stay very close to 0, implying that error and visits never improve when correct monomer interactions are heterogeneous and never degrade when incorrect monomer interactions are heterogeneous. The exception to this observation is the deviation in $\ln{\frac{\langle V \rangle}{\langle V_I \rangle}}$ observed in the small discrimination regime for  heterogeneity on incorrect monomer interactions. For heterogeneity on incorrect monomer interactions, there is tension between increased visitation in the incorrect landscape and the higher probability of traversing the correct landscape. Hence, when the correct and incorrect landscapes have a small gap to begin with (true for small discrimination), the effect of increased visitations through the incorrect landscape become more significant. 

\section{Investigations on Error Reduction Through Heterogeneity for Separating Copiers \label{app:ParamSweep2}}

Section \ref{sec:ErrorTime} revealed an interesting phenomenon, that heterogeneity on correct monomers tends to increase error rates, while heterogeneity on incorrect monomers tend to decrease error rates. We believe it would be illustrative to attempt to find levels of forward and backward heterogeneity that are (in a heuristic sense) optimal for extracting benefits from heterogeneity. 

In our fine-grained copying model, we can obtain varying forward and backward discrimination factors in the slow polymerization limit, $k_{\textnormal{pol}} \ll k_{\textnormal{bind}}$. Instead of forcing $k_{\textnormal{pol}}$ to be sequence-independent, we allow it to be a function of the added monomer pair $k_{\textnormal{pol}}(n_l,m_l)$ and set $k_{\textnormal{pol}}(n_l,m_l)e^{\Delta G_{TT}(n_l,m_l)}[M] = e^{\Delta G_{K}(n_l,m_l)}$ to obtain the following form. 

\begin{align}
\Phi^+(&n_{l-1}n_l, m_{l-1}m_l) \nonumber \\ &=  e^{\Delta G_{K}(n_l,m_l)}. \\
\Phi^-(&n_{l-1}n_l, m_{l-1}m_l) \nonumber \\ &=  e^{\Delta G_{K}(n_l,m_l)-\Delta G_{\textnormal{pol}} +\Delta G_{TT}(n_{l-1},m_{l-1}) - \Delta G_{TT}(n_l,m_l)} 
.\end{align}

\noindent Copying is thus parameterized by a scalar parameter $\Delta G_{\textnormal{pol}}$ and two matrix parameters $\Delta G_K$ and $\Delta G_{TT}$. 

Based on Section \ref{sec:ErrorTime}, we anticipate that error reduction is maximized when correct monomer pairs are kept homogeneous while the heterogeneity in the incorrect pairs is maximized, and so we make that assumption. We first consider the limit where one monomer is purely backward propensity discriminated, while the other is purely forward propensity discriminated.  Then, we gradually shift the  discrimination until one monomer has half the forward propensity discrimination (and  the other has half the backward propensity discrimination) of the other. 

\begin{table}[]
\begin{tabular}{lll}
\toprule
\textbf{Index} & $\mathbf{\Delta G_{TT}}$ & $\mathbf{\Delta G_K}$                              \\ \midrule
    i & $\begin{pmatrix}
    G_{TT}^* & G_{TT}^*\\
    0 & G_{TT}^*
    \end{pmatrix}$     
    & 
    $\begin{pmatrix}
    G_K^* & 0\\
    G_K^* & G_K^*
    \end{pmatrix}$ \\  

    ii & $\begin{pmatrix}
    G_{TT}^* & \frac{3G_{TT}^*}{4}\\
    0 & G_{TT}^*
    \end{pmatrix}$     
    & 
    $\begin{pmatrix}
    G_K^* & 0\\
    \frac{3G_K^*}{4} & G_K^*
    \end{pmatrix}$ \\

    iii & $\begin{pmatrix}
    G_{TT}^* & \frac{G_{TT}^*}{2}\\
    0 & G_{TT}^*
    \end{pmatrix}$     
    & 
    $\begin{pmatrix}
    G_K^* & 0\\
    \frac{G_K^*}{2} & G_K^*
    \end{pmatrix}$ \\
\bottomrule
\end{tabular}
\caption{Parameter sets investigated in Appendix \ref{app:ParamSweep2}.\label{table:CombinedParams}}
\end{table}

\begin{figure*}
\begin{tabular}[t]{cccc}

    \begin{subfigure}[b]{0.28\textwidth}
      \includegraphics[width=1.0\linewidth]{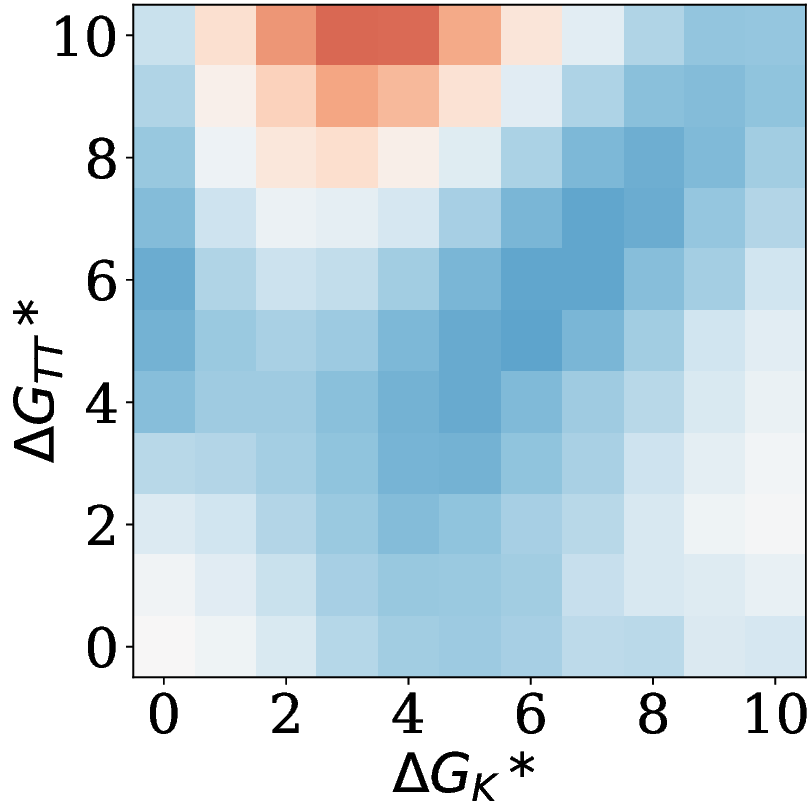}
      \caption{\label{fig:heatmapa}}
    \end{subfigure}
    \begin{subfigure}[b]{0.28\textwidth}
        \centering
      \includegraphics[width=1.0\linewidth]{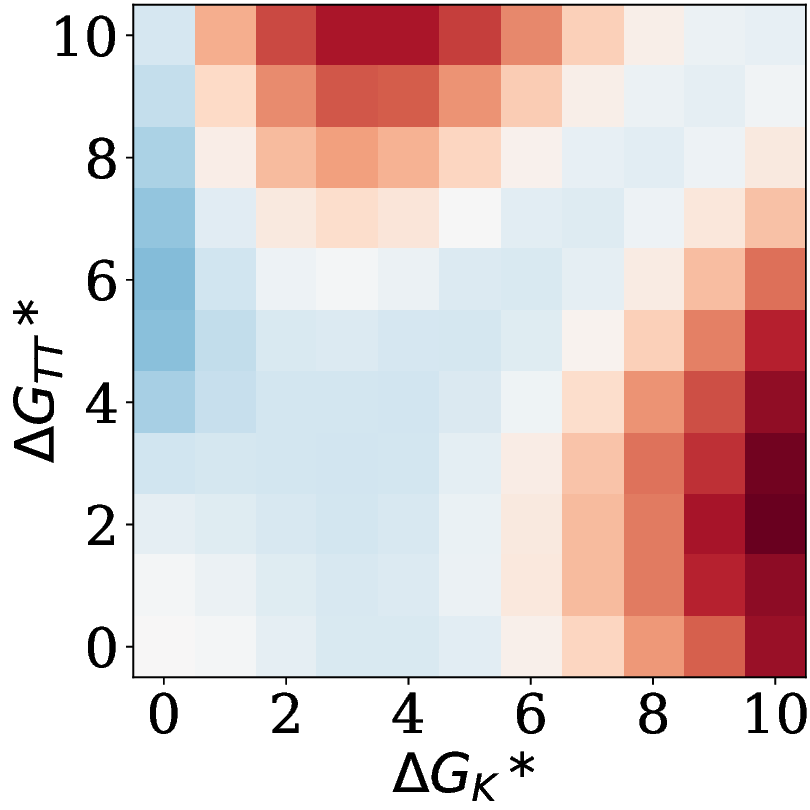}
      \caption{\label{fig:heatmapb}}
    \end{subfigure} & 
    \begin{subfigure}[b]{0.28\textwidth}
        \centering
      \includegraphics[width=1.0\linewidth]{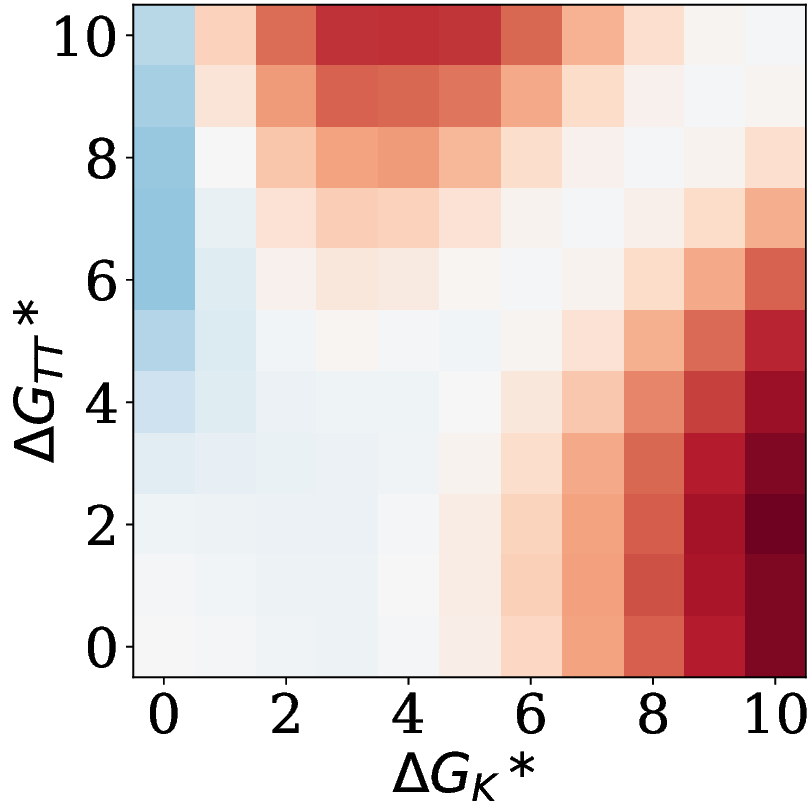}
      \caption{\label{fig:heatmapc}}
    \end{subfigure} & 
        
    \\
    \begin{subfigure}[b]{0.28\textwidth}
      \includegraphics[width=1.0\linewidth]{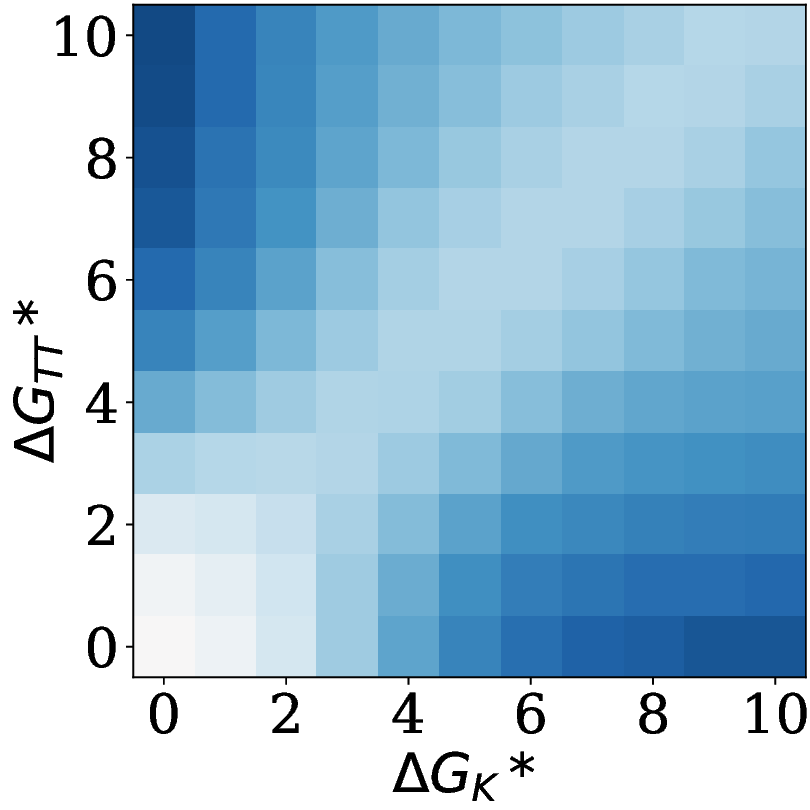}
      \caption{\label{fig:heatmapd}}
    \end{subfigure}
    \begin{subfigure}[b]{0.28\textwidth}
        \centering
      \includegraphics[width=1.0\linewidth]{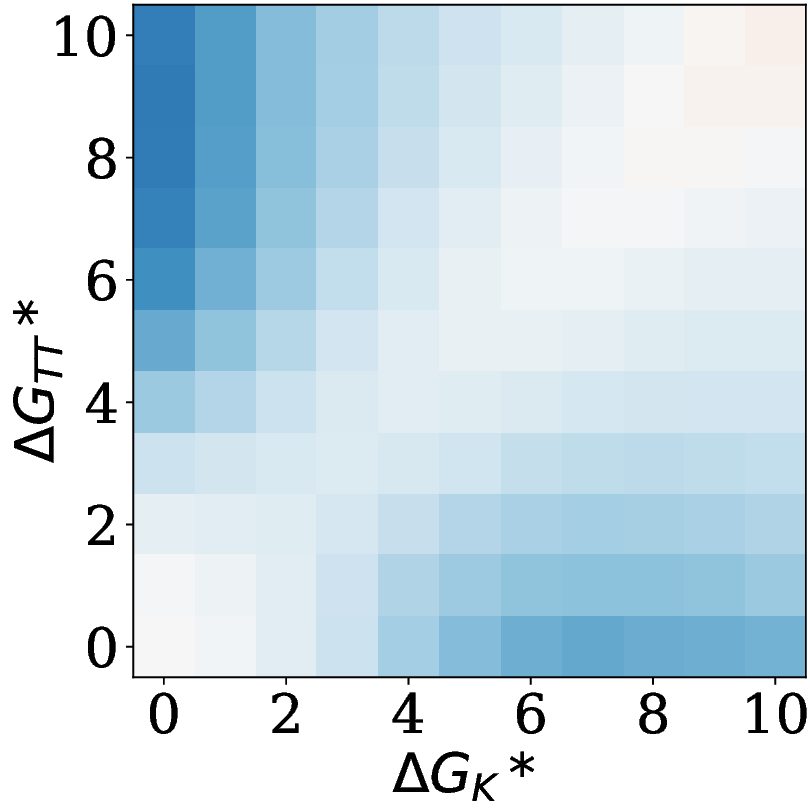}
      \caption{\label{fig:heatmape}}
    \end{subfigure} & 
    \begin{subfigure}[b]{0.28\textwidth}
        \centering
      \includegraphics[width=1.0\linewidth]{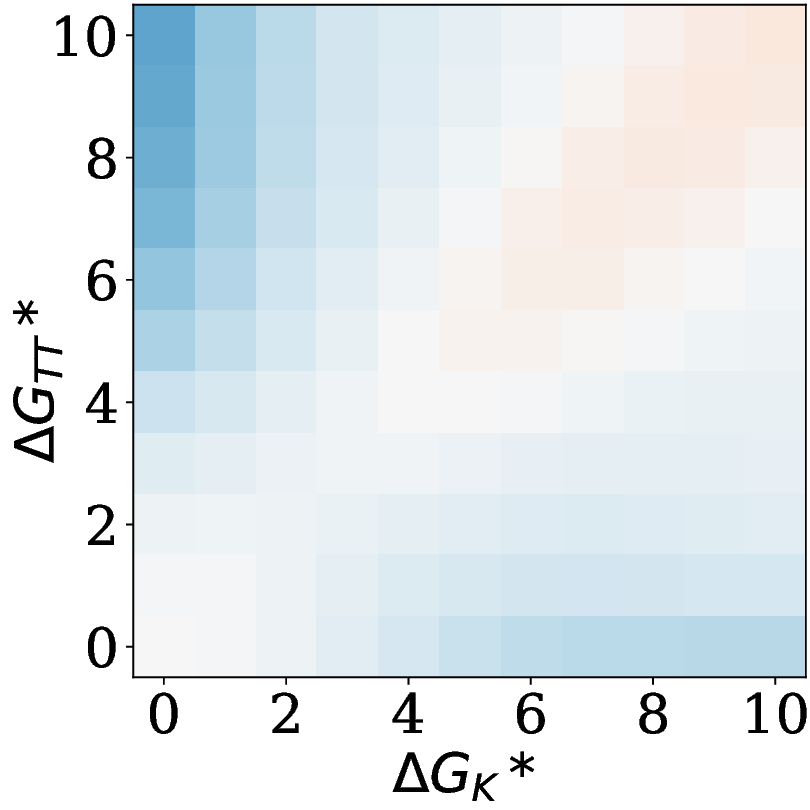}
      \caption{\label{fig:heatmapf}}
    \end{subfigure} &
    \begin{subfigure}[t]{0.14\textwidth}
          \includegraphics[width=0.85\linewidth]{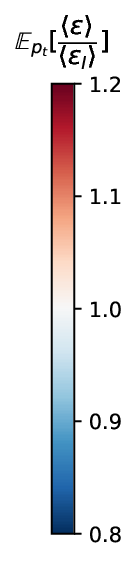}
        \end{subfigure}

\end{tabular}
\caption{ Identifying regimes where heterogeneity has large impacts on error. $\mathbb{E}_{p_t}[\frac{\langle \epsilon \rangle}{\langle \epsilon_I \rangle}]$ is plotted for parameter sets indexed $(i)$, $(ii)$ and $(iii)$ in Table \ref{table:CombinedParams}, for $\Delta G_{\textnormal{pol}} = 0$ (a, b, c) and $\Delta G_{\textnormal{pol}} = -0.3$ (d, e, f). When $\Delta G_{\textnormal{pol}} = 0$, the greatest decreases in error tend to occur when $\Delta G_K^* \approx \Delta G_{TT}^*$ for parameter set $(i)$, and $\Delta G_{K}^* = 0$ for parameter sets $(ii)$ and $(iii)$. When $\Delta G_{\textnormal{pol}} = -0.3$, the greatest decreases in error tend to occur when $\Delta G_K^* = 0$ or $\Delta G_{TT}^* = 0$.
 \label{fig:heatmap}}
\end{figure*}

We list our constrained parameter sets in Table \ref{table:CombinedParams}. For each constrained set, we allow for baseline amounts of forward and backward propensity discrimination $\Delta G_K^*$ and $\Delta G_{TT}^*$ to vary independently. Two values of $\Delta G_{\textnormal{pol}}$ are considered, $0$ and $-0.3$ (note that $-0.3$ is weaker driving).  For each parameter set, we consider the average error ratio $\mathbb{E}_{p_t}[{\frac{\langle \epsilon \rangle}{\langle \epsilon_I \rangle}}]$ (here the expectation is taken over a uniform distribution of the parameter $p_t$), a measure of how much heterogeneity helps decrease errors averaged over the content of monomer $2$ for Bernoulli templates. Sweeping over different baseline values $G_K^*$ and $G_{TT}^*$, we obtain heat maps of $\mathbb{E}_{p_t}[{\frac{\langle \epsilon \rangle}{\langle \epsilon_I \rangle}}]$, plotted in figure~\ref{fig:heatmap}.

Consider first $\Delta G_{\textnormal{pol}} = 0$. We observe that in the limit of pure opposite discrimination (i.e., one monomer is purely forward propensity discriminated, and the other purely backward propensity discriminated), the largest decreases in error tend to be obtained when backward and forward propensity discrimination is present at roughly similar amounts. However, as discrimination is shifted in parameter sets ii and iii, $\mathbb{E}_{p_t}[\frac{\langle \epsilon \rangle}{\langle \epsilon_I \rangle}]$ tends to increase in this region (Figures \ref{fig:heatmapa}-\ref{fig:heatmapc}). This response is
consistent with what we saw in Section \ref{sec:ErrorTime}, since shifting the discrimination results in a smoothing of the incorrect monomer potential. In the best cases, heterogeneity tends to decrease errors by about $10$ to $20$ percent (averaged over all $p_t$) relative to the na\"{i}ve, uncorrelated estimate of the error probability $\epsilon_I$. On the other hand, for the lower value of $\Delta G_{\textnormal{pol}} = -0.3$, beneficial effects tend to peak when one monomer is purely backward or forward propensity-discriminated, while the other experiences no discrimination at all (Figures \ref{fig:heatmapd}-\ref{fig:heatmapf}).

\section{Markov Approximations for Entropy and Information \label{app:EntropyInfo}}

Applying an $i^{th}$-order $l$-independent Markov assumption on $Y$, having marginalised over X, we get the following joint probability distribution:
\begin{align}
    p(&m_{l-i},...m_{l-1},m_l) \nonumber\\ &= \Sigma_{\mathbf{x}} p(m_{l-i},...m_{l-1},m_l|\mathbf{x})p(\mathbf{x})\nonumber\\
    &= \lim_{L \rightarrow \infty}\frac{1}{L}\Sigma_{l=1}^L p(m_{l-i},...m_{l-1},m_l|\mathbf{x}) \nonumber\\
    &= \lim_{L \rightarrow \infty}\frac{1}{L}\Sigma_{l=1}^L p(m_{l-i}|\mathbf{x}) \Pi_{j=l-i}^{l-1} p(m_{j+1}|m_{j},\mathbf{x}). \\
\intertext{We have applied a self-averaging assumption here, which follows from our $i^{th}$-order Markov assumption.  Similarly,}
p(&m_l|m_{l-i},...m_{l-1}) \nonumber\\&= \Sigma_{\mathbf{x}} p(m_l|m_{l-1},\mathbf{x})p(\mathbf{x}|m_{l-i},...m_{l-1})\nonumber\\
    &= \Sigma_{\mathbf{x}} p(m_l|m_{l-1},\mathbf{x})\frac{p(m_{l-i},...m_{l-1}|\mathbf{x})p(\mathbf{x})}{\Sigma_{\mathbf{x}} p(m_{l-i},...m_{l-1}|\mathbf{x})p(\mathbf{x})} \nonumber\\
    &= \frac{\lim_{L \rightarrow \infty}\Sigma_{l=1}^L  p(m_l|m_{l-1},\mathbf{x})p(m_{l-i},...m_{l-1},m_l|\mathbf{x})}{\lim_{L \rightarrow \infty}\Sigma_{l=1}^L p(m_{l-i},...m_{l-1},m_l|\mathbf{x})} \nonumber\\
    &=\frac{\lim_{L \rightarrow \infty}\Sigma_{l=1}^L p(m_l|m_{l-1},\mathbf{x})p(m_{l-i},...m_{l-1},m_l|\mathbf{x})}{\Sigma_{m}\lim_{L \rightarrow \infty}\Sigma_{l=1}^L p(m|m_{l-1},\mathbf{x}) p(m_{l-i},...m_{l-1},m_l|\mathbf{x})} \label{eq:stationaryEst}
.\end{align}

\noindent The final step of equation \ref{eq:stationaryEst} is based on the trivial identity $\Sigma_{m} p(m|m_{l-1},\mathbf{x}) = 1$. It is included simply to avoid small deviations in the sums of estimated probabilities away from $1$. 

\begin{figure*}
\begin{tabular}[t]{cc}
    \begin{subfigure}[t]{0.39\textwidth}
      \includegraphics[width=1.0\linewidth]{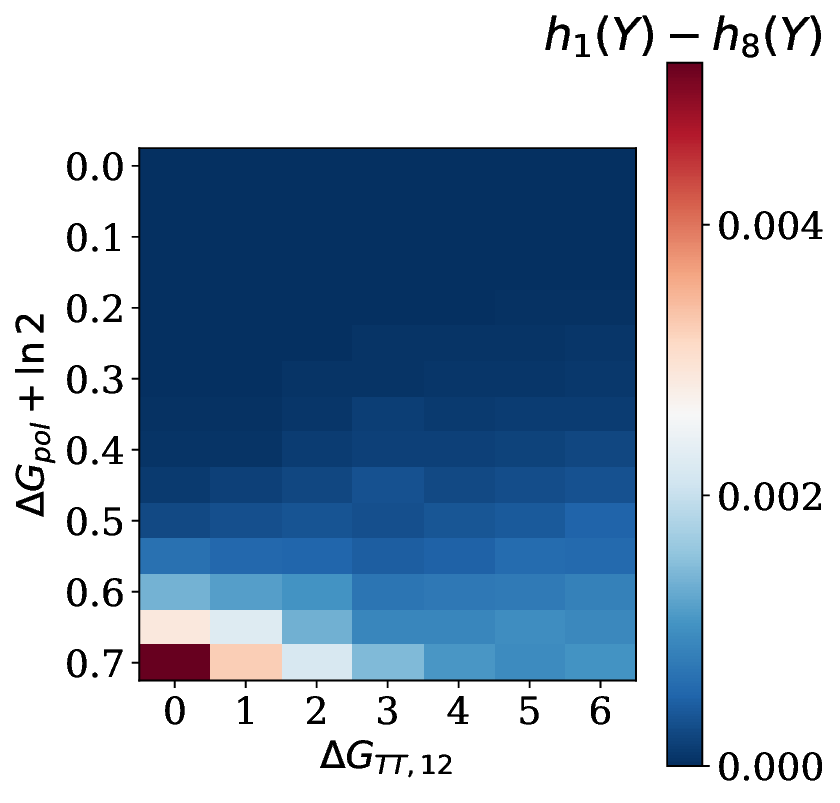}
      \caption{\label{fig:mutinfoerror.therm}}
    \end{subfigure}
    \begin{subfigure}[t]{0.39\textwidth}
        \centering
      \includegraphics[width=1.0\linewidth]{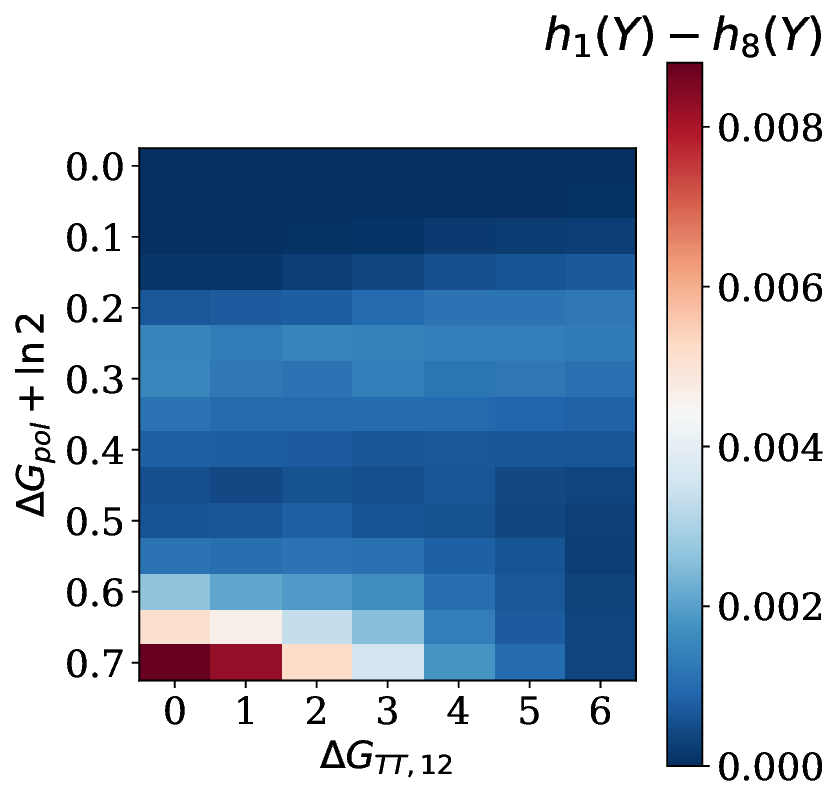}
      \caption{\label{fig:mutinfoerror.comb}}
      \end{subfigure}
      
\end{tabular}
\caption{\label{fig:mutinfoerror} Heat maps illustrating the difference in estimated entropy $h(Y)$ assuming a $1^{\rm st}$ versus $8^{\rm th}$ order Markov $Y$ for (a) backward propensity and (b) forward propensity discrimination, showing insignificant differences throughout. }
\end{figure*}

\begin{figure*}
\begin{tabular}[t]{cc}
    \begin{subfigure}[t]{0.48\textwidth}
      \includegraphics[width=1.0\linewidth]{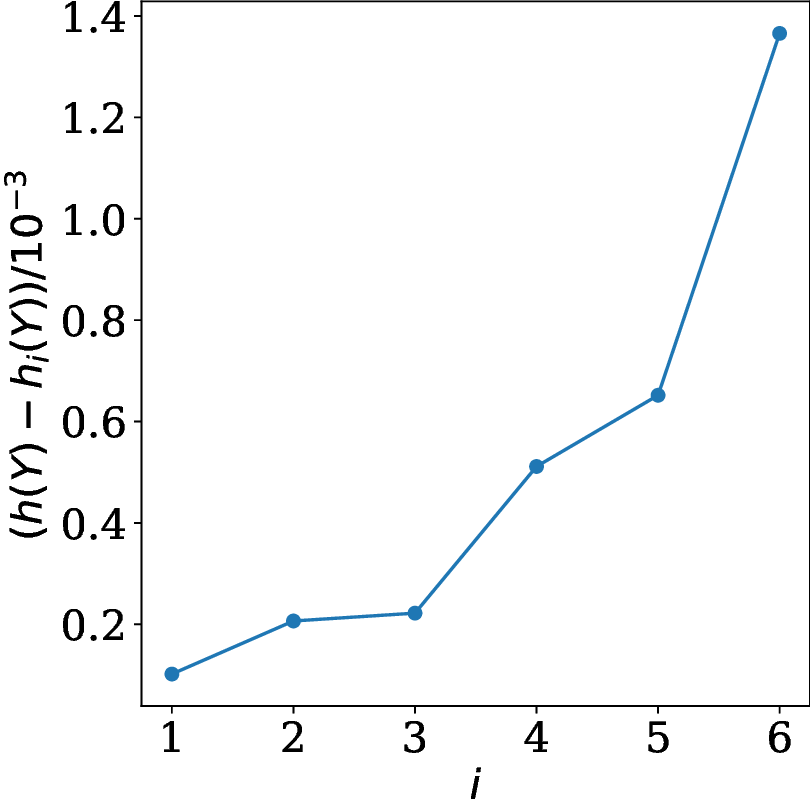}
      \caption{}
    \end{subfigure}
    \begin{subfigure}[t]{0.48\textwidth}
        \centering
      \includegraphics[width=1.0\linewidth]{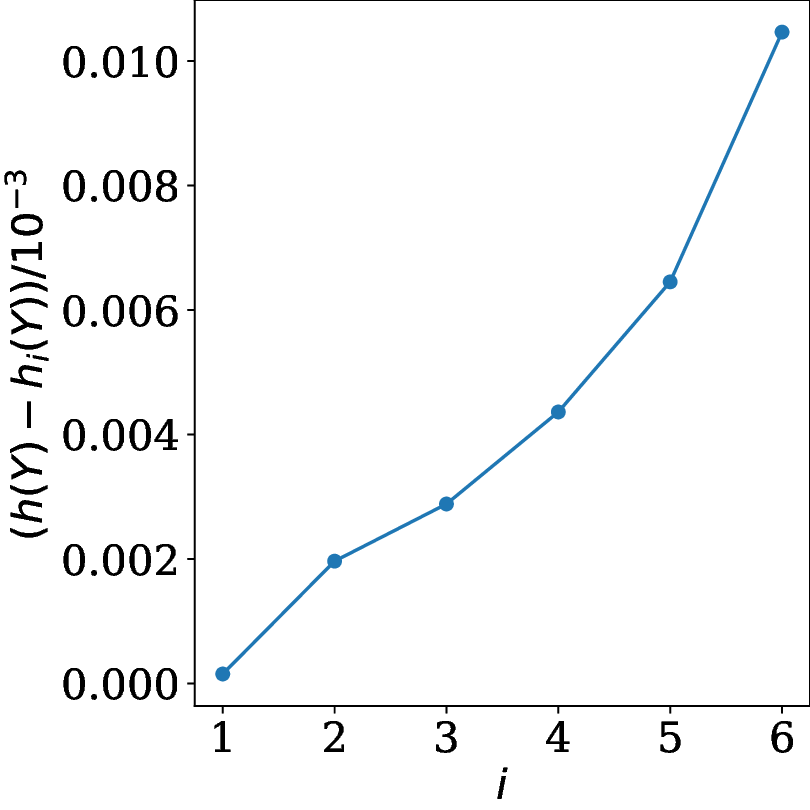}
      \caption{}
      \end{subfigure}
      
\end{tabular}
\caption{\label{fig:HomogeneousMutInfoError} Entropy estimation error for $h(Y)-h_i(Y)$ for a a homogeneous backward propensity-discriminated copying system. Estimation error is plotted for (a) $L = 10^4$ and (b) $L = 10^6$. Parameters are $\Delta G_{TT,11} = \Delta G_{TT,22} = 6$, and $p_t = 0.5$. We observe error increase due to taking higher order $i$. However, the error decreases massively going from $L = 10^4$ to  $L = 10^6$, implying that sampling is the source of this error. }
\end{figure*}

Differences in estimates of $h(Y)$ assuming a $1^{st}$ order Markov chain, $h_1(Y)$, and an $8^{th}$ order Markov chain, $h_8(Y)$, for $\Delta G_{TT,21} = 0$ and $\Delta G_{TT,11} = \Delta G_{TT,22} = 6$, maximized over $p_t$ for each point, are presented in Figure \ref{fig:mutinfoerror}. The differences are not significant, suggesting that $h_1(Y)$ is a tight estimate of $h(Y)$. Strangely, errors seem to be largest in the homogeneous case $\Delta G_{\textnormal{pol}} = 0$ and $\Delta G_{TT,12} = 0$ at $p_t = 0.5$ where $Y$ should be Markov and $h(Y)$ should be exactly $\ln{2}$ due to symmetry. However, for this parameter set it appears that sampling errors are the primary contributor to the difference $h_1(Y)-h_8(Y)$. To better illustrate the effect of sampling, we plot the estimation error $h(Y)-h_i(Y)$ for both $L = 10^4$ and $L = 10^6$ in Figure \ref{fig:HomogeneousMutInfoError}, for a a homogeneous backward propensity-discriminated copying system with $p_t = 0.5$ (note $h(Y)$ is exactly calculable: $h(Y) = \ln{2}$ in this regime). Here, $h_i(Y)$ is the entropy estimated by assuming an $i^{th}$ order Markov process. Strangely the estimation error actually increases with increasing $i$. However, as the overwhelming majority of the difference vanishes for $L = 10^6$, this estimation error is likely due to sampling (genuine errors due to correlations in $Y$ should persist in the $L \rightarrow \infty$ limit). 

Note that we estimate entropy by considering the probability of follow-up monomers conditioned on previous length $i$ strings (Equation \ref{eq:stationaryEst}). As $i$ increases for a fixed $L$, we expect the estimate of this conditional probability to become worse as there are fewer samples of each length $i$ string, leading to undersampling. This homogeneous case is unique in that we can attribute {\it all} errors to sampling; for most parameters the error in entropy is likely some combination of sampling errors (these errors likely get worse with increasing $i$) and genuine correlations in $Y$ (these errors likely get better with increasing $i$).

\bibliography{apssamp}% Produces the bibliography via BibTeX.

\end{document}